\documentclass[aps,twocolumn,prd,superscriptaddress]{revtex4-2}

\usepackage[utf8]{inputenc}

\usepackage{mathtools}
\usepackage{amsfonts}
\usepackage{mathrsfs}
\usepackage{bbm}
\usepackage{slashed}
\usepackage{tensor}

\usepackage{graphicx}
\usepackage{array}

\usepackage{placeins}
\usepackage{makecell}
\usepackage{subcaption}
\usepackage{float}

\usepackage{xspace}
\usepackage{xfrac}
\usepackage{hyperref}
\usepackage[nameinlink]{cleveref}
\usepackage{appendix}
\usepackage{units}

\usepackage{url}

\usepackage{xifthen}
\usepackage[dvipsnames]{xcolor}
\hypersetup{
	colorlinks,
	linkcolor={red!75!black},
	citecolor={blue!75!black},
	urlcolor={blue!75!black}
}

\usepackage{booktabs}
\usepackage{multirow}
\newcommand{\ra}[1]{\renewcommand{\arraystretch}{#1}}
\newcolumntype{C}{>{$}c<{$}}
\AtBeginDocument{
	\heavyrulewidth=.08em
	\lightrulewidth=.05em
	\cmidrulewidth=.03em
	\belowrulesep=.65ex
	\belowbottomsep=0pt
	\aboverulesep=.4ex
	\abovetopsep=0pt
	\cmidrulesep=\doublerulesep
	\cmidrulekern=.5em
	\defaultaddspace=.5em
}

\newcommand{\tinytext}[1]{\text{\tiny{#1}}}

\newcommand{\tierS}{{\textcolor{ForestGreen}{\large S}}}
\newcommand{\tierA}{{\textcolor{BlueViolet}{\large A}}}
\newcommand{\tierB}{{\textcolor{Turquoise}{\large B}}}
\newcommand{\tierC}{{\textcolor{BurntOrange}{\large C}}}
\newcommand{\tierD}{{\textcolor{BrickRed}{\large D}}}

\graphicspath{{./figures/}}

\newcommand{\gettitle}{Numerical RG-time integration of the effective potential: Analysis and Benchmark}

\newcommand{\getDarmstadtAffiliation}{\affiliation{Institut für Kernphysik, Technische Universität Darmstadt, D-64289 Darmstadt, Germany}}
\newcommand{\getHeidelbergAffiliation}{\affiliation{Institut für Theoretische Physik, Universität Heidelberg, Philosophenweg 16, D-69120 Heidelberg, Germany}}

\hypersetup{
	pdftitle={\gettitle},
	pdfauthor={Ihssen, Sattler, Wink},
	pdfkeywords={convection dominated}
	{functional renormalization group} {effective potential}
	{RG-time evolution} {numerical methods} {benchmark},
	bookmarksopen=true,
	bookmarksopenlevel=2,
	bookmarksnumbered=true
}

\begin{document}
\jot=0.5\baselineskip

\title{\gettitle}

\author{Friederike Ihssen}\getHeidelbergAffiliation
\author{Franz R. Sattler}\getHeidelbergAffiliation
\author{Nicolas Wink}\thanks{nicolas.wink@tu-darmstadt.de}\getDarmstadtAffiliation

\begin{abstract}
	We investigate the RG-time integration of the effective potential in the functional renormalization group in the presence of spontaneous symmetry breaking and its subsequent convexity restoration on the example of a scalar theory in $d=3$. The features of this setup are common to many physical models and our results are, therefore, directly applicable to a variety of situations.
    We provide exhaustive work-precision benchmarks and numerical stability analyses by considering the combination of different discrete formulations of the flow equation and a large collection of different algorithms.
    The results are explained by using the different components entering the RG-time integration process and the eigenvalue structure of the discrete system.
    Particularly, the combination of Rosenbrock methods, implicit multistep methods or certain (diagonally) implicit Runge-Kutta methods with exact or autodiff Jacobians proves to be very potent.
    Furthermore, a reformulation in a logarithmic variable circumvents issues related to the singularity bound in the flat regime of the potential.
	\vspace*{35pt}
\end{abstract}

\maketitle

\section{Introduction}
\label{sec:Introduction}
Phase transitions and critical phenomena are an integral part of describing physics in relativistic quantum theories, such as quantum chromodynamics, electroweak theory or condensed matter systems. The emergence of these phenomena can be best understood by following the scale dependence of the physics, starting from microscopic descriptions and coarse-graining by integrating out fluctuations to understand and quantify collective phenomena at macroscopic scales. This process is encapsulated in renormalization group (RG) methods, which are thus a suitable choice for the study of criticality and scaling since they are adapted to the physics of scales.

One realization of such methods is the functional renormalization group (fRG), which constitutes a powerful method to deal in a non-perturbative manner with the scale integration of fluctuations.
For a recent review see~\cite{Dupuis:2020fhh}, therein a detailed account of current applications is given.
Its basic premise is to introduce a regularization scale $k$ below which fluctuations are suppressed due to a momentum-dependent mass insertion, the regulator. This scale is then lowered, including successive momentum scales in a Wilsonian manner until one reaches $k\rightarrow0$ in order to obtain the full theory with all fluctuations integrated out. Although the fRG is an exact method, it is necessary to use a truncation of the flow to turn the fRG flow equation into a numerically accessible tool.

Dealing with collective phenomena and phase transitions, one may employ a derivative expansion, i.e., a truncation up to some power of momenta in the 1PI effective action $\Gamma_k[\phi]$. In spite of neglecting higher momentum dependencies, derivative expansions are usually a good choice to explore these systems in the fRG, as they allow to focus on taking into account higher interaction processes, i.e., model the effective action for high orders in the fields, see e.g.,~\cite{Balog:2019rrg, Fu:2019hdw, DePolsi:2020pjk, Fu:2022gou}.

The resulting fRG equations are often remarkably similar for many different physical systems and feature common numerical difficulties, which enables one to explore often-appearing issues related to the RG-time integration within a minimal setup, whose basic features generalize to most more elaborate theories. In this work we choose such a setup and discuss the process of spontaneous symmetry breaking in a simple scalar theory with an interaction potential which has a full field dependence without truncation.

To assess the nature of common numerical difficulties, we analyze the fRG flow of this model up to leading order in the derivative expansion, the local potential approximation (LPA).
Furthermore, in such a setup, the flow equation reduces to a single partial differential equation (PDE), though more elaborate approximations may feature a larger system of equations.

A majority of the obstacles in the numerical solution of flow equations can be explained by the presence of convexity restoration of the 1PI effective action in the regime of broken symmetry. Suppose one has an effective potential $V(\phi)$, where $\phi$ is the expectation value of a scalar field. The equation of motion of the 1PI effective action singles out the minimum $\phi_0$ of $V(\phi)$ as the solution to the equation of motion and thus the physical point. Therefore, $\phi$ also serves as an order parameter and $\phi_0 = 0$ signifies the symmetric phase and $\phi_0>0$ the symmetry broken phase.
In the symmetry broken phase in the vicinity of a second order phase transition, the potential fulfills  $V(|\phi| \leq \phi_0) = \mathrm{const}$ due to the convexity of $\Gamma_{k\rightarrow0}[\phi]$, which has to be restored during the fRG flow of the model.
This region of the potential is called the flat part.
At any finite RG-scale/regularization, however, the potential in the flat region has a finite slope which approaches zero only asymptotically, leading to technical difficulties in actual implementations. 

This paper mostly deals with the numerical issues due to convexity restoration during the fRG flow of the effective potential. To this end, we apply a heuristically driven approach to investigate the issue of choosing an appropriate RG-time integration routine.

Starting from the method of lines, it has recently been worked out on a conceptual level how the effective potential in field direction should be discretized to achieve stability, see~\cite{Grossi:2019urj, Koenigstein:2021syz, Ihssen:2022xkr} and \cite{Wink:2020tnu, Grossi:2021ksl, Koenigstein:2021rxj, Steil:2021cbu, Stoll:2021ori, Ihssen:2022xjv, Koenigstein:phd} for applications thereof. In essentially all previous works, the choice of algorithm to evolve the system in RG-time was rather arbitrary. Notable exceptions are~\cite{Pelaez:2015nsa}, where the RG-time evolution was considered in a semi-analytical approach, and \cite{Borchardt:2016pif}, where a fully implicit space-time evolution based on Chebyshev polynomials was used. The latter approach, however, is plagued by stability issues, which are especially pronounced for the case considered in this work.

Crucially, it should also be noted, that the use of explicit timestepper methods is impractical, since the wave speed in the flat region of the potential diverges exponentially fast for large RG-times~\cite{Grossi:2019urj}. Following the Courant–Friedrichs–Lewy (CFL) condition, which is necessary for both stability and validity of the results, leads to exponentially small step sizes. As any explicit method will suffer from such a bound, this issue can only be solved by using implicit methods.
The aim of this paper is to start tackling the path to an informed choice and understanding of pertinence of numerical time integrators by surveying and analyzing the performance of commonly used algorithms. 

\section{The effective potential and its discretization}
\label{sec:spatial}

To keep matters simple, we look at the $\varphi^4_3$-theory, i.e., a single scalar field with $\mathbb{Z}_2$ symmetry, in the broken phase in three Euclidean space-time dimensions.
Working with the fRG, we are primarily aiming at the calculation of the scale dependent effective potential, for details and references see \Cref{app:fRG}.
To leading order in the derivative expansion, the quantum effective action it is given by
\begin{align}
\label{eq:effActionLPA}
    \Gamma_k[\phi] = \int_x \left\{ \frac{1}{2} (\partial_\mu \phi)^2 + V(t, \rho) \right\}
\, ,
\end{align}
where $\phi=<\!\!\varphi\!\!>$ denotes the expectation value of the field, $k$ is the RG-scale and $t=-\log(k/\Lambda)$ is the RG-time relative to the initial UV scale $\Lambda$. Concerning the field dependence, the effective potential $V(t, \rho)$ must be a function of the $\mathbb{Z}_2$ invariant $\rho = \phi^2/2$ only. Finally, the integral subscript $x$ denotes integration over the entire space-time in the usual fashion.

Using the Litim regulator $R_k(p) = (k^2-p^2) \Theta(k^2-p^2)$, we obtain the flow of the effective potential by inserting the ansatz of the effective action \labelcref{eq:effActionLPA} into the Wetterich equation \labelcref{eq:wetterich} and projecting on the effective potential
\begin{align}
\label{eq:flow_effpot}
    \partial_t V(t,\rho) = - A_d \frac{k^{d+2}}{k^2 + \partial_\rho V(t,\rho) + 2 \rho\, \partial_\rho^2  V(t,\rho)}
\, .
\end{align}
The prefactor $A_d = 2/d (2\pi)^{-d} \pi^{d/2} \Gamma(d/2)^{-1}$ in \labelcref{eq:flow_effpot} collects prefactors from the Wetterich equation and the momentum integration. Its sign is positive and only depends on the space-time dimension $d$. Throughout this work we use $d=3$, but note that the qualitative statements in the broken phase are independent of this choice for any $d>2$. 

Despite this simple truncation, numerical problems related to the RG-time integration are already present, and results generalize directly to other theories and more complicated settings.

Following the ideas of \cite{Grossi:2019urj} and subsequent works, the fundamental variable is given by the derivative of the effective potential
\begin{align}
\label{eq:u_var}
    u(t,\rho) = \partial_\rho V(t,\rho)
\, .
\end{align}
For simplicity, we will omit the arguments of functions if they are obvious in the current context. The equation for $u(t,\rho)$ is now conveniently formulated using the flux
\begin{align}
\label{eq:flux}
    f(t, m^2) = - A_d \frac{k^{d+2}}{k^2 + m^2}
\, ,
\end{align}
where the square of the (curvature) mass $m^2$ is given by
\begin{align}
\label{eq:mass}
    m^2(t, \rho) = u(t, \rho) + 2\rho\, \partial_\rho u(t,\rho)
\, .
\end{align}
The flow of the derivative of the effective potential $u(t,\rho)$ is then given by
\begin{align}
\label{eq:flow_u}
    \partial_t u = \partial_\rho\, f(t, m^2)
\, .
\end{align}
For future reference we also introduce the flow equation of the squared curvature mass
\begin{align}
\label{eq:flow_msq}
    \partial_t m^2 = (1+2\rho\, \partial_\rho) \partial_\rho\, f(t, m^2)
\, .
\end{align}
Lastly, we consider the logarithm of the regularized two-point function
\begin{align}
\label{eq:log_var}
    \varpi(t,\rho) = \log(k^2 + m^2)
\, ,
\end{align}
with the corresponding flow equation given by
\begin{align}
\label{eq:flow_log}
    \partial_t \varpi = -e^{-\varpi} \left[
    (1+2\rho\partial_\rho) \partial_\rho
    \left(A_d k^{d+2} e^{-\varpi} \right)
    - 2k^2 \right]
\, .
\end{align}

These flow equations are potentially interesting as they might be beneficial for numerical applications.
The flow of the squared mass $m^2$ is closer in its form to typical advection-diffusion equations, while the log of the regularized two-point functions makes the constraint $k^2 + m^2 > 0$ for all finite $k>0$ ($t<\infty$) manifest. Such logarithmic transformations are commonly employed in chemical and biological systems for precisely the same reason, as e.g., chemical concentrations can become arbitrarily small, but must be positive.
For the initial conditions considered in this work (quartic potentials), this constraint can be shown to hold for all RG-scales $k$, see \cite{Litim:2006nn} for a detailed discussion.

\subsection{Discretization}
\label{sec:discretizations}
To keep subsequent discussions simple, we employ a straight-forward grid discretization of the field space, and resort to a first order finite difference scheme, see also~\cite{Wink:2020tnu}.

We discretize the (finite) domain on a one dimensional, ordered grid,
\begin{align}
\label{eq:basic_grid}
    \{\rho_i\} = \left\{ \rho_1 = 0, \rho_2, \dots, \rho_{N_\rho} = \rho_\tinytext{max} \right\}
\, .
\end{align}
To shorten equations in the following, we introduce the shorthand notation $u_i = u(t, \rho_i)$, $m^2_i = m^2(t, \rho_i)$ and $\varpi_i = \varpi(t, \rho_i)$.

The flux \labelcref{eq:flux} is strictly negative. Hence, we can define the first order upwind and downwind derivative operators,
\begin{align}
\label{eq:deriv_operator}
    \mathcal{D}_{u(d)} a_i = \frac{a_{i\pm 1} - a_i}{\rho_{i\pm 1} - \rho_{i}}
\, ,
\end{align}
where $a$ serves as placeholder for any variable of interest and the $+$ sign has to be taken for the upwind operator $\mathcal{D}_{u}$ and the $-$ sign for the downwind operator $\mathcal{D}_{d}$.

We start by discretizing the standard formulation \labelcref{eq:flow_u}, and subsequently derive the discretization of the mass \labelcref{eq:flow_msq} and log \labelcref{eq:flow_log} version.
Following the ideas of \cite{Grossi:2019urj}, the derivative applied to the flux \labelcref{eq:flux} in \labelcref{eq:flow_u} has to be an upwind derivative, while the derivative in the term $(1 + 2\rho\, \partial_\rho)$ of \labelcref{eq:mass} has to be a downwind derivative.

To summarize, the advection part of the equation is discretized with an upwind derivative, while the diffusive part is, taking the asymmetry of $2\rho$ into account, discretized with a central derivative.

With this at hand, we can easily write down stable discretizations of three different versions \labelcref{eq:flow_u}, \labelcref{eq:flow_msq} and \labelcref{eq:flow_log}. To that end, it is beneficial to introduce the combined derivative operator,
\begin{align}
\label{eq:d_tilde}
    \tilde{\mathcal{D}} = (1+2\rho\, \mathcal{D}_d)\mathcal{D}_u
\, ,
\end{align}
which appears in two of the three equations.

In total, the discretized flow for the derivative of the effective potential~\labelcref{eq:flow_u} is given by,
\begin{align}
\label{eq:flow_ui}
    \partial_t u_i = \mathcal{D}_u\, f(t, (1+2\rho\, \mathcal{D}_{d}) u_i)
\, .
\end{align}
Utilizing the definition of the squared mass \labelcref{eq:mass}, we obtain the discretized version of \labelcref{eq:flow_msq},
\begin{align}
\label{eq:flow_msqi}
    \partial_t m_i^2 = \tilde{\mathcal{D}}\, f(t, m_i^2)
\, .
\end{align}
Lastly, using \labelcref{eq:log_var}, we obtain the discretized version of \labelcref{eq:flow_log}, the logarithm of the regularized two-point function
\begin{align}
\label{eq:flow_logi}
    \partial_t \varpi_i = -e^{-\varpi_i} \left[
    \tilde{\mathcal{D}}
    \left(A_d k^{d+2} e^{-\varpi_i} \right)
    + 2k^2 \right]
\, .
\end{align}
At this point we would like to emphasize that the discretization of these formulations is equivalent. Therefore, the notion of stability underlying \labelcref{eq:flow_ui} implies the same stability for \labelcref{eq:flow_msqi} and \labelcref{eq:flow_logi}. This, however, only applies to the integration in RG-time if a strong stability preserving (SSP) algorithm in combination with a CFL type step size condition is used. Consequently, it does not tell us anything about the potential performance or stability of implicit algorithms with adaptive timestepping.

\subsection{Initial \& boundary conditions}
\label{sec:init_boundary}
Investigating the convexity restoration of the potential is easiest done with a quartic UV potential. Consequently, we fix the initial potential to be
\begin{align}
\label{eq:uv_potential}
    V(t=0, \rho) = m_\Lambda^2 \rho + \frac{\lambda_\Lambda}{2} \rho^2
\, .
\end{align}
The coefficient $\lambda_\Lambda$ has to be positive for the potential to be bounded from below. On the other hand, the coefficient $m_\Lambda^2$ is chosen to be negative, such that the flow starts in the symmetry broken phase. Only then, the final potential can have a flat region, because the flow is symmetry restoring, see e.g.~\cite{Litim:2006nn}. We are only interested in this regime of the theory, since it is responsible for the numerical difficulties which we are investigating.
The other branch of solutions is, from this perspective, trivial.

Throughout this paper, we omit units, since we do not identify the calculations presented here with a specific physical system. Furthermore, we employ the freedom to rescale the field to fix $\lambda_\Lambda=1$.

We fix the initial UV scale to $\Lambda=7.5$, which is sufficiently large to incorporate the full dynamics related to convexity restoration. Lastly, to stay deeply within the symmetry broken regime in the limit of vanishing regularization ($t\to\infty$), we set $m^2_\Lambda = -2.5$.

Since we are interested in the flattening of the potential, which is entirely dominated by the infrared part of the flow, we will not pay any attention to any potential UV issues, such as RG consistency, see~\cite{Braun:2018svj} for more details thereon. This procedure is fully compatible with usual calculations in effective low-energy effective theories.

With these parameters, the minimum of the effective potential is initially at $\rho_0(t=0) = 2.5$ and freezes in at $\rho_0(t\to\infty) = 2.296$. To fully incorporate the associated dynamics, we chose $\rho_\tinytext{max} = 7.5$ and $N_\rho = 256$. These parameters correspond to realistic choices that we would use in associated studies of the phase structure of the theory, i.e., these are parameters of typical calculations.

Lastly, we have to specify the boundary conditions to complete the description of the discretization.
The left boundary of the computational domain is always located at $\rho = 0$, hence we need to specify the downwind derivate at this boundary.
In the current setup, we do not expect any (nor can they appear) irregularities at, $\rho = 0$, and the combination $2\rho\mathcal{D}_d$ can safely be set to zero.
In certain cases the situation might be more complicated, see~\cite{Koenigstein:2021syz} for a detailed discussion.
At the right boundary, $\rho_\text{\tiny max}$, the flow is sufficiently suppressed and upwinding does not matter~\cite{Grossi:2019urj}.
Hence, we can replace the upwind derivative at the boundary with a downwind derivative.
In summary,
\begin{equation}
\begin{aligned}
\label{eq:boundary_cond}
    2\rho\, \mathcal{D}_{d} &\rightarrow 0\phantom{\mathcal{D}_{d}} \quad \text{at}\ \rho = 0 \\
    \mathcal{D}_u &\rightarrow \mathcal{D}_{d}\phantom{0} \quad \text{at}\ \rho = \rho_\text{\tiny max}
\, .
\end{aligned}  
\end{equation}
The discussion above fully specifies the discretization of the field domain and results in a system of ordinary differential equations (ODEs).

\begin{figure*}[t]
	\centering
	\begin{subfigure}[t]{0.48\textwidth}
		\centering
		\includegraphics[width=\linewidth]{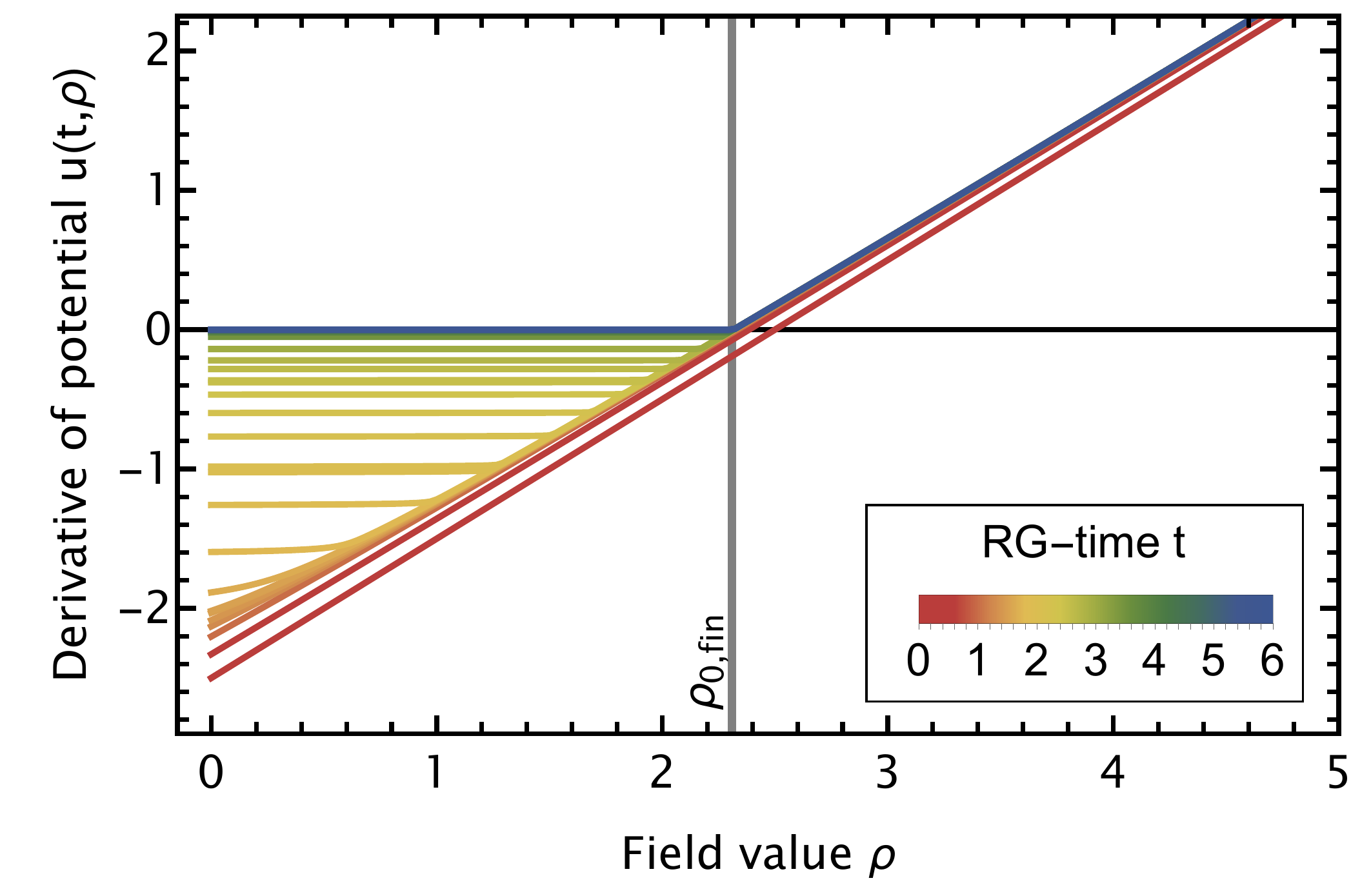}
	\end{subfigure}	~
	\begin{subfigure}[t]{0.48\textwidth}
		\centering
		\includegraphics[width=\linewidth]{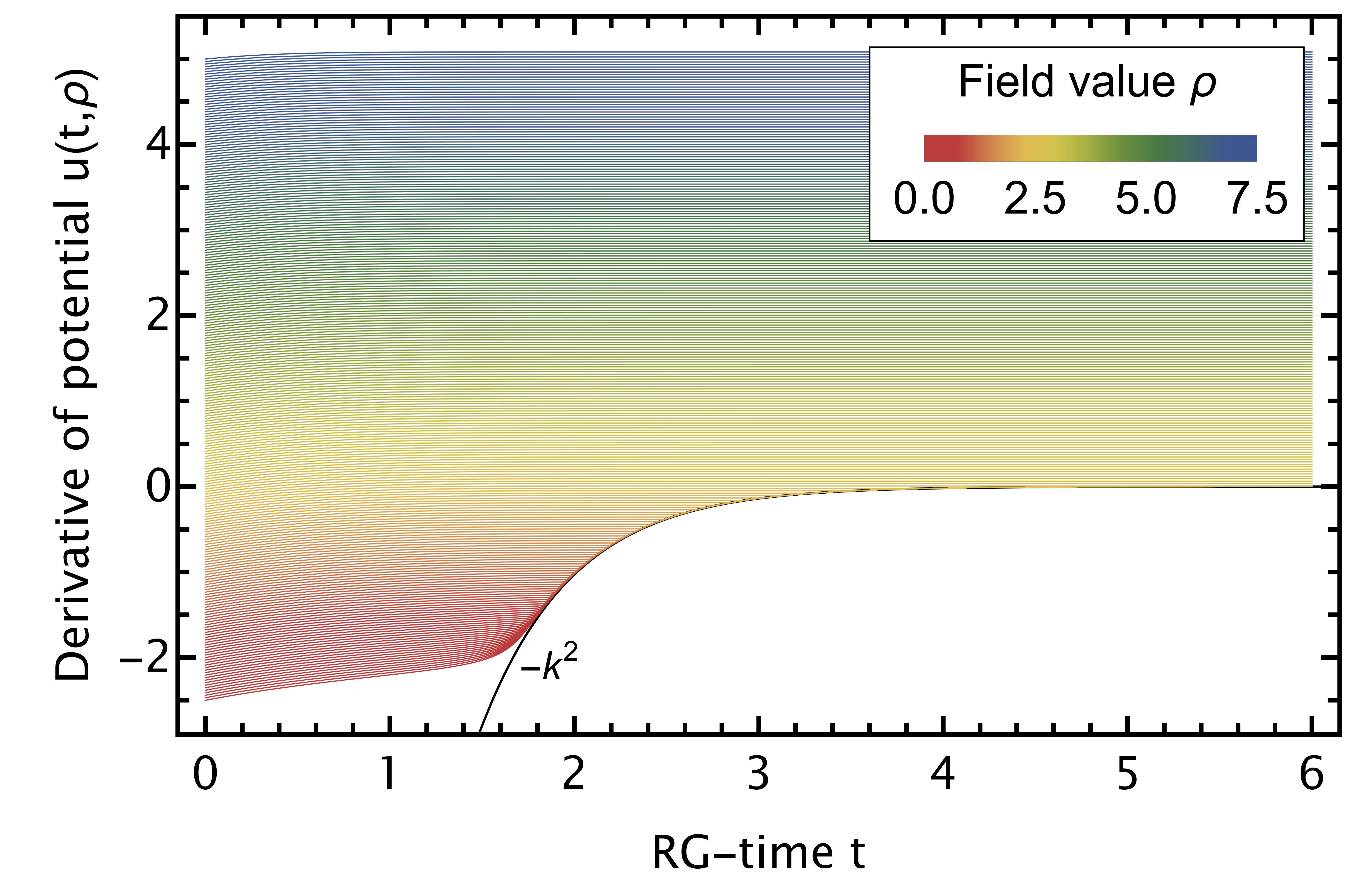}
	\end{subfigure}%
	\caption{Solution of the standard formulation \labelcref{eq:flow_ui} in the ($\rho$, $t$)-plane. In the left panel, the linearly interpolated $\rho$-dependent solution is shown for different RG-times $t$, while in the right panel the RG-time evolution of individual grid points is shown.}
	\label{fig:std_sol_lines}
\end{figure*}
%
\begin{figure*}[t]
	\centering
	\begin{subfigure}[t]{0.48\textwidth}
		\centering
		\includegraphics[width=\linewidth]{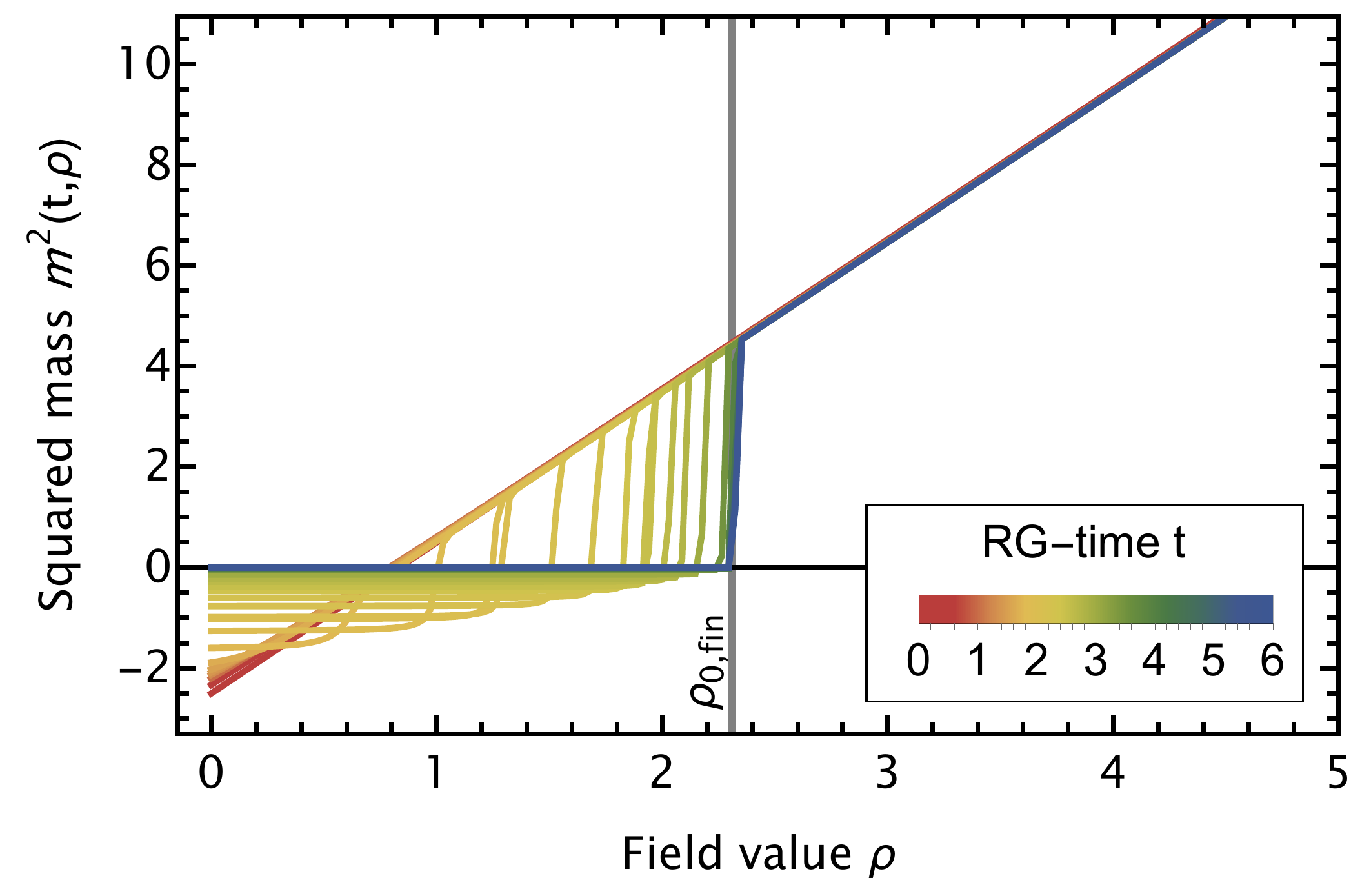}
	\end{subfigure}	~
	\begin{subfigure}[t]{0.48\textwidth}
		\centering
		\includegraphics[width=\linewidth]{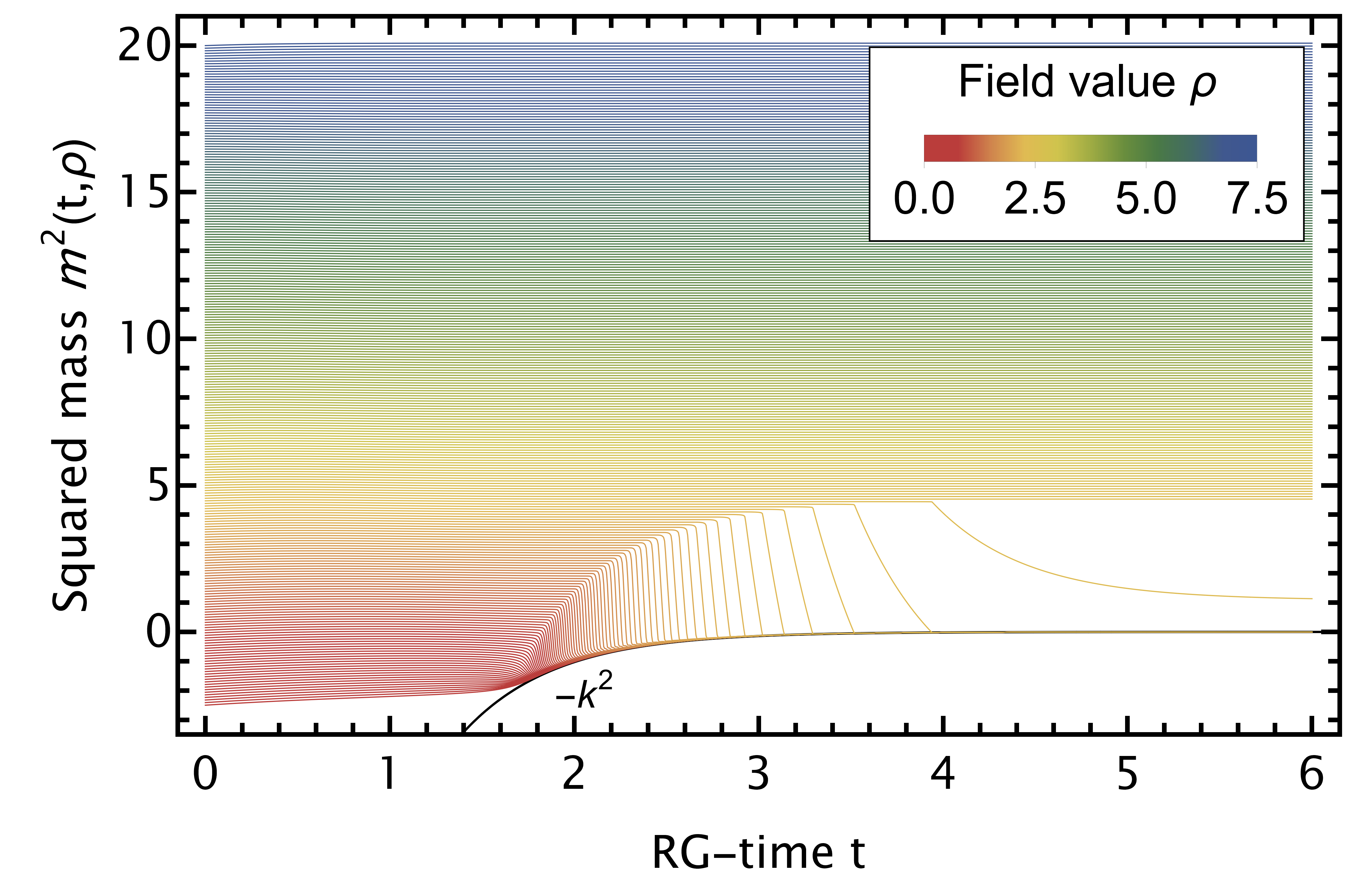}
	\end{subfigure}%
	\caption{Solution of the mass formulation \labelcref{eq:flow_msqi} in the ($\rho$, $t$)-plane. In the left panel, the linearly interpolated $\rho$-dependent solution is shown for different RG-times $t$, while in the right panel the RG-time evolution of individual grid points is shown. The shock is smeared due to the diffusive character of the spatial discretization. The effect thereof, a single grid point gets frozen inside the jump, can be clearly seen in the right panel.}
	\label{fig:msq_sol_lines}
\end{figure*}
%
\begin{figure*}[t]
	\centering
	\begin{subfigure}[t]{0.48\textwidth}
		\centering
		\includegraphics[width=\linewidth]{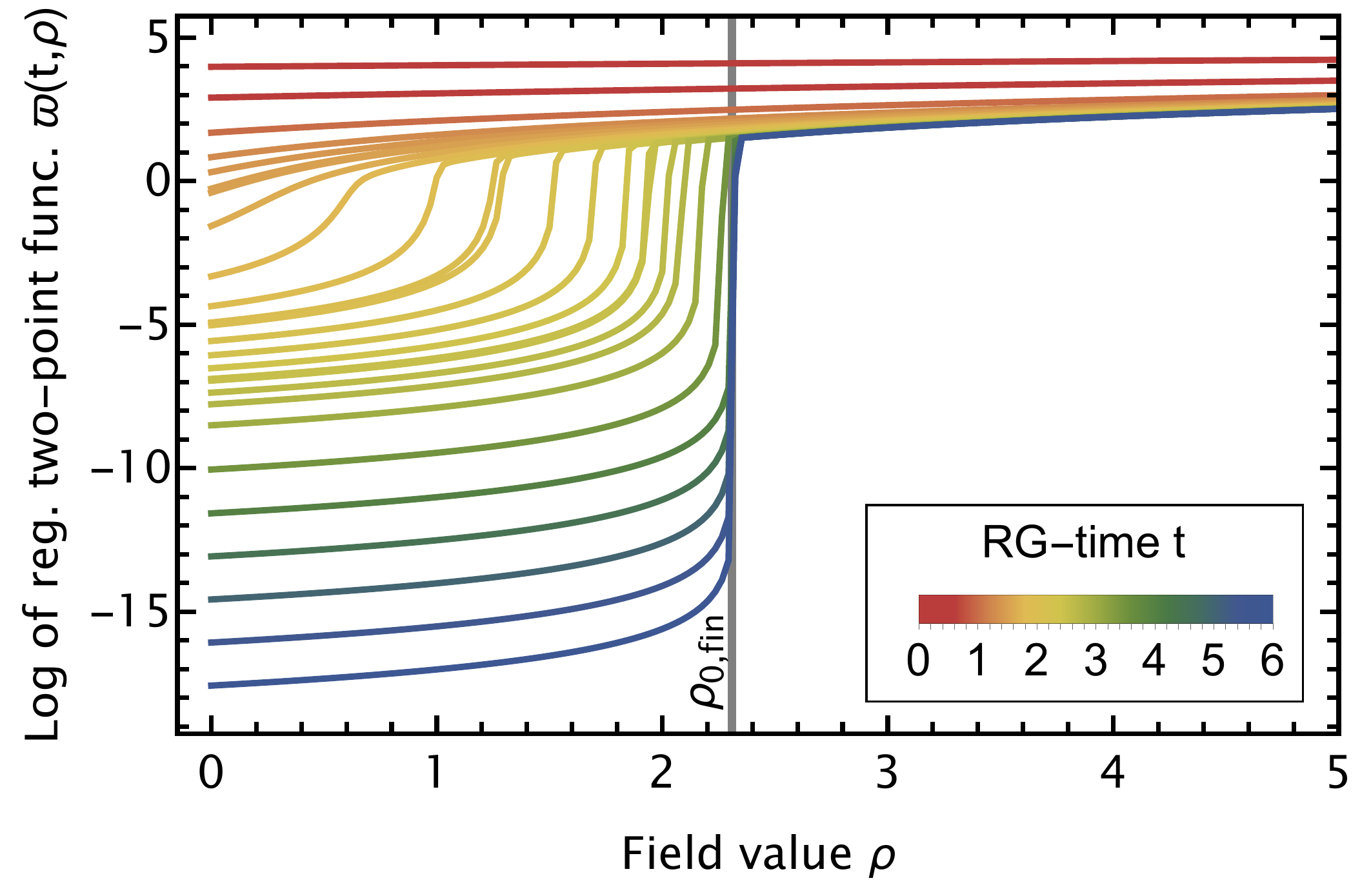}
	\end{subfigure}	~
	\begin{subfigure}[t]{0.48\textwidth}
		\centering
		\includegraphics[width=\linewidth]{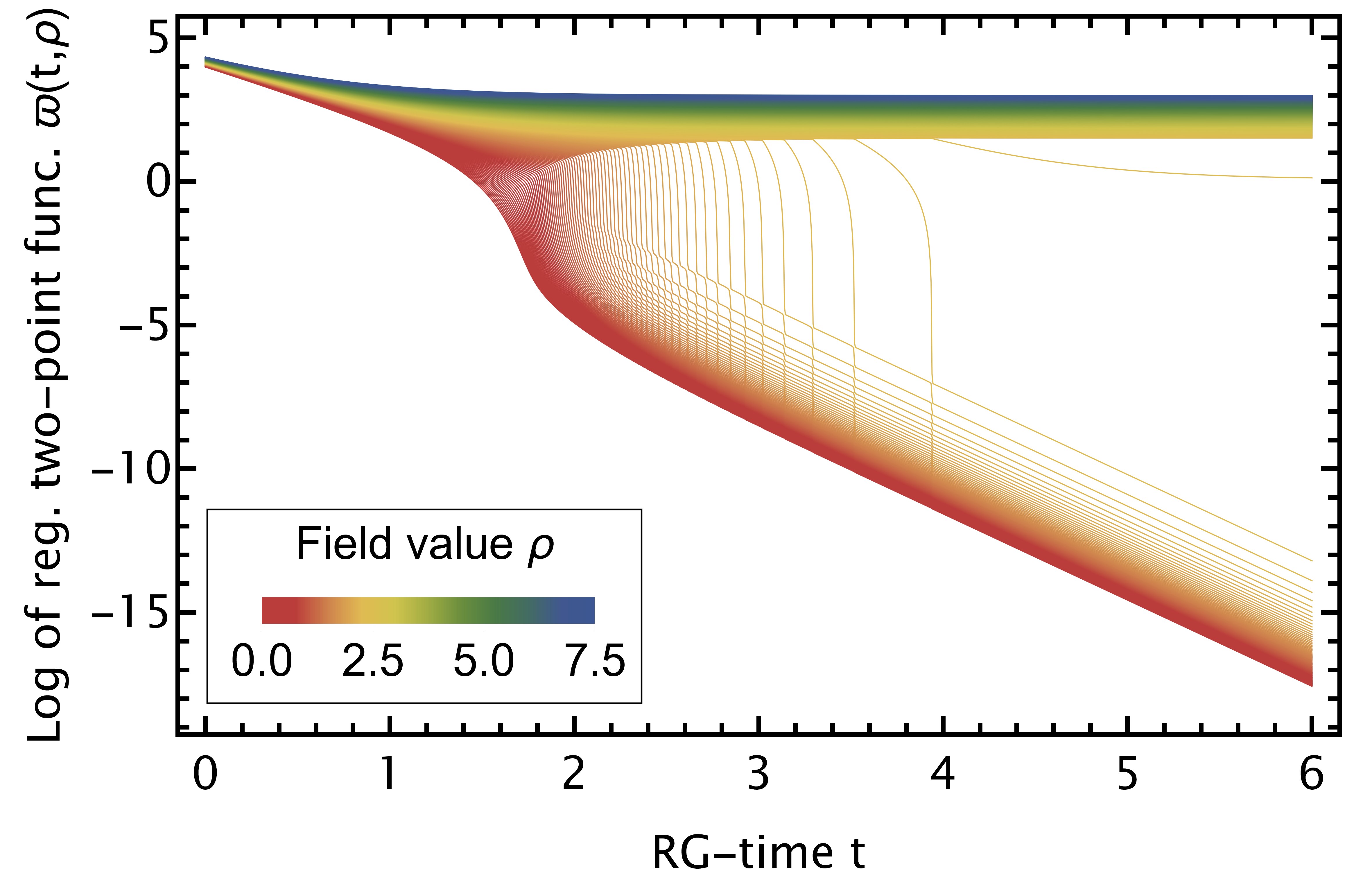}
	\end{subfigure}%
	\caption{Solution of the log formulation \labelcref{eq:flow_logi} in the ($\rho$, $t$)-plane. In the left panel, the linearly interpolated $\rho$-dependent solution is shown for different RG-times $t$, while in the right panel the RG-time evolution of individual grid points is shown. The shock is smeared due to the diffusive character of the spatial discretization. The effect thereof, a single grid point gets frozen inside the jump, can be clearly seen in the right panel.}
	\label{fig:log_sol_lines}
\end{figure*}
%
\begin{figure*}[t]
	\centering
	\begin{subfigure}[t]{0.48\textwidth}
		\centering
		\includegraphics[width=\linewidth]{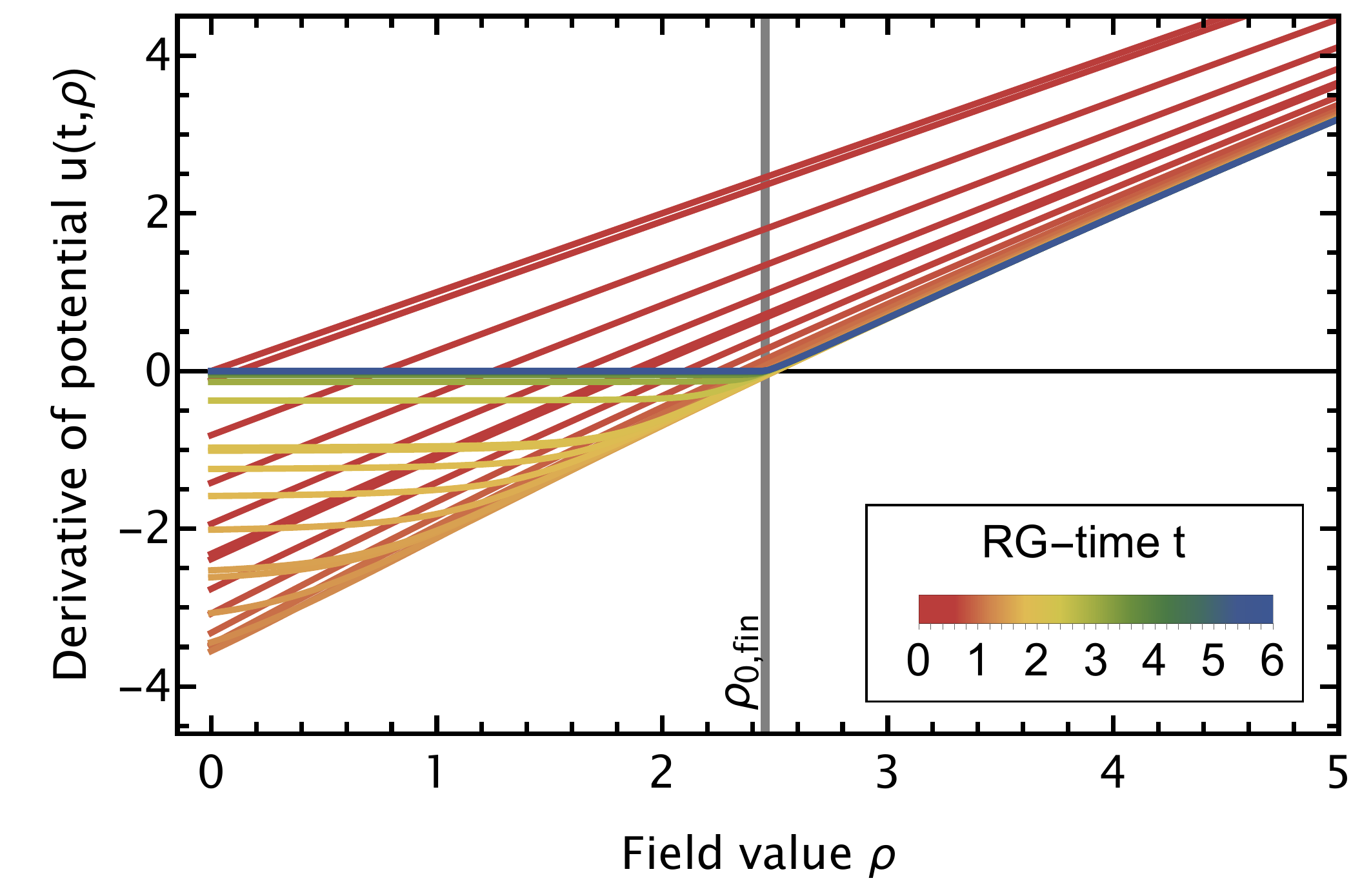}
	\end{subfigure}	~
	\begin{subfigure}[t]{0.48\textwidth}
		\centering
		\includegraphics[width=\linewidth]{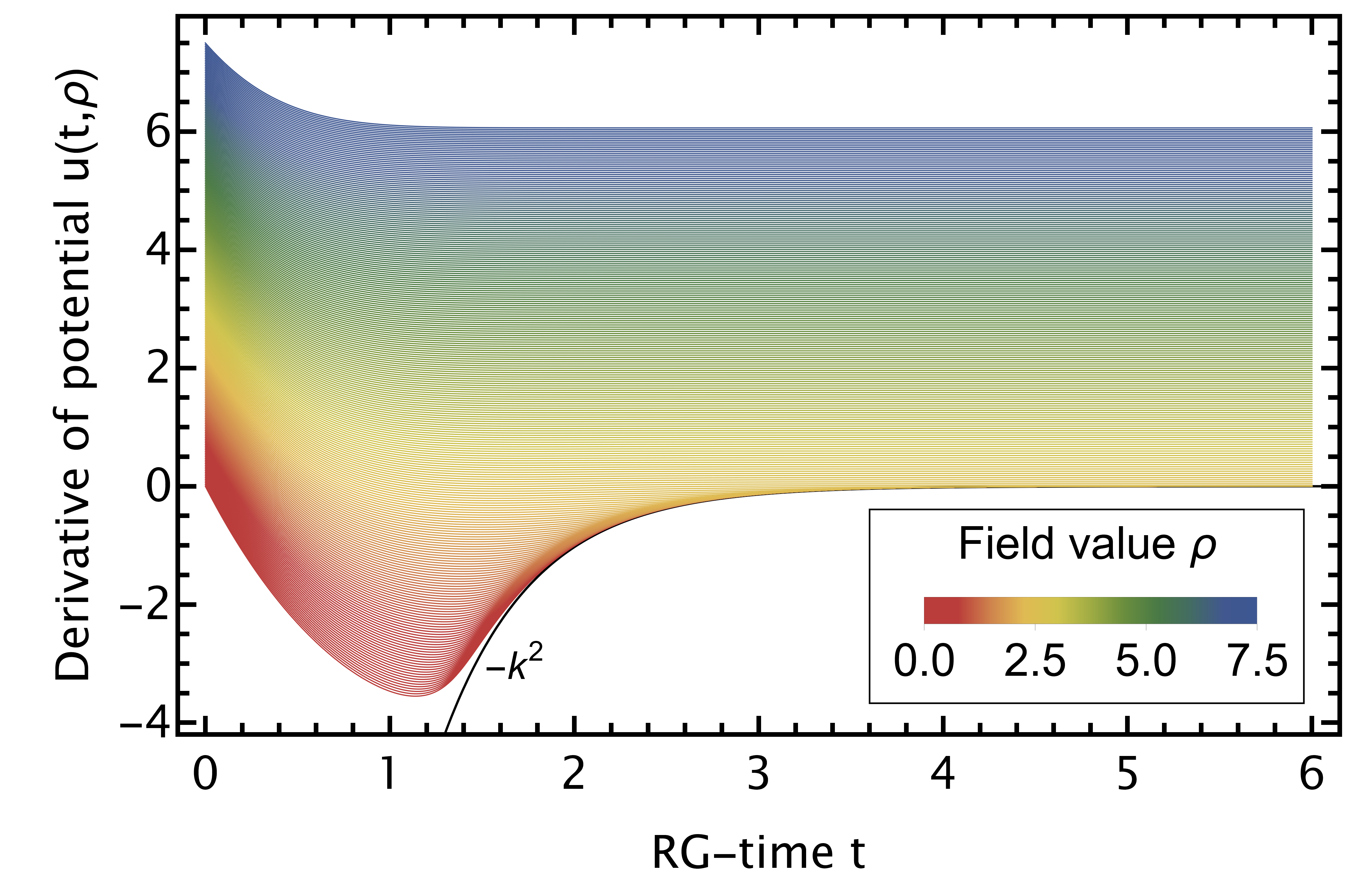}
	\end{subfigure}%
	\caption{Solution of the standard formulation in the case with fermionic extension \labelcref{eq:flow_effpotQM} in the ($\rho$, $t$)-plane. In the left panel, the linearly interpolated $\rho$-dependent solution is shown for different RG-times $t$, while in the right panel the RG-time evolution of individual grid points is shown.}
	\label{fig:QM_sol_lines}
\end{figure*}

\subsection{Fermionic extension}
\label{sec:fermions}
In principle, the general statements of this work should hold in general, since the mechanisms of convection dominated evolution are independent of the theory.
In order to test if our analysis is indeed robust against variations in the flow equation, we extend the system for this subsection with an additional quark-like term and increase the number of scalar fields from N=1 to N=4. The ansatz for the effective action \labelcref{eq:effActionLPA} changes to
\begin{align}
\label{eq:effActionLPAQM}
    \Gamma_k[\phi] = \int_x 
    \Big\{& \Bar{\psi}_{a,f} i\gamma_\mu\partial_\mu \psi_{a,f} + \frac{1}{2} (\partial_\mu \phi)^2 \\ \nonumber
    &+ \frac{h}{\sqrt{2}} \Bar{\psi}_{a,f}(i\sigma + \gamma_5 \vec\tau \cdot \vec\pi) \psi_{a,f} + V(t, \rho) \Big\}
\, ,
\end{align}
where the quark fields $\psi_{a,f}$ carry color indices $a=1,2,3$ and flavor indices $f=1,2$, while the Dirac indices are suppressed for simplicity. In the current context, the scalar field $\phi = (\sigma, \vec{\pi})$ has four components, in contrast to the $\mathbb{Z}_2$ theory of the previous section with one component. The scalars and fermions are coupled through a Yukawa interaction $h$, where we used the Pauli matrices $\tau_i$. For a detailed discussion of this model, called the Quark-Meson model, see e.g., \cite{Jungnickel:1995fp, Schaefer:2004en, Alkofer:2018guy}. This extension is an oft-used low energy effective theory of QCD which allows to make predictions about the phase structure of QCD in low-energy settings and thus constitutes another physically relevant model.

From the effective action \labelcref{eq:effActionLPAQM} we get, with the Litim regulator, a similar flow equation as for the $\mathbb{Z}_2$ case
\begin{align}
\label{eq:flow_effpotQM}
    &\partial_t V(t,\rho) = - A_d k^{d+2} \Bigg( \frac{3}{k^2 + \partial_\rho V(t,\rho)}   \\ \nonumber
    &+ \frac{1}{k^2 + \partial_\rho V(t,\rho) + 2 \rho\, \partial_\rho^2  V(t,\rho)} 
    - \frac{24}{k^2 + h^2 \rho}
    \Bigg)
\, .
\end{align}
Compared to \labelcref{eq:flow_effpot}, we get an additional term for the Goldstone degrees of freedom, similar to the one of the radial mode but purely advective, i.e., it contains no diffusion. Furthermore, there is a fermionic contribution, which acts effectively as a source term in the equation, see~\cite{Wink:2020tnu, Grossi:2021ksl, Stoll:2021ori, Koenigstein:phd} for previous studies with fermions. 

In the current Quark-Meson case, we follow the same logic as for the $\mathbb{Z}_2$ theory and work with the derivative of the effective potential~\labelcref{eq:u_var}.
The discretization of the equation then proceeds in the exact same fashion as outline in \Cref{sec:discretizations}, and leads to an equation almost identical to \labelcref{eq:flow_ui},
\begin{align}
\label{eq:flow_ui_QM}
    \partial_t u_i = \mathcal{D}_u\, f_\text{QM}(t, u_i, (1+2\rho\, \mathcal{D}_{d}) u_i)
\, ,
\end{align}
where the QM flux reads
\begin{align}
\label{eq:flux_QM}
    f(t, m_\pi^2, m_\sigma^2) = - A_d k^{d+2} \Bigg( \frac{3}{k^2 + m_\pi^2} &+ \frac{1}{k^2 + m_\sigma^2} \\ \nonumber
                                &- \frac{24}{k^2+h^2\rho} \Bigg)
\, ,
\end{align}
where we defined the pion and sigma masses as 
\begin{align}
    m^2_\pi(t,\rho) &= u(t,\rho) \\ \nonumber
    m^2_\pi(t,\rho) &= u(t,\rho) + 2\rho \partial_\rho u(t,\rho)
    \, .
\end{align}
To keep matters simple, we will only consider the standard formulation \labelcref{eq:flow_ui_QM}, i.e., directly discretizing the derivative of the effective potential.

Similarly as in the $\mathbb{Z}_2$-theory, we aim for a realistic setting, which also means choosing physically sensible initial conditions.
In the current Quark-Meson model the fermionic degrees of freedom, i.e., quarks,  drive the symmetry breaking and the initial conditions are usually chosen to be in the symmetric phase. This can be achieved by setting $m_\Lambda = 0$ and leaving the remaining parameters untouched. \\
Furthermore, the Yukawa coupling is an additional parameter not present in the $\mathbb{Z}_2$-theory, which we fix to $h=1.5$ in order to make the symmetry breaking scale comparable to the one in the $\mathbb{Z}_2$-theory case.

\subsection{Typical solutions}
\label{sec:typ_solution}

Before diving into the analysis of different numerical schemes for the RG-time integration, it is helpful to discuss the qualitative features of the solution.
We discuss the solution in the range $t\in[0,6]$. At the final RG-time $t=6$ the flow is effectively frozen and is almost indistinguishable from the $t\to\infty$ limit. This will be discussed further in \Cref{sec:jacobian_EV}.

In the case of the standard formulation \labelcref{eq:flow_ui} the qualitative behavior is straightforward and is shown in \Cref{fig:std_sol_lines}. For small RG-times $t$, the linear shape of $u(t, \rho)$ stays roughly intact and gets shifted only slightly. At $t\approx 1.5$ the smallest values, located around $\rho=0$, get comparable to the singularity bound $-k^2$, cf.~\labelcref{eq:flow_u}, and the convexity restoration dynamics set in. This is particularly well visible if the RG-time evolution of the individual grid points are visualized, shown in right panel of \Cref{fig:std_sol_lines}. At this point, the zero crossing, which determines the solution of the equation of motion, starts to freeze in exponentially fast. In the remaining RG-time evolution, the region at field values smaller than the zero crossing $\rho \leq \rho_0(t)$ gets pushed by the singularity bound towards zero.

The challenging aspect in these calculations is usually that the solution of the equation of motion is located at the boundary of the flat region. This makes a precise resolution of the region around $\rho_{0,\tinytext{fin}} = \rho_0(t\to\infty)$ difficult, as the point of interest itself becomes non-analytic and quantities, such as the mass, cf.~\labelcref{eq:mass}, contain one-sided derivates.
The non-analyticity, in the presence of the singularity bound in the equation, necessitates the use of a diffusive scheme, regularizing the non-analyticity.

This problem is particularly well visible when considering the discretization in terms of the mass~\labelcref{eq:flow_msqi}.
The solution thereof is shown in \Cref{fig:msq_sol_lines}.
In this case, the non-analyticity at $\rho_{0,\tinytext{fin}}$ turns into a jump, which becomes sharper and sharper as the RG-time becomes larger.
From the right panel in \Cref{fig:msq_sol_lines}, it is evident that there is one grid point inside the jump, reflecting its diffusive nature.
Please note that this grid point survives the $t\to\infty$ limit and that the details of the diffusive smearing of the shock strongly depend on the chosen spatial discretization.
Furthermore, this point is also contained in the solution of the standard formulation \labelcref{eq:flow_ui}, being equivalent on the discrete level. In the mass formulation \labelcref{eq:flow_msqi}, it is also quite well visible that each grid point in the flat regime goes through a rapid change, underlining the difficult RG-time evolution.

Lastly, we consider the log formulation~\labelcref{eq:flow_logi} of the problem. In such a formulation, the singularity bound is not manifest in the equations, as $\varpi(t,\rho) = \log(k^2 + m^2) \in \mathbb{R}$, in contrast to $u(t,\rho) > -k^2$, solving the issue of the singularity bound. The price to pay is that the points in the flat region tend towards negative infinity, as seen in \Cref{fig:log_sol_lines}.

This lets the height of the jump of $m^2$ diverge linearly with the RG-time, which may lead to additional complications, particularly, concerning the stability of the spatial discretization in the vicinity of the jump. Similarly to the case of the mass formulation above, there is only one point inside the jump discontinuity which is interpreted in the same manner.

A further peculiarity can be seen when considering the RG-time evolution of the individual grid points in the right panel of \Cref{fig:log_sol_lines}.
Every time a grid point enters the flat regime, every other point in the flat region is shifted to its predecessor, leading to a stair-like behavior. In the other formulations, this is not visible, as it happens entirely in the flat region. Nevertheless, such a behavior is certainly relevant in order to properly resolve the flat region and to avoid the singularity bound.

Finally, we would like to note that there is no known, stable discretization scheme that avoids the issues or qualitative features discussed above.

For completeness, we have also included the visualization of the solution in the case of added fermions at this point in \Cref{fig:QM_sol_lines}. A discussion thereof can be found in \Cref{sec:res_QM}.

All solutions discussed in this section correspond to our reference solutions. They are calculated as close to machine precision as possible, details thereon are given in \Cref{sec:work-precision}.

\section{Framework for numerical experiments}
\label{sec:framework}
In this section, we describe the implementation details and general setup we have used to analyze and benchmark the RG-time integration of the sets of ODEs resulting from the discussion in \Cref{sec:discretizations}.

The solution of ODEs with an implicit algorithm requires explicit choices for a couple of different components:
\begin{itemize}
    \item Algorithm
    \item Non-linear solver
    \item Linear solver
    \item Timestep controller
    \item Error norms
    \item Determination of the Jacobian
\end{itemize}
Note that not all algorithms require all the other components, e.g., Rosenbrock methods have a Newton iteration step built-in and consequently do not require a non-linear solver. Nevertheless, all components can be discussed on a relatively general footing, since we found many patterns to be independent of the concrete algorithm in use.

To discuss all these different aspects, we found the \texttt{DifferentialEquations.jl}~\cite{rackauckas2017differentialequations} package most suitable. All our tests are written in the Julia language~\cite{Julia-2017} and available on \href{https://github.com/NicolasW1/Numerical-RG-time-integration-of-the-effective-potential-Analysis-and-Benchmark}{GitHub}~\cite{github_link}.

\subsection{Components required for numerical solution}
\label{sec:components}
In the following, we discuss the different components and how they affect our benchmark process or the possibility to solve the system in general:
\subsubsection{Algorithm}
The first and most prominent component in the numerical solution is the choice of a timestepping algorithm itself.
Broadly, timestepping algorithms fall into two categories, implicit and explicit schemes, with numerous candidates in both categories, as well as the possibility of hybrid schemes to choose from.
However, in the present context it is a simple choice. 

To see this, consider that in the flat regime of the potential, the local wave speed grows exponentially fast (see~\cite{Grossi:2019urj} and \cref{sec:jacobian_EV}; note that the wave speed is defined as the largest eigenvalue of the local Jacobian).
As a consequence, the maximum step size in RG-time becomes exponentially small by virtue of the CFL condition, i.e.,
\begin{equation}\label{eq:cfl}
    \Delta t < \frac{C \Delta x}{\alpha} \sim e^{-b t}
    \, ,
\end{equation}
with $C$ being the Courant number of the problem and $b$ the scale exponent of the wave speed $\alpha$.

While a straight-forward CFL condition as in \labelcref{eq:cfl} only holds exactly in the infinite N limit of the O(N)-theory, cf.~\cite{Grossi:2019urj}, the flat part of the potential is convection dominated and thus \labelcref{eq:cfl} provides a good approximation also in the present case.
Furthermore, the presence of diffusive terms usually decreases the allowed step size, worsening the situation.
In summary, the required exponentially small timestep size renders explicit timestepper infeasible, and we will focus on implicit algorithms only.

This still leaves us with a vast landscape of implicit algorithms.
For a detailed introduction see e.g., \cite{brenan1995numerical, atkinson2011numerical}, here we only state their rough features.
The popular family of Runge-Kutta schemes splits into two different classes.

The first and most commonly used class of algorithms are the Diagonally Implicit Runge-Kutta (DIRK) methods, it includes well-known algorithms such as Implicit Euler or the Trapezoid/Implicit Midpoint algorithm used in the Crank–Nicolson scheme. The second class contains the Fully Implicit Runge-Kutta (FIRK) schemes.
While in DIRK schemes for each internal stage the system of equations can be solved subsequently, in FIRK schemes a coupled system of equations has to be solved.
To be more precise, in explicit RK schemes each stage is obtained from previous stages only, while in DIRK methods the right-hand side of each stage can contain itself.
In FIRK methods, the right-hand side may contain all stages.
As a direct consequence, DIRK schemes are a lot easier and cheaper to implement and are typically also faster, if they capture the stiffness of the system.
Consequently, DIRK methods are the most commonly implemented implicit methods in toolboxes for PDEs.

The next big class of algorithms under investigation in this work are the Rosenbrock methods, including their derived Rosenbrock-W extensions.
They incorporate the Jacobian directly in the timestep update, essentially performing the step of a Newton iteration, and they can be seen as an extension of the DIRK methods.
As a consequence, for non-autonomous equations also the RG-time gradient of the flux appears in the formula.
They show impressive stability and accuracy properties for a wide range of problems, but their implementation is comparatively tedious.
In schemes collected as Rosenbrock-W, the Jacobian and time gradient are not updated at every step, but in a lazy manner, depending on the algorithm.

The last big class of implicit algorithms are the implicit multistep schemes. Compared to the RK schemes, no internal stages are evaluated, but the information from previous steps is used in the update step. Some of the most used libraries to solve stiff systems of ODEs, including the \texttt{SUNDIALS} library~\cite{hindmarsh2005sundials, gardner2022sundials} or the \texttt{ODEPACK} package~\cite{hindmarsh1983odepack}, focus on implicit multistep methods. Both of the aforementioned libraries are also included in our study via their Julia interface~\cite{rackauckas2017differentialequations}.
A big advantage of implicit multistep methods is that they easily allow for adaptive choice of order, greatly improving performance.

To conclude this subsection on algorithms, we would like to mention and comment on some other classes of algorithms.

The first of these are implicit extrapolation methods, which are included in the survey plots \Cref{fig:std_WP_all} to \Cref{fig:log_tmax_all}. We found them in general to be considerably underperforming compared to other algorithms. However, one appealing feature is their prospects regarding parallelizability, see e.g.,~\cite{elrod2022parallelizing}.

Another class of algorithms that we deemed promising when starting this project were exponential integrators, see e.g.,~\cite{hochbruck2010exponential}. These algorithms are built around the idea to split the equation in a linear part, which is integrated exactly, and a non-linear part which can be handled explicitly. We deemed this promising, since at asymptotically large RG-times all values are simply exponentially decaying in the standard formulation, c.f.~\Cref{fig:log_sol_lines}.
However, we could not find a suitable splitting of the equation, nor did automatic splitting based on a linearization at each timestep, as implemented in~\cite{rackauckas2017differentialequations}, work, i.e., reach a final RG-time of at least $t=6$ in any formulation.
These shortcomings might be overcome in the future with implicit exponential integrators. However, this would include an implicit algorithm again, making the extra effort questionable.

For completeness, we also mention that we tried the algorithms offered by the \texttt{IRKGaussLegendre.jl} package~\cite{antonana2017reducing}, but found them unsuitable due to the general failure of functional iteration, commented on below.

\subsubsection{Non-linear solver}
\label{sec:non_linear_solver}
Most implicit algorithms require the solution of a non-linear system of equations, with the notable exception of Rosenbrock methods, which incorporate a Newton solver iteration step already.
In the context of solving systems of ODEs numerically, there are essentially two commonly employed methods:
\paragraph*{Newton iteration}
This includes the classical (quasi-)Newton methods for solving systems of equations.
We found this class of methods to be the only one that works in practice.
Particularly, \texttt{NLNewton} in the ODE Julia framework~\cite{rackauckas2017differentialequations} worked well, and was insensitive to changes of numerical tuning parameters.

\paragraph*{Functional iteration}
Unfortunately, we were unable to get neither direct iteration nor Anderson accelerated variants thereof to work properly, i.e., to converge to sufficient accuracy within a reasonable number of iterations.
This can, at least partially, be explained by the peculiar eigenvalue spectrum of the Jacobian in the flat region, which will be discussed further below in \Cref{sec:jacobian_EV}.

\subsubsection{Linear solver}
\label{sec:linear_solver}
All RG-time evolution algorithms, which can seriously be considered in the present context, require the solution of systems of linear equations at each RG-timestep.
In our set-up, a plethora of different methods is available via the interface to the Julia package \texttt{LinearSolve.jl}.
We found that all commonly used algorithms converge, but that there are significant differences concerning their performance.
LU decomposition based algorithms, and in particular their sparse variants, performed extremely well.
Krylov subspace and QR decomposition based algorithms on the other hand performed significantly worse, increasing the runtime by several orders of magnitude.
This can, again, be explained by the peculiar eigenvalue spectrum of the Jacobian, discussed in \Cref{sec:jacobian_EV}. In the current context, this connection is particularly obvious in the QR algorithm, where the rate of convergence is proportional to the ratio of eigenvalues.

\subsubsection{RG-timestep controller}
\label{sec:controller}
Adaptive control over the RG-timestep size is in general a very attractive feature, for the obvious reasons.
Additionally, in fRG applications, it is common to redo calculations for changing initial conditions (either to tune initial conditions or to scan parameter spaces for phase structures) and the characteristics of the flow can change substantially.
Manual control over the RG-timestep size is rather inefficient in these scenarios, making it particularly relevant in fRG flows.

The most commonly options as controller of the adaptive step size are of control loop type, i.e., Integral, Proportional-Integral or Proportional-Integral-Derivative controller.
We have tested all three options, and additionally Gustafsson Acceleration, which are the four variants implemented in~\cite{rackauckas2017differentialequations}.
We found no noteworthy difference and simply chose the default option of the individual RG-timestep algorithms.

\subsubsection{Error norms}
\label{sec:error_norms}
Both for the purpose of checking if a Newton solver has converged, as well for choosing the size of an adaptive timestep, an error norm of the residual is used. For large problems, where a LU decomposition is not feasible, such a norm is also needed to check if an iterative algorithm has converged. 

Also in view of \Cref{sec:jacobian_EV}, with regards to stability it is important to ensure that the error norm takes into account spatially local accuracy.
To be more explicit, consider absolute and relative precision limits $s_\text{abs}$ and $s_\text{rel}$.
For a residual vector $r$ and a solution vector before the iteration ($u$) and after ($\tilde{u}$) on a grid of size $N$, we could define the normalized error using an $l_2$-norm as 
\begin{equation}
    e_{l_2} = n_{l_2}\frac{\lVert r\rVert_2}{s_\text{abs} + s_\text{rel}\,\text{max}(\lVert u\rVert_2, \lVert\tilde{u}\rVert_2)}
    \, ,
\end{equation}
with $n_{l_p} = N^\frac{1}{p}$. Generically, this is the standard kind of error norm implemented in many discretization frameworks for finite difference and finite elements.
However, such a definition can allow locally large errors in cases where locally strong dynamics occur, which quickly destabilize the system.

Therefore, an error-norm which evaluates the relative precision locally is more pertinent,
\begin{equation}\label{eq:local_err}
    e_\text{loc} = n_{l_p}\left\lVert\frac{r_i}{s_\text{abs} + s_\text{rel}\,\text{max}(|u_i|,|\tilde{u}_i|)}\right\rVert_p
    \, .
\end{equation}
In \texttt{DifferentialEquations.jl} and in the \texttt{SUNDIALS} suite, the error norm~\labelcref{eq:local_err} is implemented with $p=2$, which is sufficient for the problems considered in this work. However, we would like to comment that in the presence of spatially localized phenomena with fast dynamics we found it necessary to choose $p=\infty$ in order to ensure stability of the evolution.

\subsubsection{(Automatic) Jacobian construction}
\label{sec:jacobian_construction}
Lastly, we have to consider the construction and approximation of the Jacobian, required for most timestep algorithms.
Firstly, it should be noted that spatial discretizations dealing with convection dominated PDEs usually have sparse Jacobians.
Utilizing this structure, i.e., working with sparse matrices, is of course important, and closely related to the discussion about linear solvers, cf.~\Cref{sec:linear_solver}.
In practice, there are three different options to calculate the Jacobian:
\begin{itemize}
    \item Automatic differentiation (autodiff)
    \item Exact/Symbolic calculation
    \item Finite differences
\end{itemize}
We found that the usage of Jacobians which are accurate on the level of the working precision, greatly benefit the numerical RG-time integration.
This is achieved by the first two options.

The first option, i.e., automatic differentiation, is an option to obtain the Jacobian at working precision, by repeated application of the chain rule at the level of the compiler. In the Julia language, this technique is readily available via \texttt{ForwardDiff.jl}~\cite{RevelsLubinPapamarkou2016} and the calculation of Jacobians is automatically supported in \texttt{DifferentialEquations.jl}.
We found this the most convenient setup and used it as default throughout this work for all Julia based solvers, i.e., all except the \texttt{SUNDIALS} (\texttt{CVODE\_BDF}) and \texttt{ODEPACK} (\texttt{lsoda}) routines.
In these cases, we resorted to their internal default.

The second option, implementing the explicit Jacobian, performed very similar to the case of automatic differentiation, but comes with the additional overhead of deriving it.
Sometimes, this task can be efficiently automated, making it also a very viable option.
Particularly, if automatic differentiation is not an option.

Lastly, we found the practical usability of finite difference to be strongly dependent on the heuristics employed by different libraries.
The external libraries, mentioned above, performed very well, utilizing their own approximation heuristics.
On the other hand, we did not manage to get the Julia based solver combined with finite differences, i.e.,~using \texttt{FiniteDiff.jl}, to work at all.
We would like to add, that we have previously made similar experiences in Mathematica, leading us to the stated conclusion.

\subsection{Setup of the numerical experiments}
All benchmark runs measuring execution times were run five times to avoid outliers, the reported value corresponds to the median.
In all cases, we found the median to be very close to the minimum, confirming the absence of outliers.
The implementation is available at \href{https://github.com/NicolasW1/Numerical-RG-time-integration-of-the-effective-potential-Analysis-and-Benchmark}{GitHub}~\cite{github_link}.
The study was performed on a laptop with Windows 11 operating system, an Intel Core i7-11800H CPU with a maximal capacity of 4.60GHz, and 32GB RAM. To facilitate comparison, no forms of parallelization were used, all calculations were restricted to a single core and thread. All floating-point operations were performed in the double-precision floating-point format.
The final RG-time is chosen as $t=6$, unless specified otherwise. At $t=6$ the flow is safely in the asymptotic regime, cf. \Cref{sec:jacobian_EV}.

\section{Results}
\label{sec:results}
This section presents our analysis and results, focusing on the RG-time evolution solver algorithm. The other components were found to be generic regarding their effect on the numerical RG-time integration, and are discussed in the previous section, i.e., \Cref{sec:framework}.

\subsection{Work-Precision}
\label{sec:work-precision}
%
\begin{figure*}[t]
	\centering
	\begin{subfigure}[t]{0.48\textwidth}
		\centering
		\includegraphics[width=\linewidth]{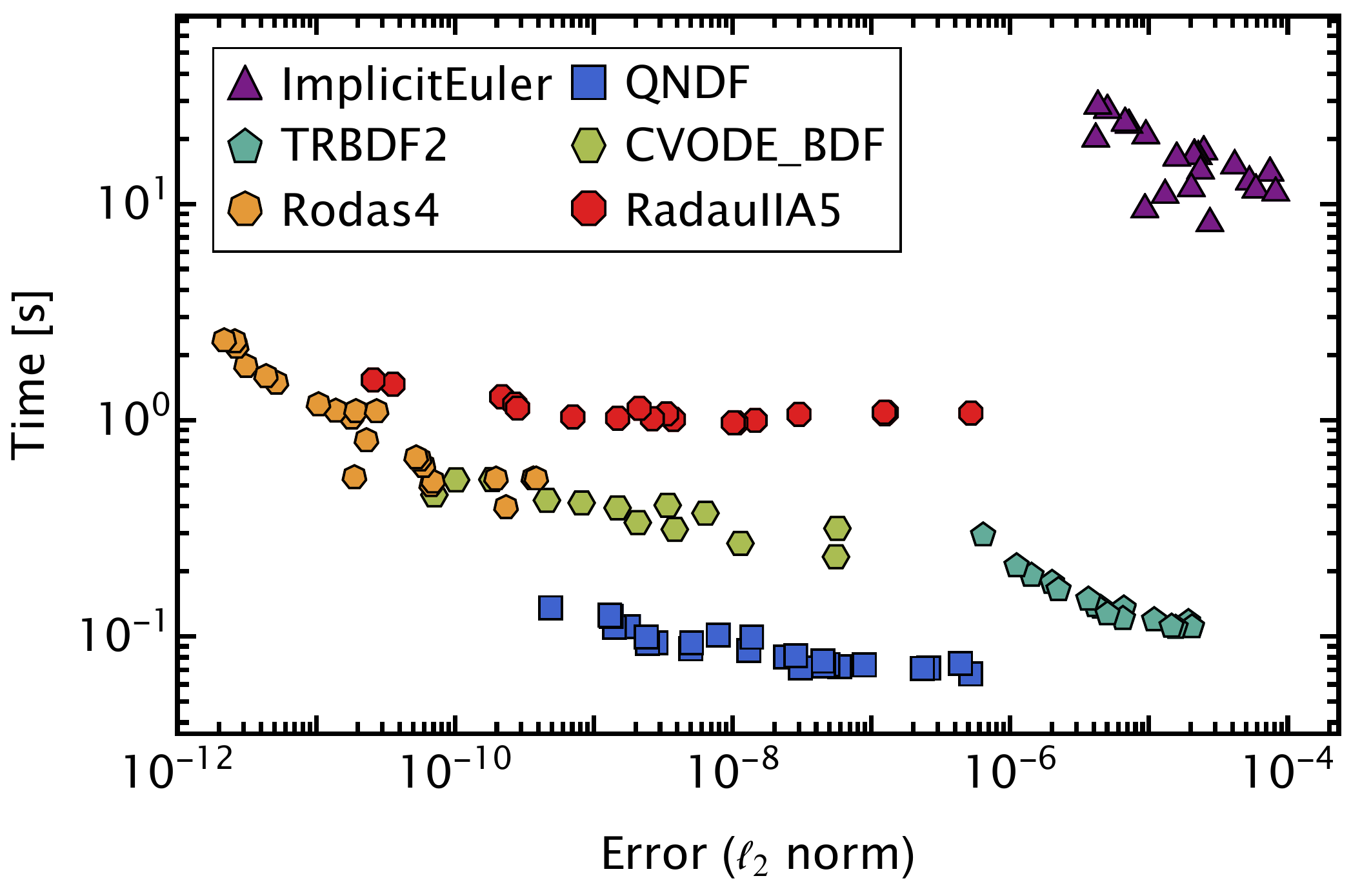}
		\caption{Standard formulation}
		\label{fig:std_wp}
	\end{subfigure}	~
	\begin{subfigure}[t]{0.48\textwidth}
		\centering
		\includegraphics[width=\linewidth]{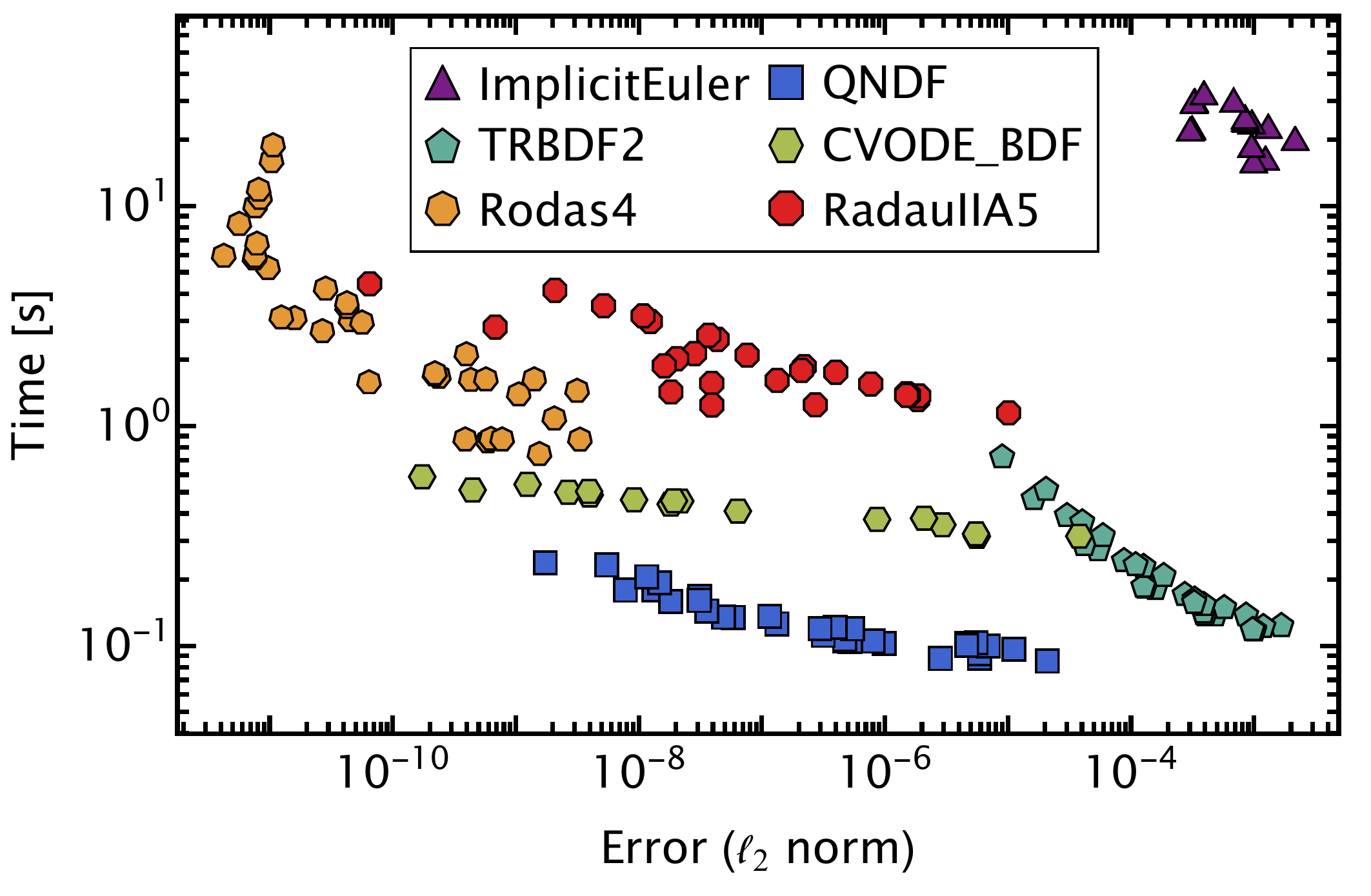}
		\caption{Mass formulation}
		\label{fig:msq_wp}
	\end{subfigure}
	\begin{subfigure}[t]{0.48\textwidth}
		\centering
		\includegraphics[width=\linewidth]{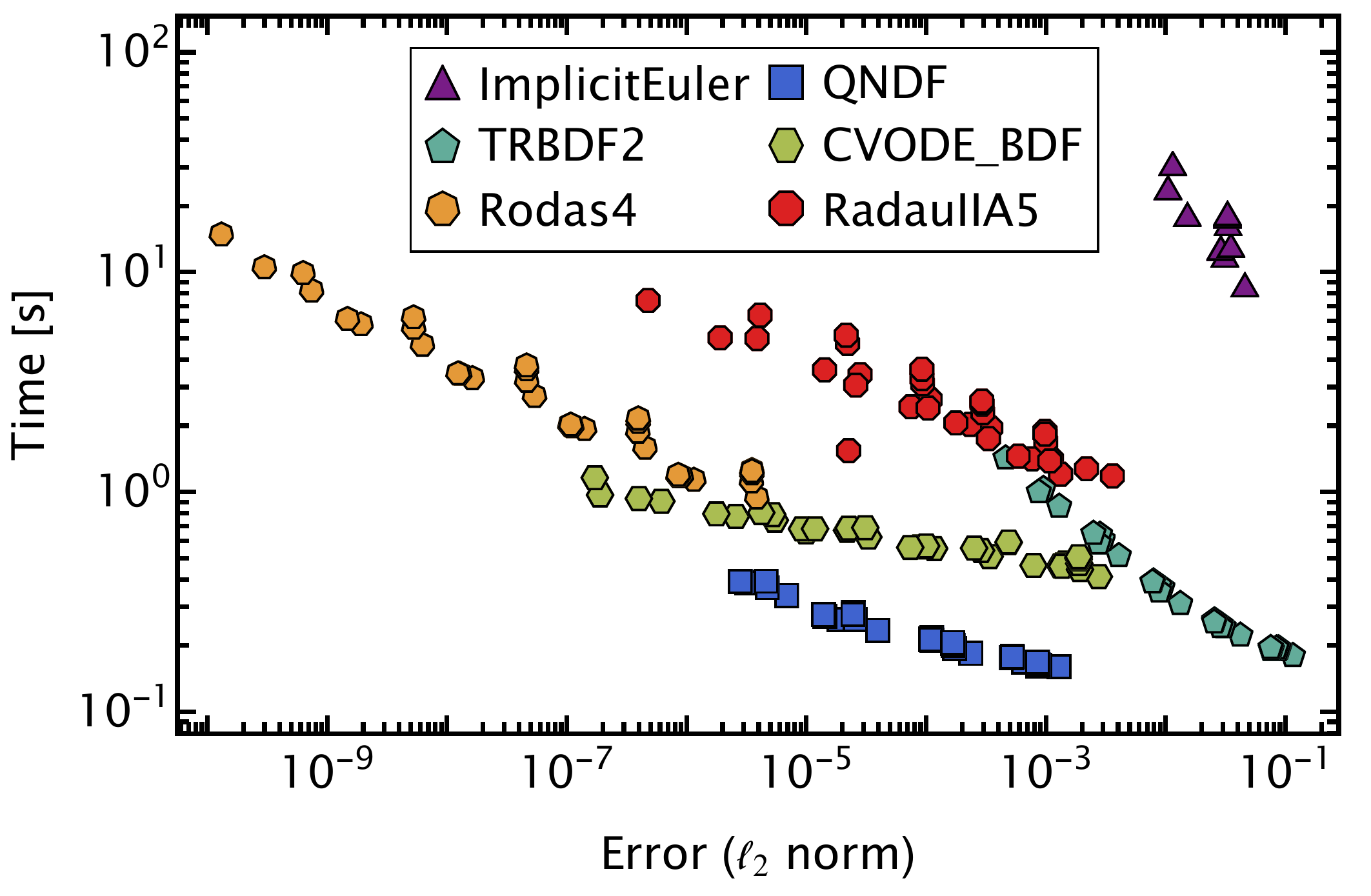}
		\caption{Log formulation}
		\label{fig:log_wp}
	\end{subfigure}	~
	\begin{subfigure}[t]{0.48\textwidth}
		\centering
		\includegraphics[width=\linewidth]{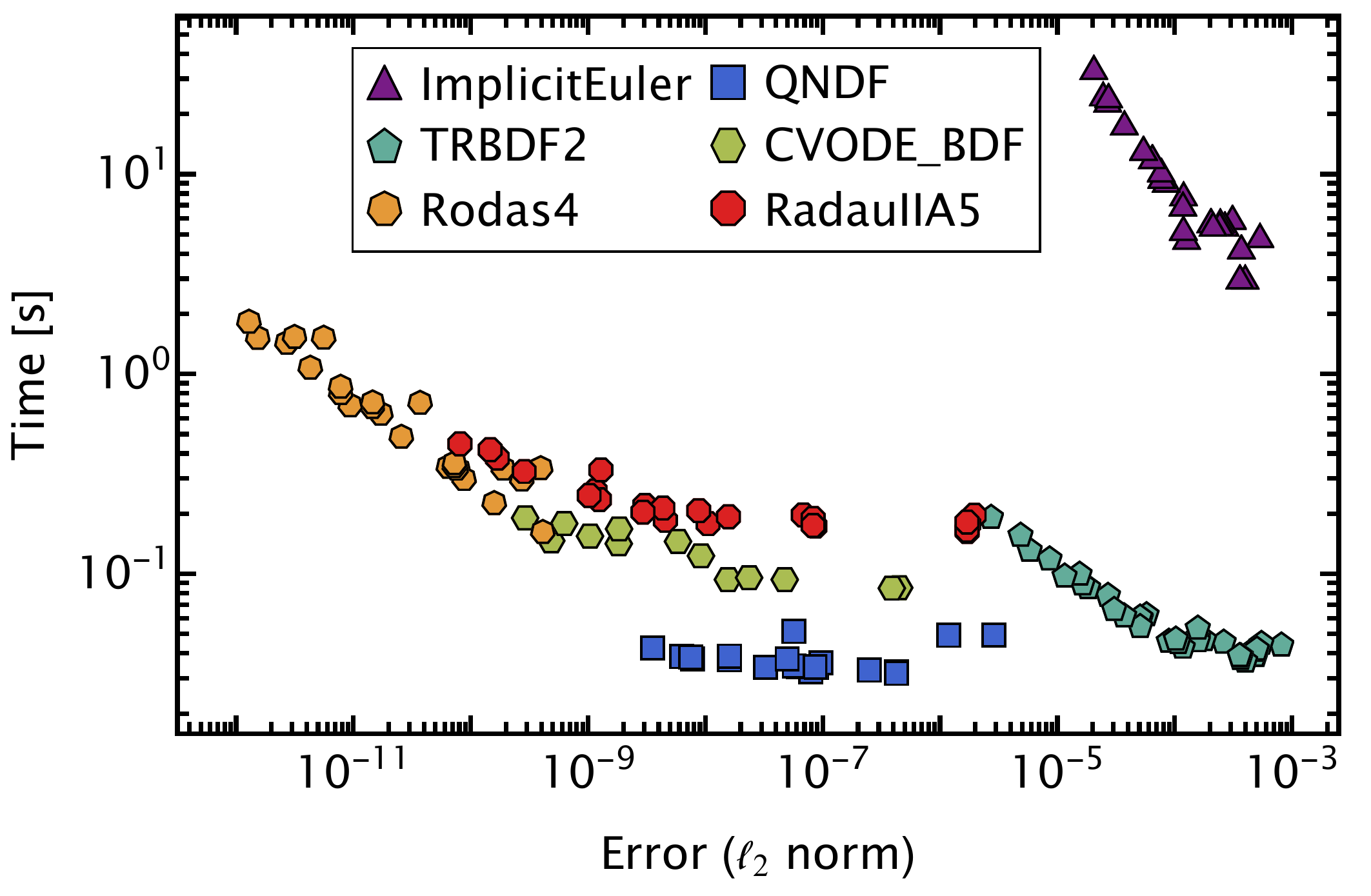}
		\caption{Standard formulation of the fermionic extension}
		\label{fig:qm_wp}
	\end{subfigure}%
	\caption{Work-precision diagrams for all three formulations, the fermionic case and selected algorithms. Noteworthy are the very efficient nature of the Implicit Multistep algorithm \texttt{QNDF}, the high accuracy of the Rosenbrock algorithm \texttt{Rodas4} and the poor performance of \texttt{ImplicitEuler}. The complete survey is depicted in \Cref{fig:std_WP_all}, \Cref{fig:msq_WP_all}, \Cref{fig:log_WP_all} and a detailed discussion is given in \Cref{sec:work-precision}.}
	\label{fig:wp}
\end{figure*}
%

For solving flow equations in everyday applications, the most important quantity measuring the performance is the work-precision relation, i.e., how does the accuracy behave with computing time.
Therefore, we surveyed 49 different algorithms, collected in \Cref{app:overview}.
Beforehand, a reference solution was generated using the \texttt{KenCarp58} algorithm. For this reference solution, the absolute and relative accuracy goals were chosen as $10^{-15}$, the highest accuracy feasible with double-precision floating-point numbers.
By comparing different algorithms, i.e.,~\texttt{RadauIIA5}, \texttt{QNDF} and \texttt{Rodas4}, at this target accuracy, we found the maximally actual achievable accuracy to be roughly $10^{-12}$, measured in the $\ell_2$-norm.
Therefore, we excluded all results in the survey with $\lVert a_i - a^\tinytext{ref}_i\rVert_{\ell_2} < 1.25\cdot10^{-12}$.
Slight remnants of the resulting saturation effect when investigating the work-precision relation are still visible, particularly in the results for the mass formulation \labelcref{eq:flow_msqi}.
\begin{figure*}[t]
	\centering
	\includegraphics[width=0.48\textwidth]{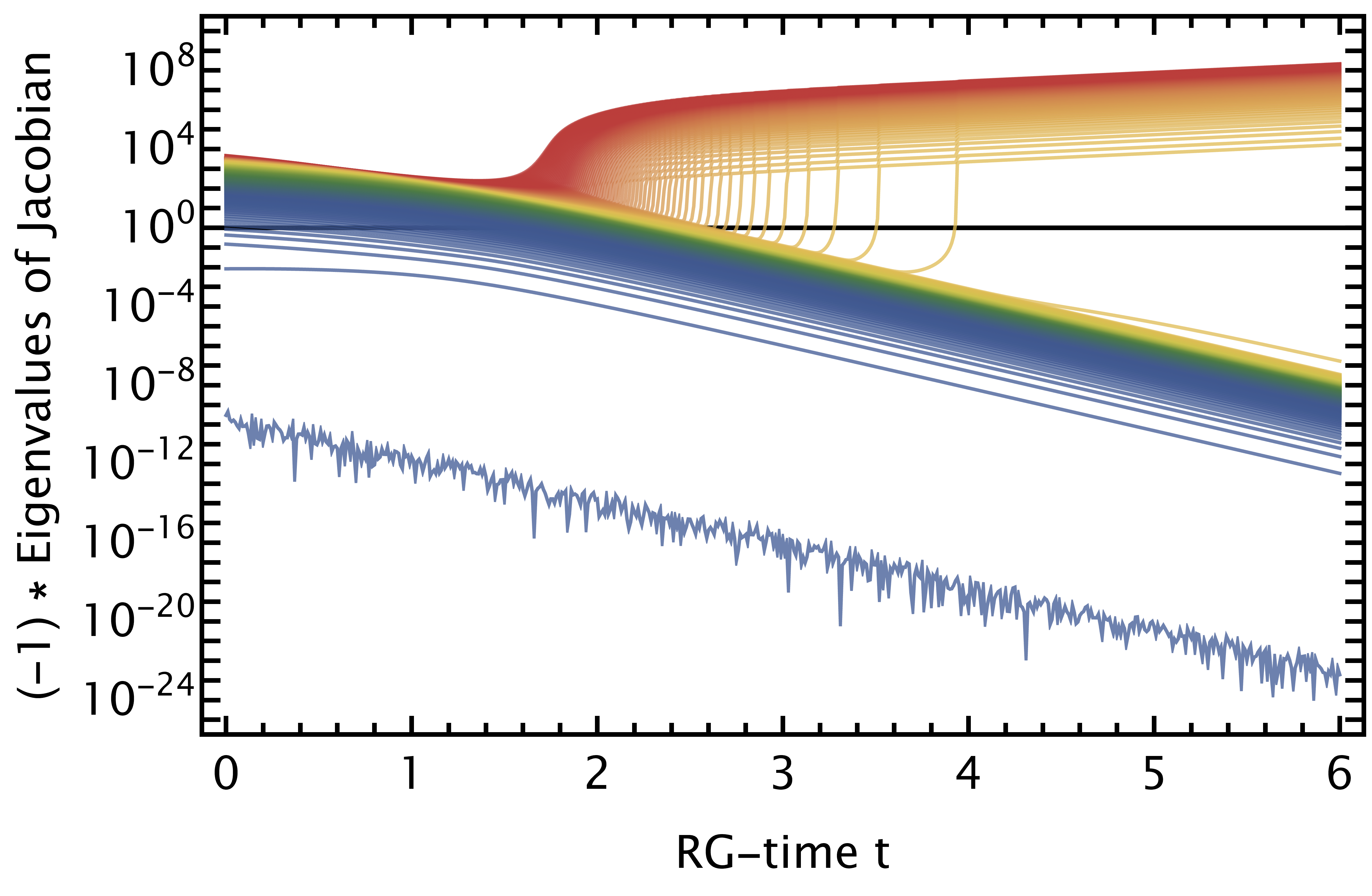}
	\caption{RG-time evolution of the negative eigenvalues of the Jacobian.
	The significantly suppressed, wiggly line can be attributed to the boundary condition at large field values, i.e., the $\rho_\tinytext{max}$ boundary.
	The remaining eigenvalues split in two categories, the ones associated with the flat regime, which shown exponential growth, and the remaining ones, which show exponential suppression. A detailed discussion of the implications thereof is given in \Cref{sec:jacobian_EV}.}
	\label{fig:evs}
\end{figure*}

In order to investigate the work-precision relation, we scanned all possible combinations of absolute and relative target accuracies in the range $10^{-8}\ldots10^{-12}$ in steps of one ($\log_{10}$ scale).
We excluded all runs that did not reach the final RG-time $t=6$, which was mostly the case at lower accuracy goals and explains the lower scan range of $10^{-8}$.
The full survey for work-precision relations are collected in \Cref{fig:std_WP_all} for the standard formulation \labelcref{eq:flow_ui}, in \Cref{fig:msq_WP_all} for the mass formulation \labelcref{eq:flow_msqi} and in \Cref{fig:log_WP_all} for the log formulation \labelcref{eq:flow_logi}.
The general performance is similar for all three formulations, hence we refrain from separated discussions for each formulation.

The DIRK algorithms, the \texttt{KenCarp} family, except \texttt{KenCarp5}, and \texttt{TRBDF2} are looking very promising.
Particularly, \texttt{KenCarp3} and \texttt{TRBDF2} are showing a stellar performance when taking into account that they are low-order methods. 
It is also noteworthy that the popular choices \texttt{ImplicitEuler} and \texttt{Trapezoid} performed extremely poor.

The Rosenbrock methods are relatively independent of the specific choice of algorithm.
While they do require significantly more time to push the high precision boundary, they shine through their remarkable stability and over-performance in actually achieved accuracy.

Rosenbrock-W methods are running on average slightly faster, but their decreased stability dims their attractiveness.

We turn to implicit multistep methods, for the purpose of this discussion, including \texttt{SUNDIALS}' \texttt{CVODE\_BDF} and \texttt{ODEPACK}'s \texttt{lsoda}. While being unable to push toward really small accuracies, the variable order BDF implementations, i.e.,~\texttt{QBDF}, \texttt{QNDF} and \texttt{CVODE\_BDF}, are attractive due to their very short run-times. \texttt{FBDF} performed similar, but performed worse than \texttt{QBDF} and \texttt{QNDF} due to reduced stability, i.e., requiring a significantly smaller minimal allowed step size.

Turning towards the FIRK methods, we found \texttt{RadauIIA5} to be very stable and well-performing, while \texttt{RadauIIA3} was one of the worst performing algorithms that still managed to solve the system.
We could not find an obvious explanation for the drastic difference between the two very similar algorithms. We suspect it to be related to the lack of a pseudo-stability region for \texttt{RadauIIA3}, which will be discussed in \Cref{sec:infinite_RG-time}.

Finally, the family of implicit extrapolation algorithms performed significantly worse than the other families of algorithms, rendering them irrelevant for normal applications to flow equations.

For the reader's convenience, a collection of well performing algorithms, and \texttt{ImplicitEuler} for reference, is shown in \Cref{fig:wp}. The depicted algorithms, i.e., \texttt{QNDF},  \texttt{TRBDF2},  \texttt{Rodas4}, \texttt{RadauIIA5} and \texttt{CVODE\_BDF}, can be considered excellent choices when being confronted with a novel problem/equation in the context of derivative expansions in flow equations.

\subsection{Jacobian eigenvalues}
\label{sec:jacobian_EV}
Further insight into the difficulties encountered inside the flat region can be gained by investigating the Jacobian of the system.
For simplicity, we restrict ourselves to the standard formulation \labelcref{eq:flow_ui}. The results are, however, directly applicable to all three formulations.
The (negative) eigenvalues of the (discrete) Jacobian are shown in \Cref{fig:evs} as a function of RG-time.
Some peculiarities of the system become immediately obvious.
Firstly, each grid point in the flat region comes with its own exponentially growing eigenvalue.
Please note that this statement is entirely dependent on the spatial discretization, but the situation is similar in all known, stable discretizations.
Secondly, their exponential growth sets in subsequently.
The eigenvalue which can be associated to the grid point inside the kink/jump is the last one to show dynamic behavior. It shows similar growth to the other points in the flat region, but freezes in the non-flat region. This marks a unique RG-time where extrapolation, similar to the one discussed in \cite{Grossi:2019urj}, can be performed safely.
Lastly, the non-flat region of the potential starts freezing in exponentially fast, once the flow becomes smaller than the local mass scale. This is the expected behavior of RG flows.
Please note that the wiggly eigenvalue which is significantly more suppressed than all other eigenvalues is related to the boundary condition \labelcref{eq:boundary_cond} at $\rho_\tinytext{max}$.

To summarize, the eigenvalues split in two groups, an exponentially growing one in the flat region and an exponentially suppressed one in the non-flat region.
Consequently, there is an exponentially growing gap between these two groups. This significantly reduces the performance of all algorithms whose rate of convergence is related to the ratio of eigenvalues, cf.~\Cref{sec:framework}.
Additionally, it hints toward potential design principles for new spatial discretizations, i.e., reducing the number of exponentially growing eigenvalues per evaluation node in the flat region. Hereby, the insights from \cite{Pelaez:2015nsa} might be particularly useful, providing linearized equations in the flat region.

The exponential increase of the Jacobian eigenvalues also shows that the system becomes increasingly susceptible to large errors and instabilities due to perturbations. Such perturbations can quickly arise due to numerical errors done in previous timesteps, making high precision during the timestepping an important prerequisite for stability.


\subsection{Stability in the infinite RG-time limit}
\label{sec:infinite_RG-time}
%
\begin{figure*}[ht]
	\centering
	\begin{subfigure}[t]{0.48\textwidth}
		\centering
		\includegraphics[width=\linewidth]{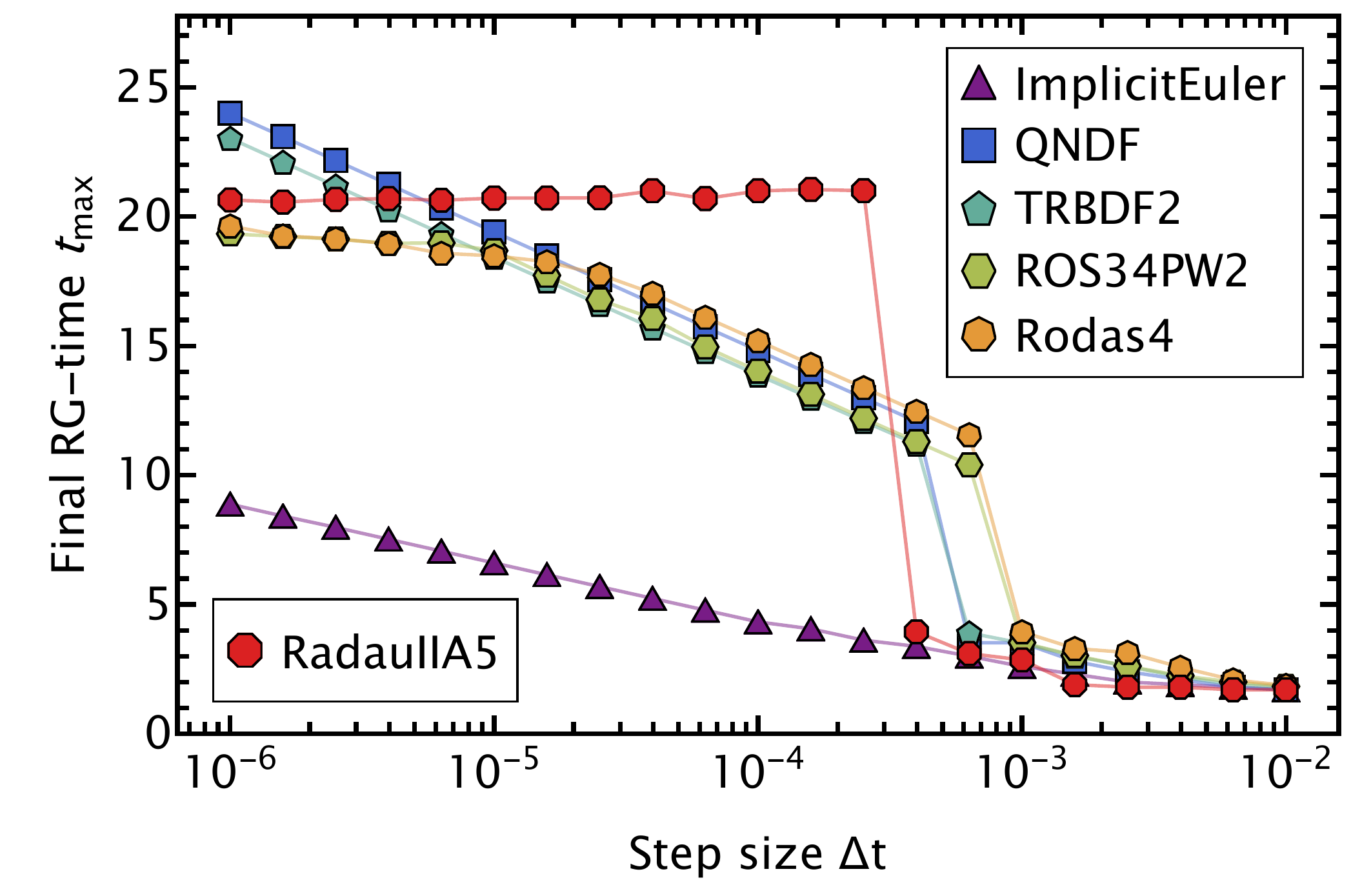}
		\caption{Standard formulation}
		\label{fig:std_tmax}
	\end{subfigure}	~
	\begin{subfigure}[t]{0.48\textwidth}
		\centering
		\includegraphics[width=\linewidth]{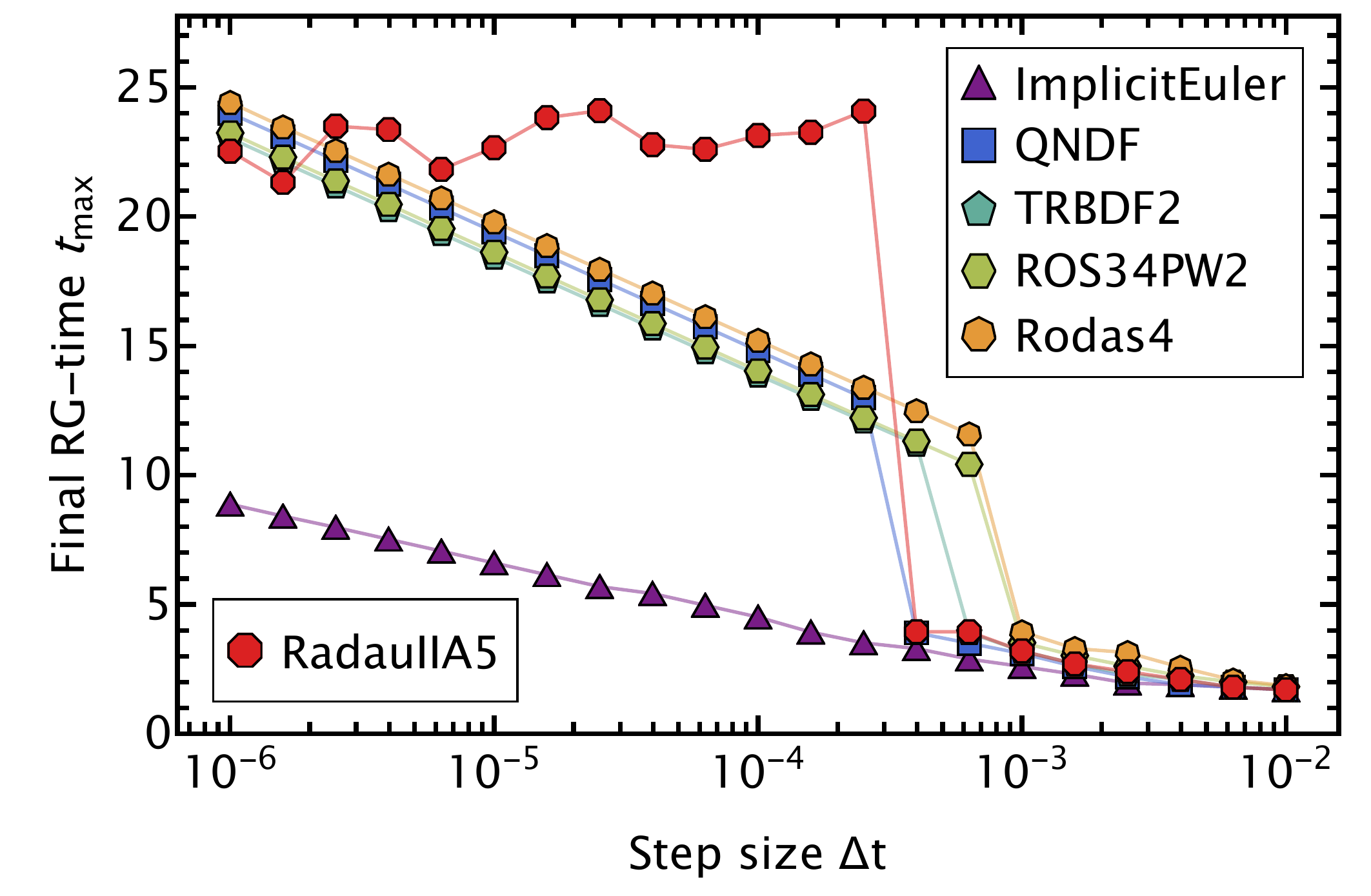}
		\caption{Mass formulation}
		\label{fig:msq_tmax}
	\end{subfigure}
	\begin{subfigure}[t]{0.48\textwidth}
		\centering
		\includegraphics[width=\linewidth]{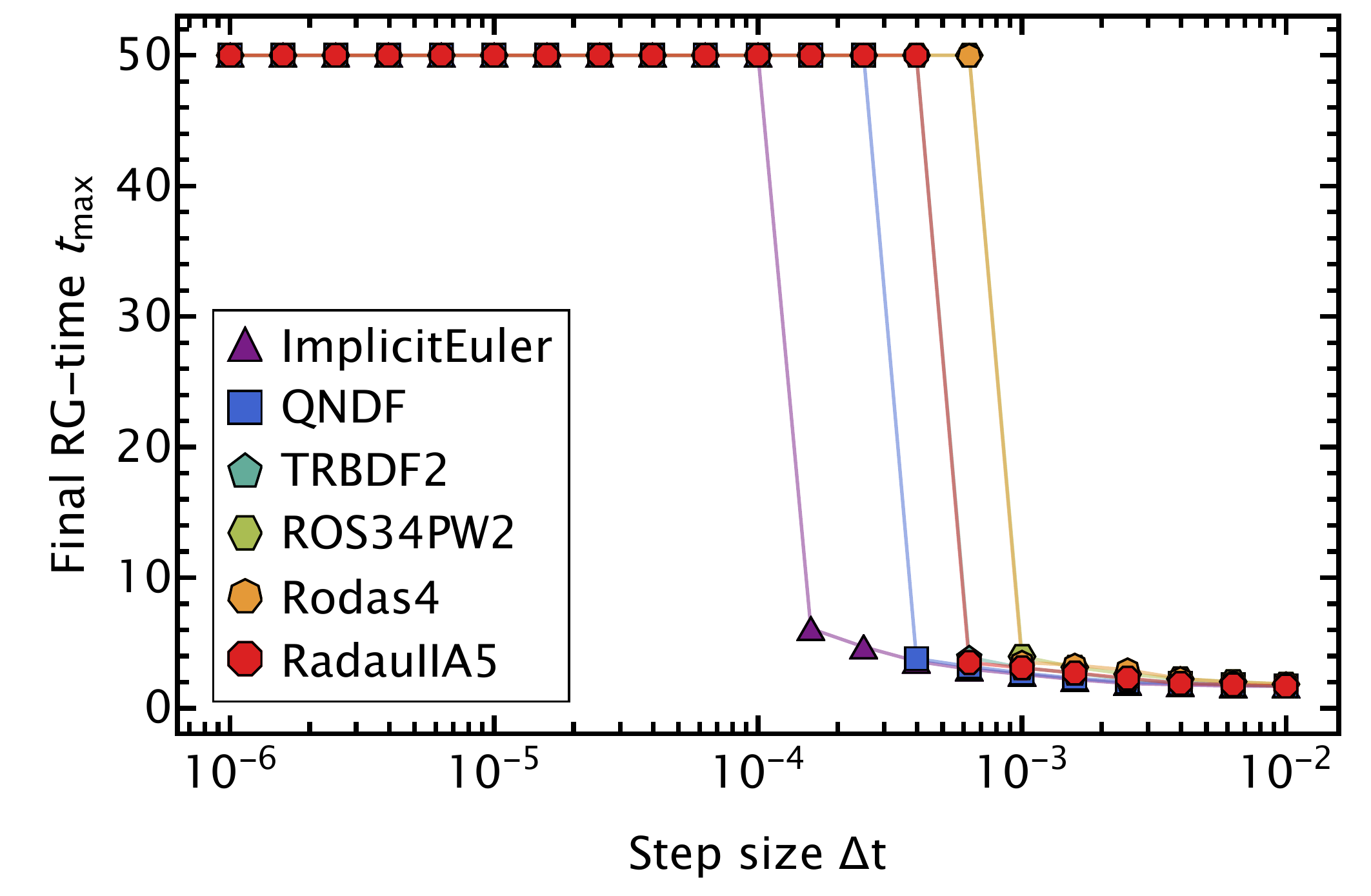}
		\caption{Log formulation}
		\label{fig:log_tmax}
	\end{subfigure}	~
	\begin{subfigure}[t]{0.48\textwidth}
		\centering
		\includegraphics[width=\linewidth]{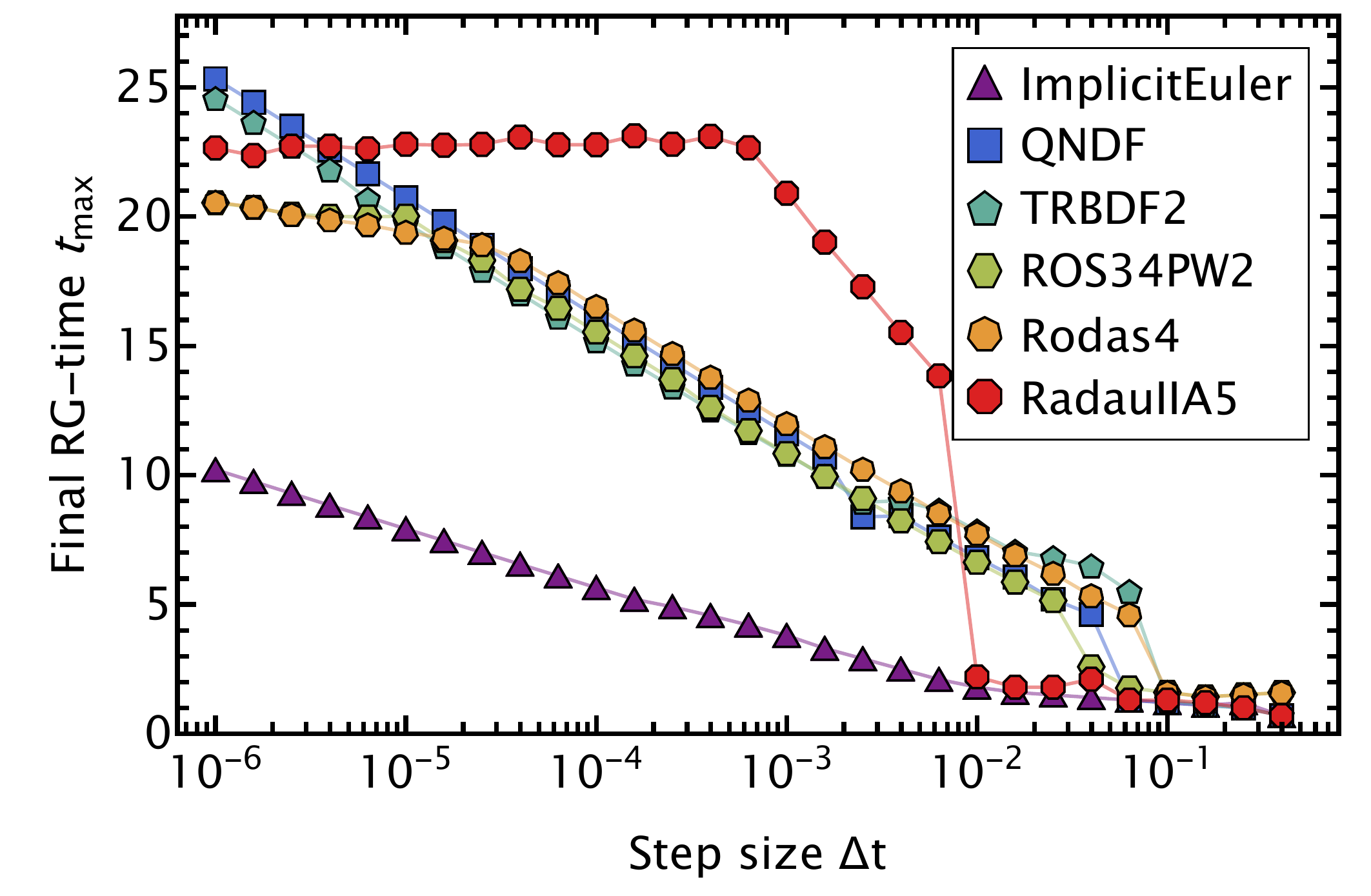}
		\caption{Standard formulation of the fermionic extension}
		\label{fig:qm_tmax}
	\end{subfigure}%
	\caption{RG-time stability for all three formulations, the fermionic case and selected algorithms. Noteworthy are the notion of pseudo stability in the standard~\labelcref{eq:flow_ui} (including the fermionic case~\labelcref{eq:flow_ui_QM}) and mass formulations \labelcref{eq:flow_msqi} and the proper stability in the logarithmic formulation \labelcref{eq:flow_logi}. The complete survey is depicted in \Cref{fig:std_tmax_all}, \Cref{fig:msq_tmax_all}, \Cref{fig:log_tmax_all} and a detailed discussion is given in \Cref{sec:infinite_RG-time}.}
	\label{fig:tmax}
\end{figure*}
%

There is another aspect when solving fRG equations in everyday applications that is highly relevant:
The long RG-time stability of the RG-time integration.
In order to test this faithfully, we switch to fixed RG-timestep size evolution, where the step size is denoted by $\Delta t$.
The flow is integrated until we detect an inconsistency/instability in the flow, i.e., either the solution becomes non-monotonic or the positivity bound is violated.
The last successful RG-timestep then determines $t_\tinytext{max}$.
Ideally, there would be some maximal step size $\Delta t_\tinytext{cr}$, for which all smaller RG-timesteps $\Delta t < \Delta t_\tinytext{cr}$ provide stable evolution, i.e., $t_\tinytext{max}\to\infty$.
In practice, we cannot integrate until $t_\tinytext{max}\to\infty$.
Therefore, we chose the maximal cutoff at $t=50$, which is the same for all practical purposes.
The external libraries are excluded for technical reasons in this survey, we do not expect them to perform as well as the pure Julia implementations, due to the lack of auto differentiation in the calculation of the Jacobian.

The full survey of results can be found in \Cref{fig:std_tmax_all} for the standard formulation \labelcref{eq:flow_ui}, in \Cref{fig:msq_tmax_all} for the mass formulation \labelcref{eq:flow_msqi} and in \Cref{fig:log_tmax_all} for the log formulation \labelcref{eq:flow_logi}.
The selection of algorithms is depicted in \Cref{fig:tmax}.
There are some immediate observations:
Firstly, the only formulation which shows proper stability is the log formulation \labelcref{eq:flow_logi}.
At $\Delta t \approx 10^{-3} \ldots 10^{-3}$ the final RG-time becomes independent of the step size.
Secondly, the other two formulations do not exceed $t_\tinytext{max} \approx 20\ldots25$, i.e., they do not feature stability.
However, at roughly the same $\Delta t$, where the log formulation shows stability, they feature a sharp increase in $t_\tinytext{max}$.
This sharp increase precisely makes the difference between breaking down during the flatting of the potential and being able to properly resolving this essential part of the evolution.
We call this notion \textit{pseudo stability}.
Luckily, a large part of the investigated algorithms, particularly the Rosenbrock and implicit multistep methods, posses this property.
The final RG-time reached by most algorithms in the standard and mass formulation is limited by double precision in the (non-)linear solver part.
This can be easily seen by estimating the relevant orders of magnitude in their discrete equations.

Furthermore, for the method \texttt{RadauIIA5} the jump in the final RG-time is so large, that we cannot differentiate between pseudo stability and proper stability of the algorithm.
Due to the notion severe size of the jump in the mass formulation, c.f.~\Cref{fig:msq_tmax}, one might suspect that the method is stable, but limited by finite numerical precision.
However, this statement has to be contrasted with the observations of the fermionic case, discussed in \Cref{sec:res_QM}.

Particularly, the log formulation, despite its infinitely growing gap in the solution, get quite appealing due to the RG-time stability.
We would like to note that the generalization thereof to more complicated theories, e.g., a $O(N)$-theory with $N>1$, is ambiguous and has to be worked out.

In general, the notion of pseudo stability hints towards the existence of two different stability criteria.
One being related to stability with respect to non-analyticities, i.e., the kink in the current example.
The other one being related to the existence of a singularity bound and can be circumvented by the log formulation.

\subsection{Fermionic extension of model}
\label{sec:res_QM}
The results presented so far have all been obtained in the $\mathbb{Z}_2$-theory.
Here we briefly discuss results of the theory with fermions introduced in \Cref{sec:fermions}.
For simplicity, we restrict ourselves to the standard formulations, the extensions of other formulations, particularly the log one, will be discussed elsewhere.

A typical RG-time evolution of the derivative of the effective potential is shown in \Cref{fig:QM_sol_lines}.
Most notably, during the initial phase of the flow, the fermionic contributions increase the expectation value of the field.
However, as soon as the flattening of the potential sets in, the evolution becomes very similar to the $\mathbb{Z}_2$ case.
Additionally, the non-analyticity seems softer, i.e., it manifests itself in the discrete system over a larger range in RG-time.

The work-precision result, shown in \Cref{fig:qm_wp} for selected algorithms, is very similar to the pure $\mathbb{Z}_2$ case.
This was to be expected, due to the very similar final solution.

The infinite RG-time analysis, however, reveals that the requirements for pseudo stability are reached at already much larger timesteps ($\approx 10^{-2}$), which can easily be explained by the softer non-analyticity.
Furthermore, the \texttt{RadauIIA5} algorithm clearly shows only pseudo stable behavior.
This might indicate that the observed behavior in the $\mathbb{Z}_2$ case was also only pseudo stability, limited by finite numerical precision, or special to this particular case.

\section{Conclusion and outlook}
\label{sec:conclusion}
In this work, we investigated the numerical RG-time integration of fRG flow equations as encountered in a derivative expansion of a scalar theory in three spacetime dimensions with a $\mathbb{Z}_2$ symmetry.
This example displays the prototypical behavior of a flat region of the effective potential with an accompanying non-analyticity at its boundary, marking the expectation value of the theory.

Investigating three different formulations of the equation, we found that certain families of algorithms, combined with (almost) exact Jacobians, are suitable to perform the RG-time integration to desired accuracy.
An overview of our heuristic opinion is collected \Cref{tbl:overview}.

Interestingly, a considerable number of algorithms feature the notion of pseudo stability, i.e., there exists a maximum step size which allows for properly resolving the non-analyticity at the minimum of the potential.
For smaller step sizes, the only limiting factor is the accuracy to resolve the singularity bound.

However, the log formulation, introduced in~\labelcref{eq:flow_log}, circumvents this issue.
This allows for integration to arbitrarily large RG-times, i.e., small RG-scales.

This immediately highlights future goals:
Firstly, to understand the different notions of stability arising from non-analyticities and from the singularity bound.
The discussion at the end of \Cref{sec:infinite_RG-time} could be a good starting point for a more formal investigation of the equation.
Secondly, to understand the generalization of the log formulation and its feasibility in more general theories.


\section*{Acknowledgements}
\label{Acknowledgments}

We would like to thank Eduardo Grossi, Adrian Königstein, Jan M. Pawlowski, Fabian Rennecke and Lorenz von Smekal for discussions and work on related topics.

FI acknowledges funding by the Studienstiftung des deutschen Volkes.
FRS acknowledges funding by the GSI Helmholtzzentrum für Schwerionenforschung.
NW acknowledges support by the Deutsche Forschungsgemeinschaft (DFG, German Research Foundation) – Project number 315477589 – TRR 211 and by the State of Hesse within the Research Cluster ELEMENTS (Project ID 500/10.006).

\appendix
\begin{table*}[ht]
\hspace{-100pt}
\begin{minipage}[t]{\columnwidth}
\vspace{0pt}
	\centering
	\ra{1.1}
	\begin{tabular}{l c c c}
		\toprule
		Name & Ref & Order  &  Rating \\
		\midrule\midrule
		\multicolumn{4}{c}{DIRK} \vspace{1mm} \\ 
		\midrule
		\texttt{ImplicitEuler} & & 1 & \tierC \\
		\texttt{Trapezoid} & & 2 & \tierC \\ 
		\texttt{TRBDF2} & \cite{hosea1996analysis} & 2 & \tierS \\
		\texttt{SDIRK2} & \cite{hindmarsh2005sundials} & 2 & \tierC \\
		\texttt{Kvaerno3} & \cite{kvaerno2004singly} & 3 & \tierC \\
		\texttt{KenCarp3} & \cite{kennedy2001additive} & 3 & \tierS \\
		\texttt{Cash4} & \cite{hindmarsh2005sundials} & 4 & \tierD \\
		\texttt{Hairer4} & \cite{wanner1996solving} & 4 & \tierA \\ 
		\texttt{Hairer42} & \cite{wanner1996solving} & 4 & \tierA \\
		\texttt{Kvaerno4} & \cite{kvaerno2004singly} & 4 & \tierB \\
		\texttt{KenCarp4} & \cite{kennedy2001additive} & 4 & \tierS \\
		\texttt{KenCarp47} & \cite{kennedy2019higher} & 4 & \tierS \\
		\texttt{Kvaerno5} & \cite{kvaerno2004singly} & 5 & \tierC \\
		\texttt{KenCarp5} & \cite{kennedy2001additive} & 5 & \tierB \\
		\texttt{KenCarp58} & \cite{kennedy2019higher} & 5 & \tierS \\
		\midrule\midrule
		\multicolumn{4}{c}{FIRK} \vspace{1mm} \\ 
		\midrule
		\texttt{RadauIIA3} & \cite{hairer1999stiff} & 3 & \tierC \\
		\texttt{RadauIIA5} & \cite{shampine1997matlab} & 5 & \tierS \\
		\midrule\midrule
		\multicolumn{4}{c}{Implicit multistep} \vspace{1mm} \\
		\midrule
		\texttt{QNDF1} & \cite{shampine1997matlab} & 1 & \tierB \\
		\texttt{QBDF1} & \cite{shampine1997matlab} & 1 & \tierB \\
		\texttt{ABDF2} & \cite{celaya2014implementation} & 2 & \tierB \\
		\texttt{QNDF2} & \cite{shampine1997matlab} & 2 & \tierA \\
		\texttt{QBDF2} & \cite{shampine1997matlab} & 2 & \tierA \\
		\texttt{QNDF} & \cite{shampine1997matlab} & v & \tierS \\ 
		\texttt{QBDF} & \cite{shampine1997matlab} & v & \tierS \\
		\texttt{FBDF} & \cite{shampine2002solving} & v & \tierA \\
		\bottomrule
	\end{tabular}
\end{minipage} ~
\begin{minipage}[t]{\columnwidth}
\vspace{0pt}
	\centering
	\ra{1.1}
	\begin{tabular}{l c c c}
		\toprule
		Name & Ref & Order  &  Rating \\
		\midrule\midrule
		\multicolumn{4}{c}{Rosenbrock} \vspace{1mm} \\ 
		\midrule
		\texttt{ROS3P} & \cite{lang2001ros3p} & 3 & \tierB \\
		\texttt{Rodas3} & \cite{wanner1996solving} & 3 & \tierB \\ 
		\texttt{RosShamp4} & \cite{shampine1982implementation} & 4 & \tierB \\
		\texttt{Veldd4} & \cite{van1984d} & 4 & \tierA \\
		\texttt{Velds4} & \cite{van1984d} & 4 & \tierA \\
		\texttt{GRK4T} & \cite{kaps1979generalized} & 4 & \tierA \\
		\texttt{GRK4A} & \cite{kaps1979generalized} & 4 & \tierA \\
		\texttt{Ros4LStab} & \cite{wanner1996solving} & 4 & \tierA \\ 
		\texttt{Rodas4} & \cite{wanner1996solving} & 4 & \tierA \\
		\texttt{Rodas42} & \cite{wanner1996solving} & 4 & \tierA \\
		\texttt{Rodas4P} & \cite{steinebach1995order} & 4 & \tierA \\
		\texttt{Rodas4P2} & \cite{steinebach2020improvement} & 4 & \tierA \\
		\texttt{Rodas5} & \cite{di1993rodas5} & 5 & \tierA \\
		\midrule\midrule
		\multicolumn{4}{c}{Rosenbrock-W} \vspace{1mm} \\ 
		\midrule
		\texttt{Rosenbrock23} & \cite{shampine1997matlab} & 2 & \tierC \\
		\texttt{Rosenbrock32} & \cite{shampine1997matlab} & 3 & \tierD \\ 
		\texttt{ROS34PW1a} & \cite{rang2005new} & 3 & \tierB \\
		\texttt{ROS34PW1b} & \cite{rang2005new} & 3 & \tierB \\
		\texttt{ROS34PW2} & \cite{rang2005new} & 3 & \tierA \\
		\texttt{ROS34PW3} & \cite{rang2005new} & 3 & \tierA \\
		\midrule\midrule
		\multicolumn{4}{c}{External libraries} \vspace{1mm} \\ 
		\midrule
		\texttt{CVODE\_BDF} & \cite{hindmarsh2005sundials, rackauckas2017differentialequations} & v & \tierS \\
		\texttt{lsoda} & \cite{hindmarsh1983odepack, petzold1983automatic, radhakrishnan1993description, rackauckas2017differentialequations} & v & \tierA \\
		\midrule\midrule
		\multicolumn{4}{c}{Implicit extrapolation} \vspace{1mm} \\ 
		\midrule
		\footnotesize{\texttt{ImplicitEulerExtrapolation}} & \cite{rackauckas2017differentialequations} & v & \tierC \\
		\footnotesize{\texttt{ImplicitDeuflhardExtrapolation}} & \cite{rackauckas2017differentialequations} & v & \tierD \\ 
		\footnotesize{\texttt{ImplicitHairerWannerExtrapolation}} & \cite{rackauckas2017differentialequations} & v & \tierC \\
		\bottomrule
	\end{tabular}
\end{minipage}
	\caption{Overview of all methods explored in this work. Their names are given as used in the \texttt{DifferentialEquations.jl} package~\cite{rackauckas2017differentialequations}. If the order of the algorithms is variable and may be changed during runtime, it is denoted as "v". Additionally, we give a subjective rating of each algorithm for the application at hand, which should serve as guidance only. The ratings are explained in \labelcref{eq:tier_rating_alg}, with \tierS\phantom{a}being the best and \tierD\phantom{a}the worst.}
    \label{tbl:overview}
\end{table*}
\section{Functional Renormalization Group}
\label{app:fRG}
We briefly summarize the central ideas of the functional renormalization group (fRG) and introduce our notation, for reviews see \cite{Dupuis:2020fhh} and references therein. The RG describes the evolution of a QFT along the (momentum) RG-scale $k$. It interpolates between the microscopic theory at small scales (large $k$) and the macroscopic theory at large scales (small $k$), by successively integrating out fluctuations. The fRG provides a modern functional implementation of this idea,
\begin{align}
\label{eq:wetterich}
    \partial_t \Gamma_k[\phi] = \frac{1}{2} \mathrm{Tr}\ \left[\left(\Gamma^{(2)} + R_k \right)^{-1}\, \partial_t R_k\right]
\, ,
\end{align}
the Wetterich equation~\cite{Wetterich:1992yh}. In \labelcref{eq:wetterich}, the RG-time $t$ is given by
\begin{align}
\label{eq:RG-time}
    t=-\ln{\frac{k}{\Lambda}}
\, ,
\end{align}
with some (ultraviolet) reference scale $\Lambda$, which is usually used as initialization scale of the flow.
In \labelcref{eq:RG-time} we would like to point out the extra minus sign compared to most fRG literature, which turns the RG-time evolution positive instead of negative.
$\Gamma_k[\Phi]$ in \labelcref{eq:wetterich} is the quantum effective action (QEA) at a given RG-scale $k$, which has to be truncated to yield a finite set of equations, c.f.~\Cref{sec:discretizations}.
$R_k$ denotes the regulator, which acts like a mass term $\sim k^2$ and hence renders loops infrared finite, while its decay in momentum space renders diagrams via its RG-time derivative in \labelcref{eq:wetterich} ultraviolet finite.
Lastly, the trace $\mathrm{Tr}$ collects summation/integration over indices/arguments, which includes an integration over spacetime and summation/integration over internal indices, which might be present, depending on the theory under investigation.

The QEA $\Gamma_k[\Phi]$ approaches the classical action $S[\Phi]$ of the theory for $t\to-\infty$ ($k\to\infty$).
On the other hand, the regularization is removed in the limit $t\to\infty$ ($k\to0$), where the physical theory, in a given truncation, is recovered.
This RG evolution is provided by \labelcref{eq:wetterich}, which typically results in a set of coupled PDEs.
A common truncation scheme is the standard derivative expansion, whose leading order is called \textit{local potential approximation} (LPA). In this approximation, only the effective potential enters as RG-scale dependent quantity, and one is left with a single PDE.

\section{Overview solver}
\label{app:overview}
\Cref{tbl:overview} collects all algorithms explored extensively in this work. The tables collect the name of the method, as implement in~\cite{rackauckas2017differentialequations}, their reference(s) and their order (with v=variable).
Additionally, we list a rating, which reflects our subjective view on the performance of the algorithm for the problem at hand.
Hereby, we included the work-precision performance for all three formulations \labelcref{eq:flow_ui}, \labelcref{eq:flow_msqi} and \labelcref{eq:flow_logi}, as well as the fermionic extension of the problem \labelcref{eq:flow_ui_QM}.
The RG-time stability at fixed step size, detailed in \Cref{sec:infinite_RG-time}, also played a significant role.
Finally, when in doubt how to rank an algorithm, we have also considered the ease of implementation.
For example, \texttt{TRBDF2} is comparatively easy to implement, while Rosenbrock methods add another level of complexity.
The rating scheme is as follows:
\begin{equation}
\begin{aligned}
\label{eq:tier_rating_alg}
    &\text{\tierS} \quad \text{Best performance, little to no drawbacks} \\
    &\text{\tierA} \quad \text{Overall very good performance} \\
    &\text{\tierB} \quad \text{Works well, but some drawbacks} \\
    &\text{\tierC} \quad \text{Converges, but inefficient} \\
    &\text{\tierD} \quad \text{Major convergence/stability issues}
\end{aligned}
\end{equation}
We would like to stress again that this rating is subjective and based on our experience while completing this study.
Nevertheless, we hope that it might be a helpful starting point for choosing an algorithm.


\begin{figure*}[h!]
	\centering
	\begin{subfigure}[t]{0.45\textwidth}
		\centering
		\includegraphics[width=\linewidth]{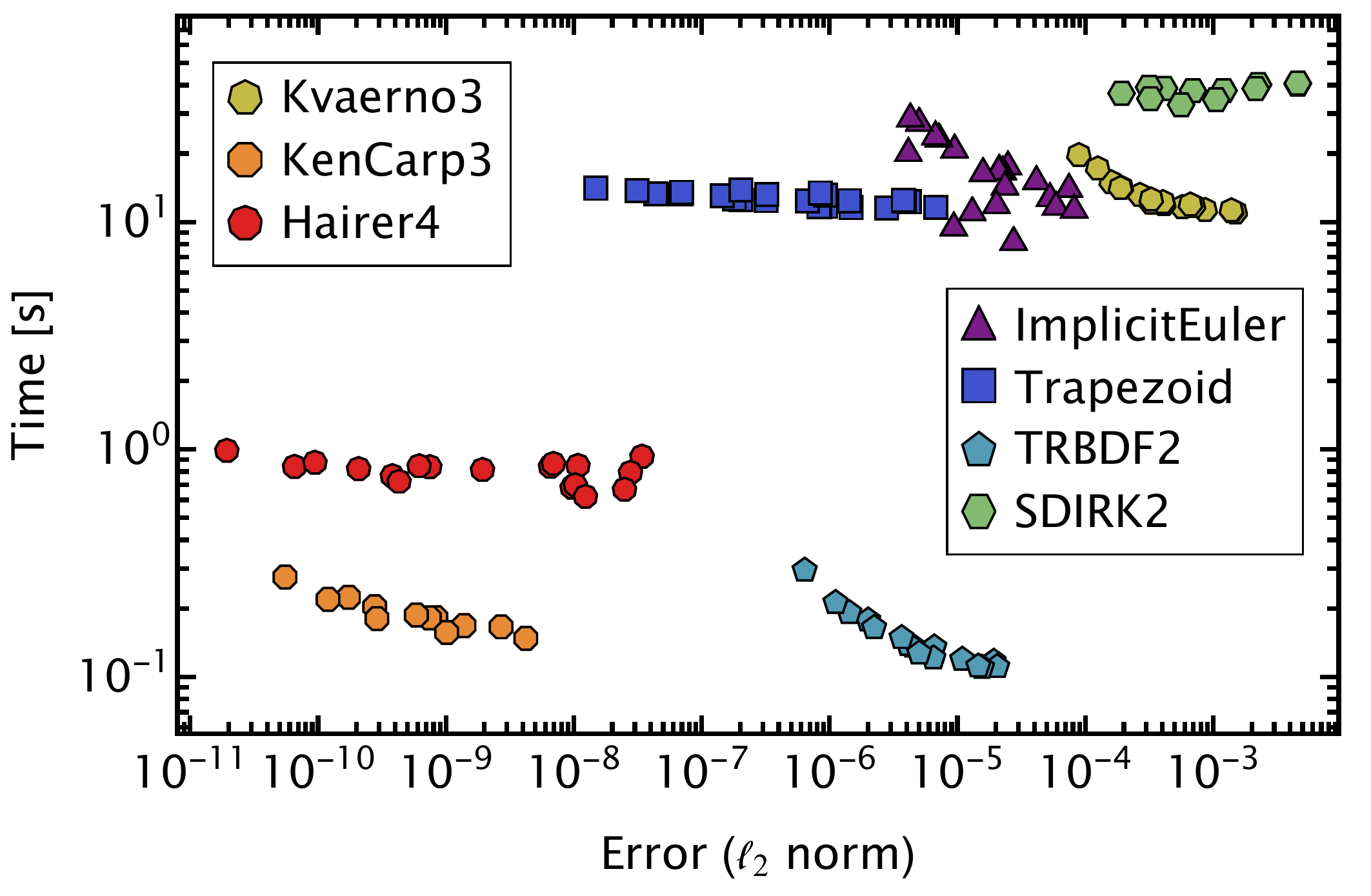}
		\caption{DIRK 1}
		\label{fig:std_WP_DIRK1}
	\end{subfigure}	~
	\begin{subfigure}[t]{0.45\textwidth}
		\centering
		\includegraphics[width=\linewidth]{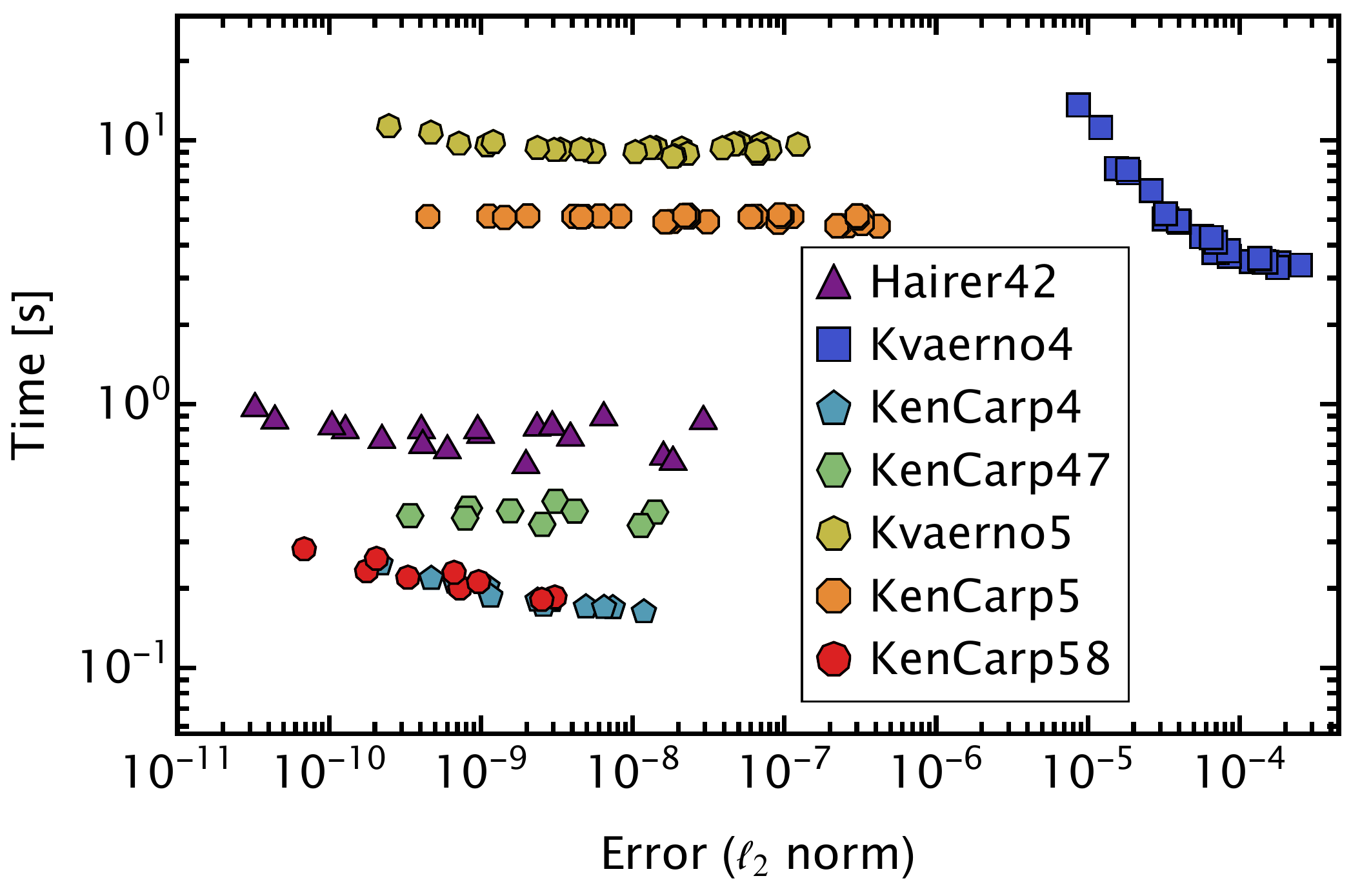}
		\caption{DIRK 2}
		\label{fig:std_WP_DIRK2}
	\end{subfigure}
	\begin{subfigure}[t]{0.45\textwidth}
		\centering
		\includegraphics[width=\linewidth]{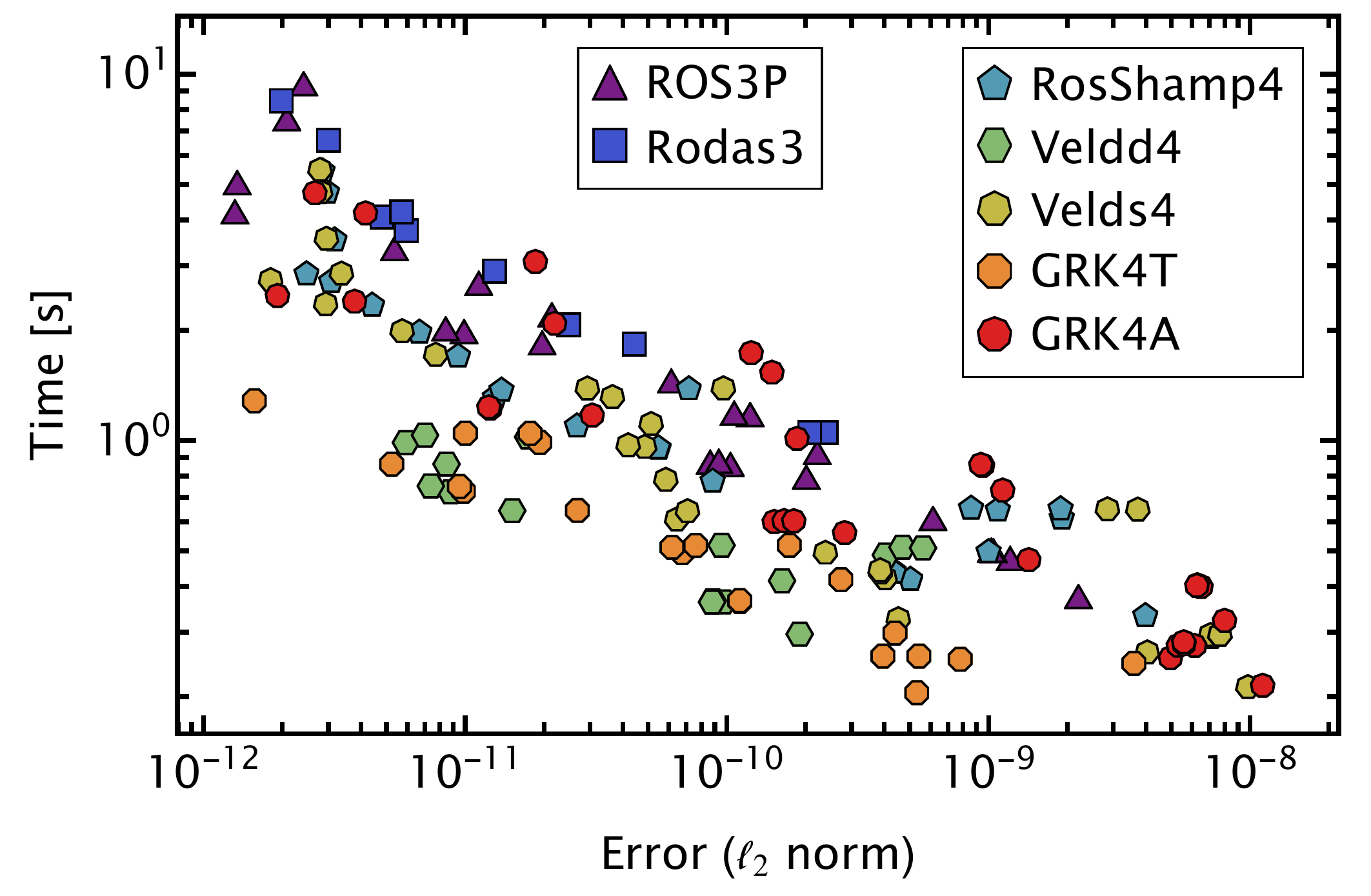}
		\caption{Rosenbrock 1}
		\label{fig:std_WP_RB1}
	\end{subfigure}	~
	\begin{subfigure}[t]{0.45\textwidth}
		\centering
		\includegraphics[width=\linewidth]{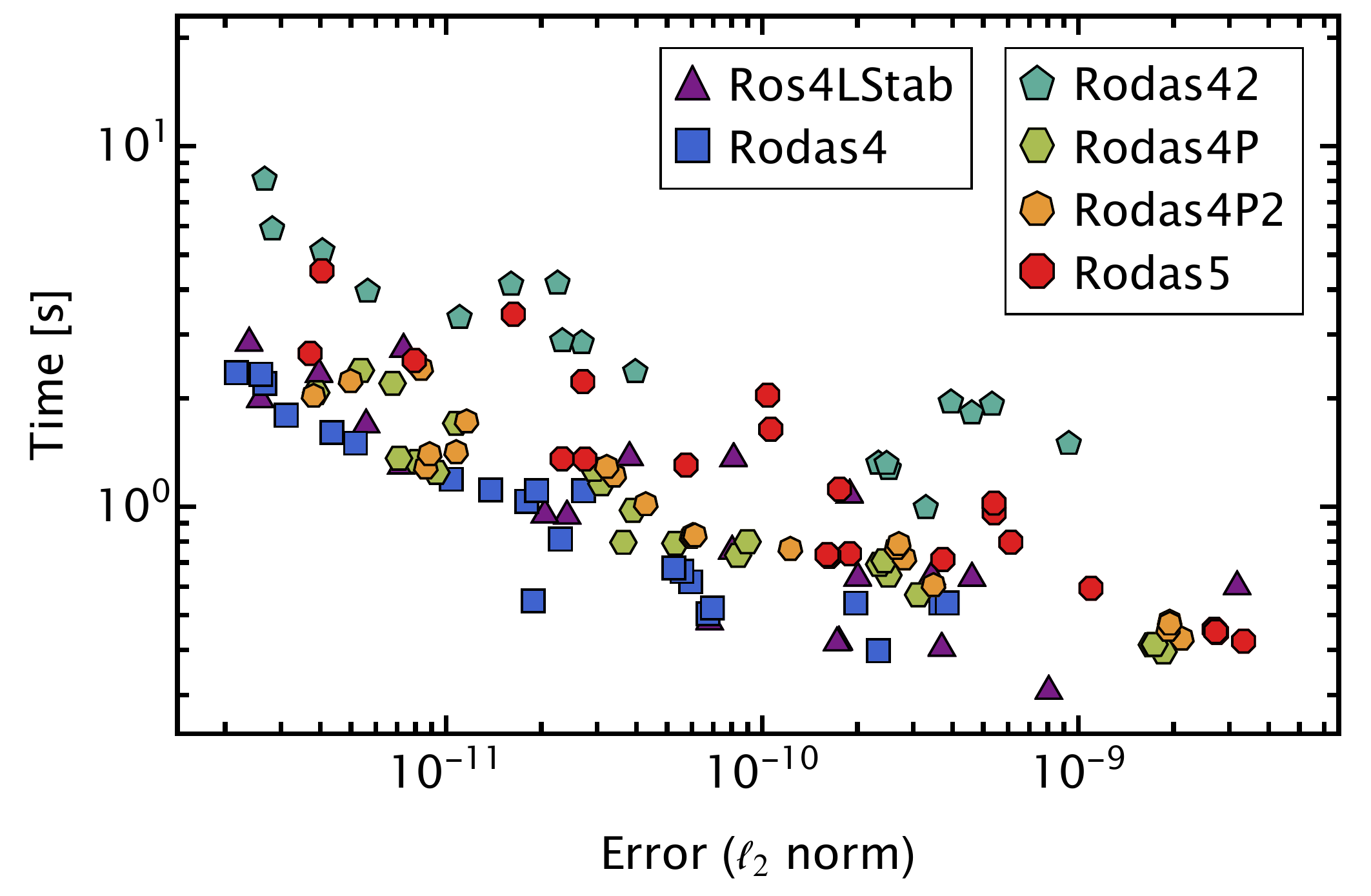}
		\caption{Rosenbrock 2}
		\label{fig:std_WP_RB2}
	\end{subfigure}
		\begin{subfigure}[t]{0.45\textwidth}
		\centering
		\includegraphics[width=\linewidth]{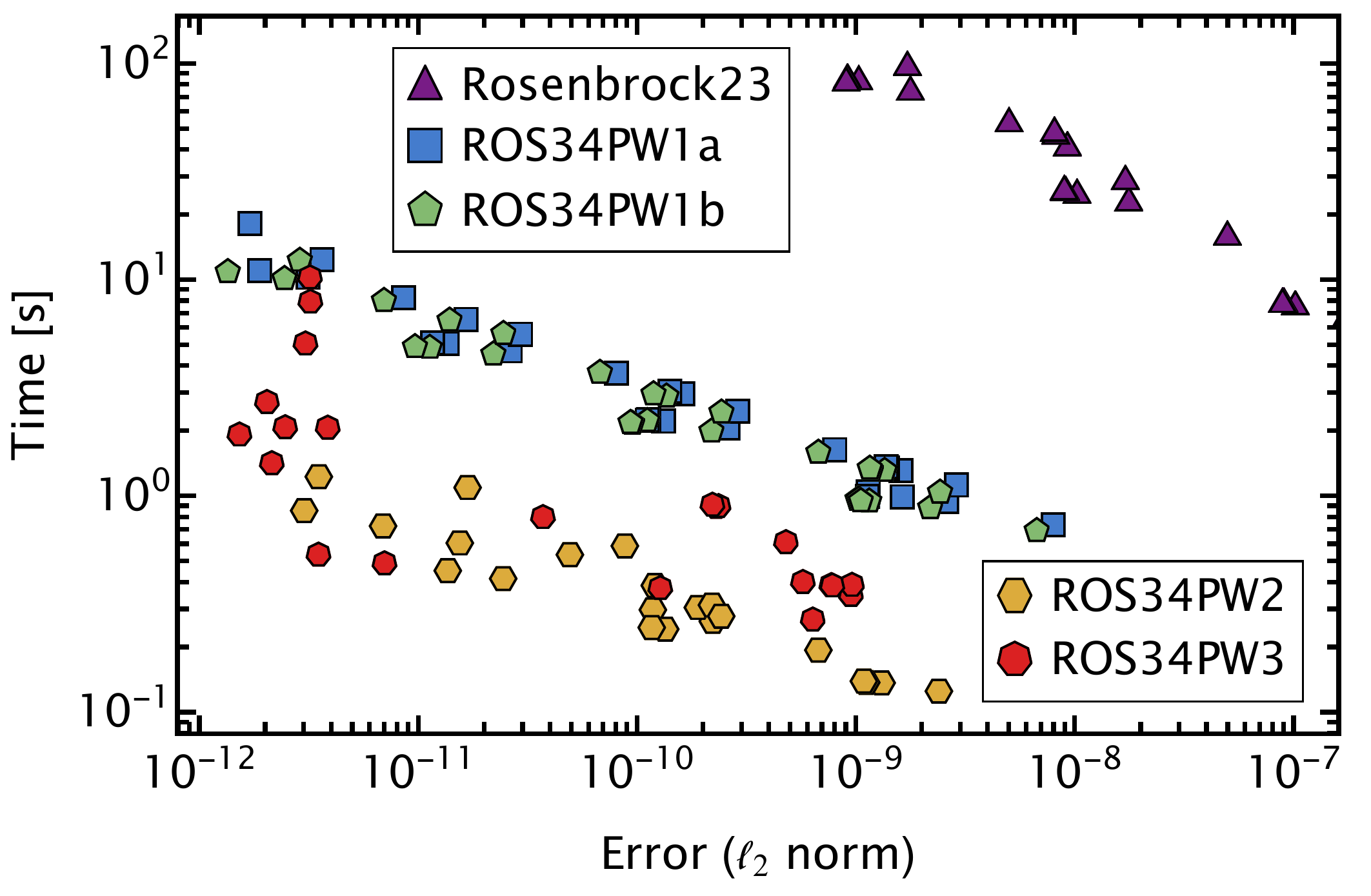}
		\caption{Rosenbrock-W}
		\label{fig:std_WP_RBW}
	\end{subfigure}	~
	\begin{subfigure}[t]{0.45\textwidth}
		\centering
		\includegraphics[width=\linewidth]{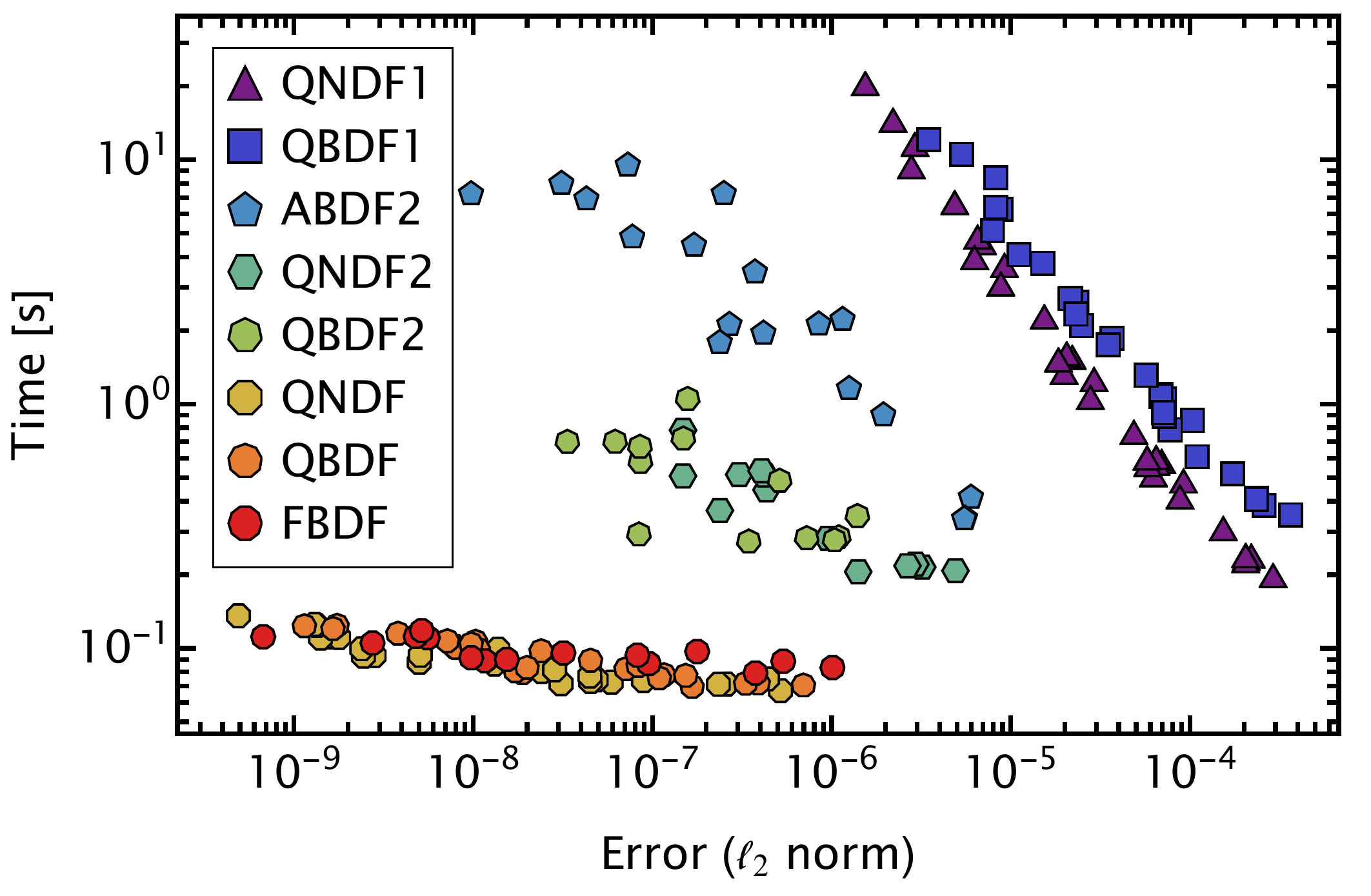}
		\caption{Implicit multistep}
		\label{fig:std_WP_IM}
	\end{subfigure}
	\begin{subfigure}[t]{0.45\textwidth}
		\centering
		\includegraphics[width=\linewidth]{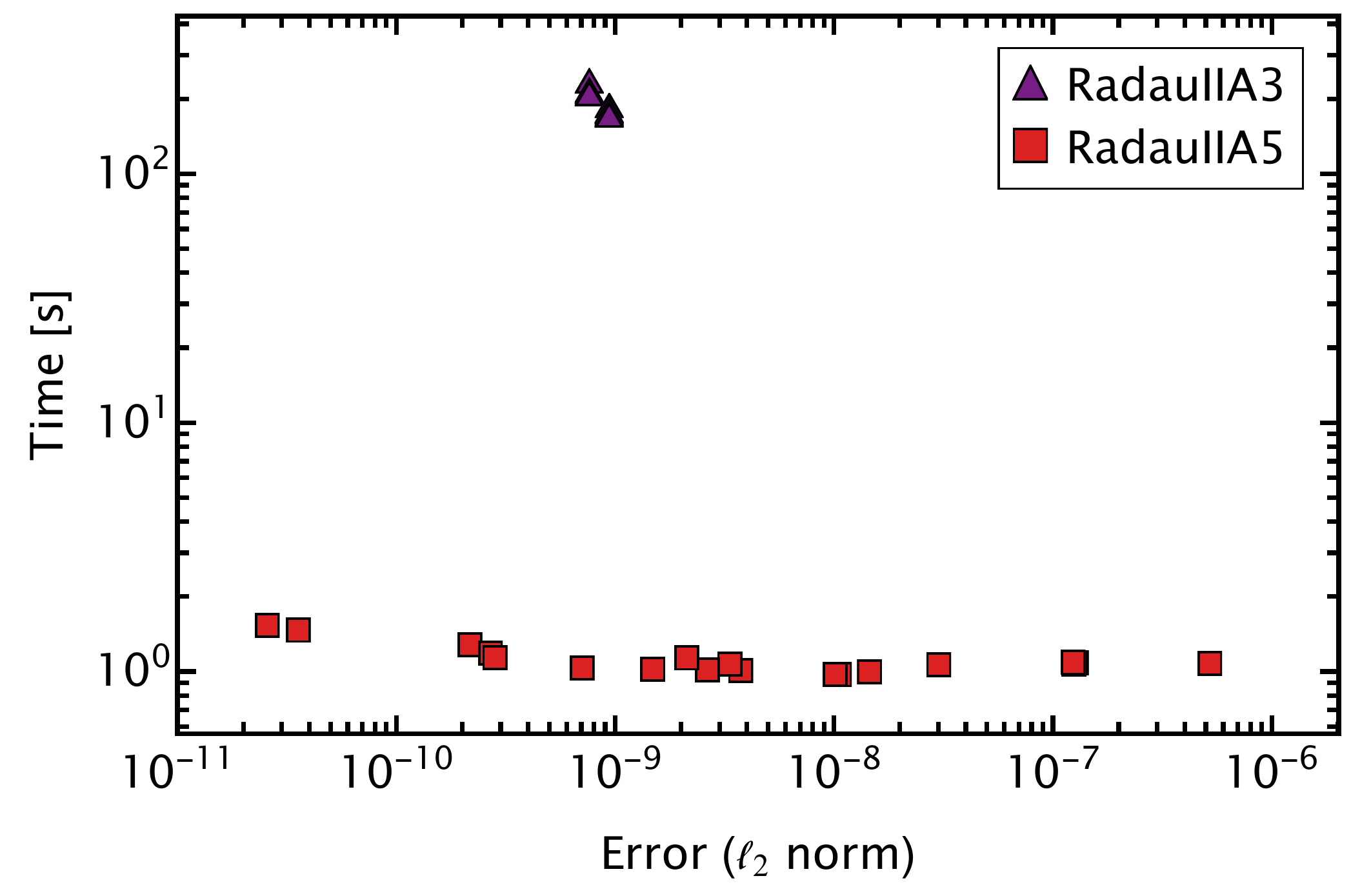}
		\caption{RadauII}
		\label{fig:std_WP_RadauII}
	\end{subfigure}	~
	\begin{subfigure}[t]{0.45\textwidth}
		\centering
		\includegraphics[width=\linewidth]{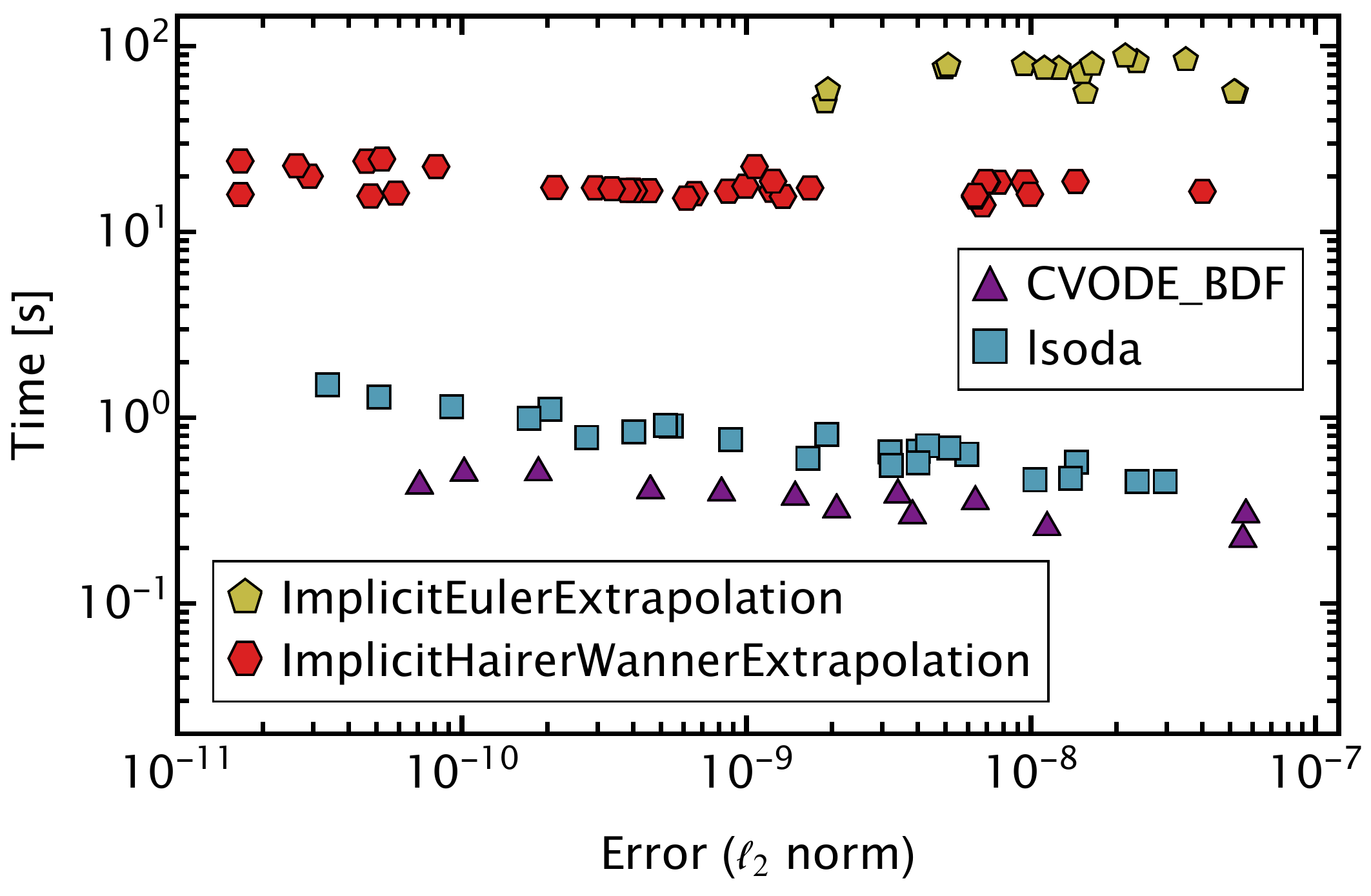}
		\caption{External libraries \& implicit extrapolation}
		\label{fig:std_WP_ext}
	\end{subfigure}%
	\caption{Work-precision survey for the standard formulation \labelcref{eq:flow_ui}.}
	\label{fig:std_WP_all}
	\vspace{-30pt} 
\end{figure*}
%
\begin{figure*}[h!]
	\centering
	\begin{subfigure}[t]{0.45\textwidth}
		\centering
		\includegraphics[width=\linewidth]{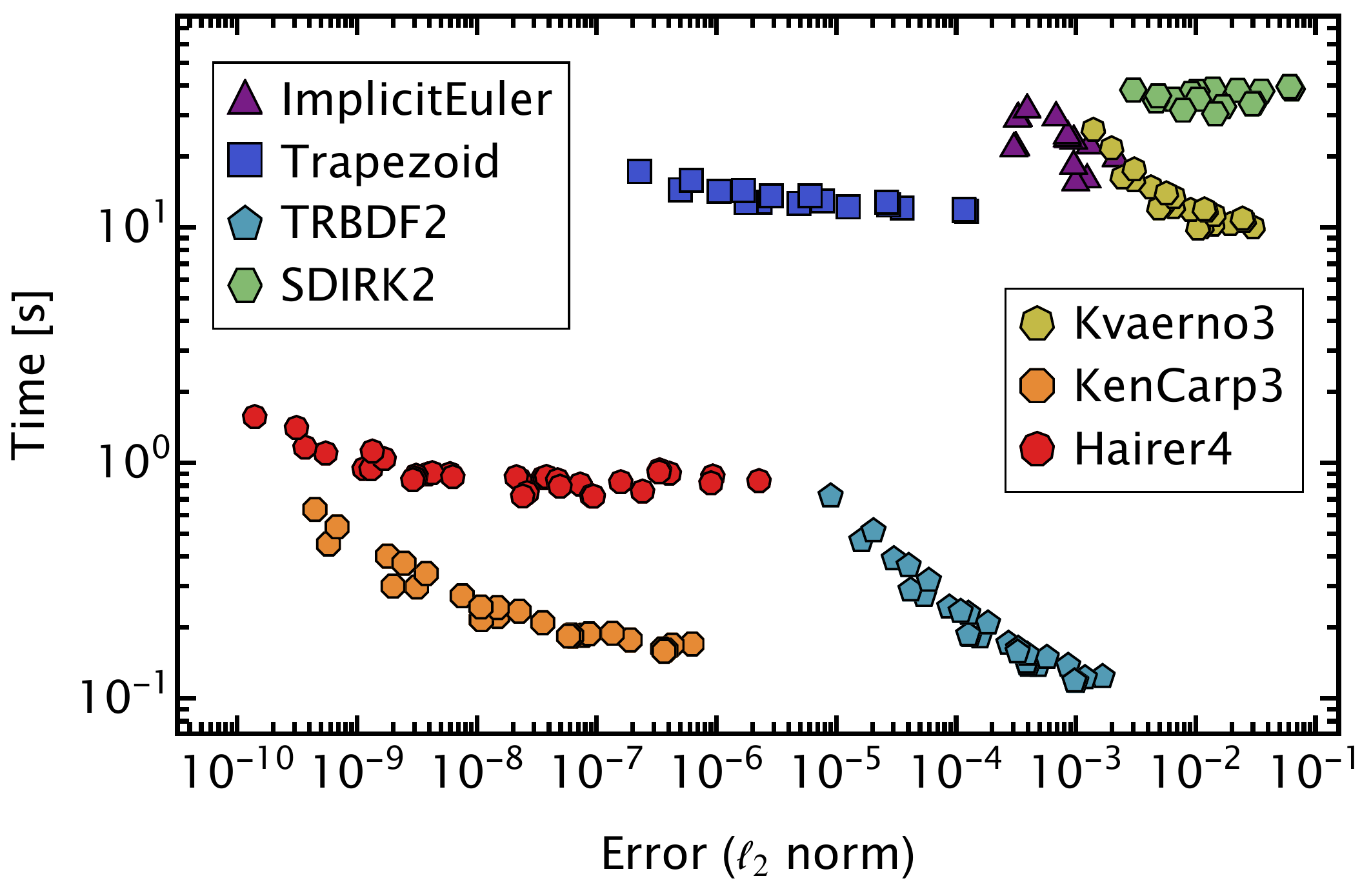}
		\caption{DIRK 1}
		\label{fig:msq_WP_DIRK1}
	\end{subfigure}	~
	\begin{subfigure}[t]{0.45\textwidth}
		\centering
		\includegraphics[width=\linewidth]{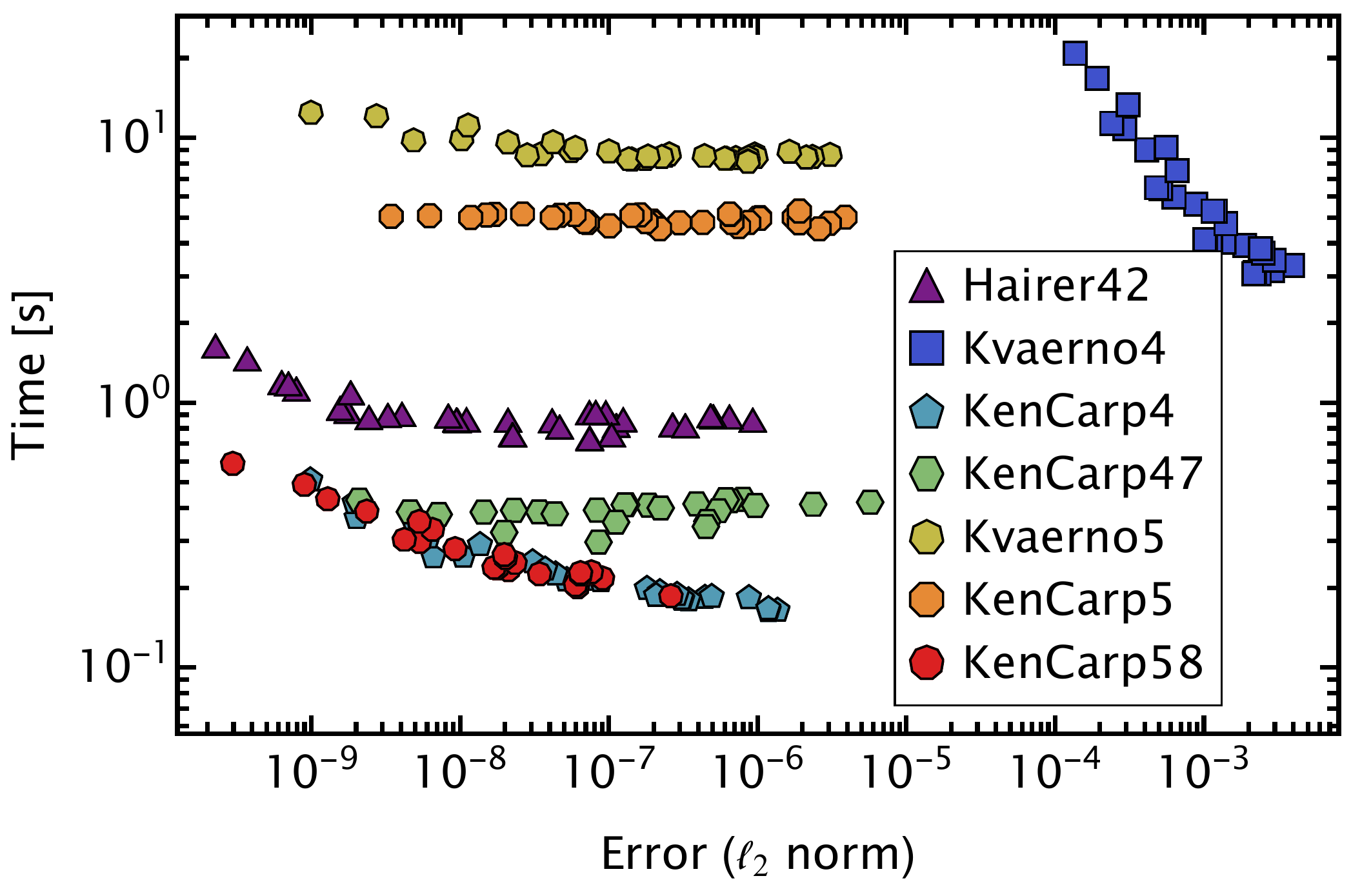}
		\caption{DIRK 2}
		\label{fig:msq_WP_DIRK2}
	\end{subfigure}
	\begin{subfigure}[t]{0.45\textwidth}
		\centering
		\includegraphics[width=\linewidth]{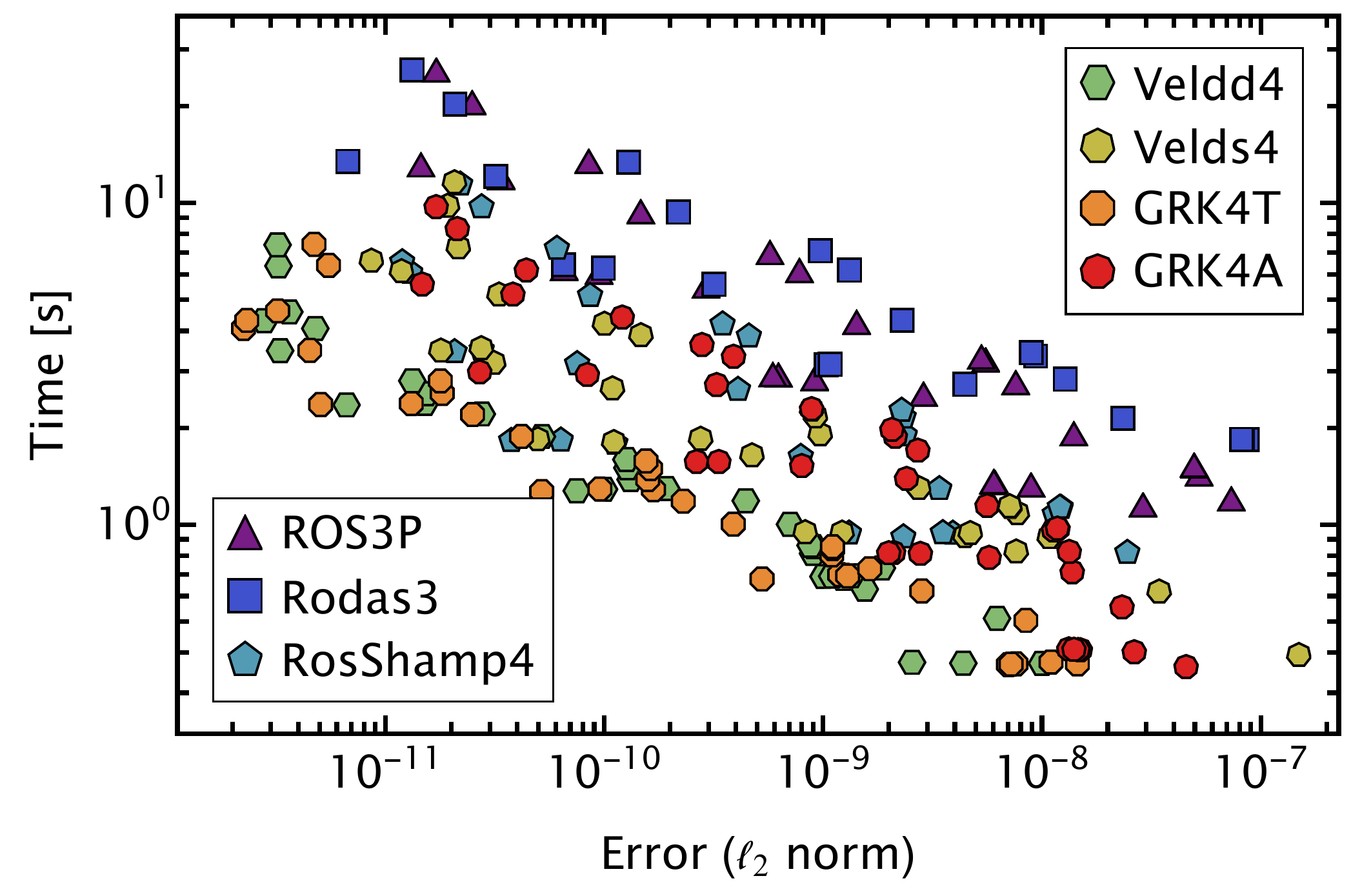}
		\caption{Rosenbrock 1}
		\label{fig:msq_WP_RB1}
	\end{subfigure}	~
	\begin{subfigure}[t]{0.45\textwidth}
		\centering
		\includegraphics[width=\linewidth]{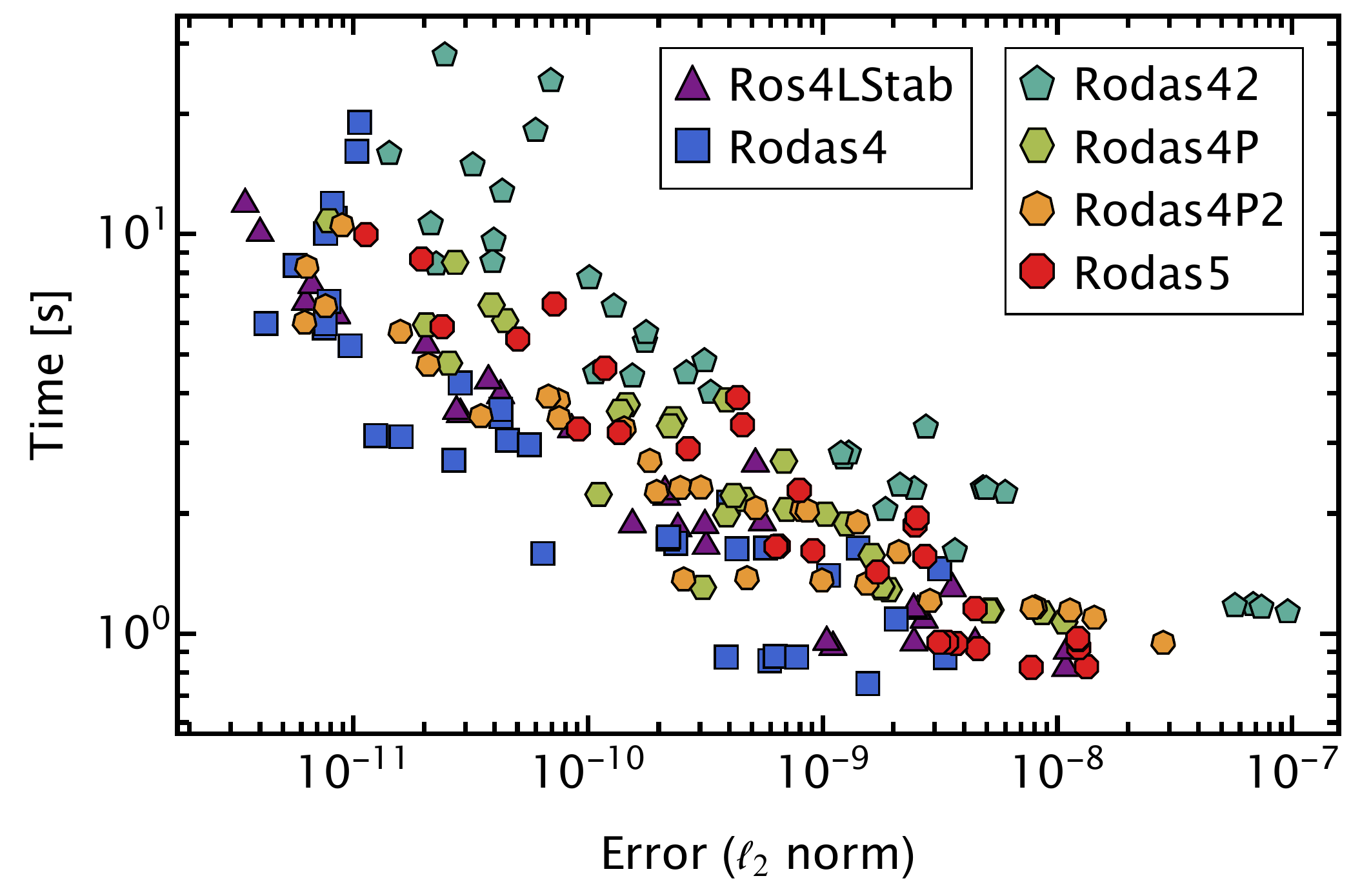}
		\caption{Rosenbrock 2}
		\label{fig:msq_WP_RB2}
	\end{subfigure}
		\begin{subfigure}[t]{0.45\textwidth}
		\centering
		\includegraphics[width=\linewidth]{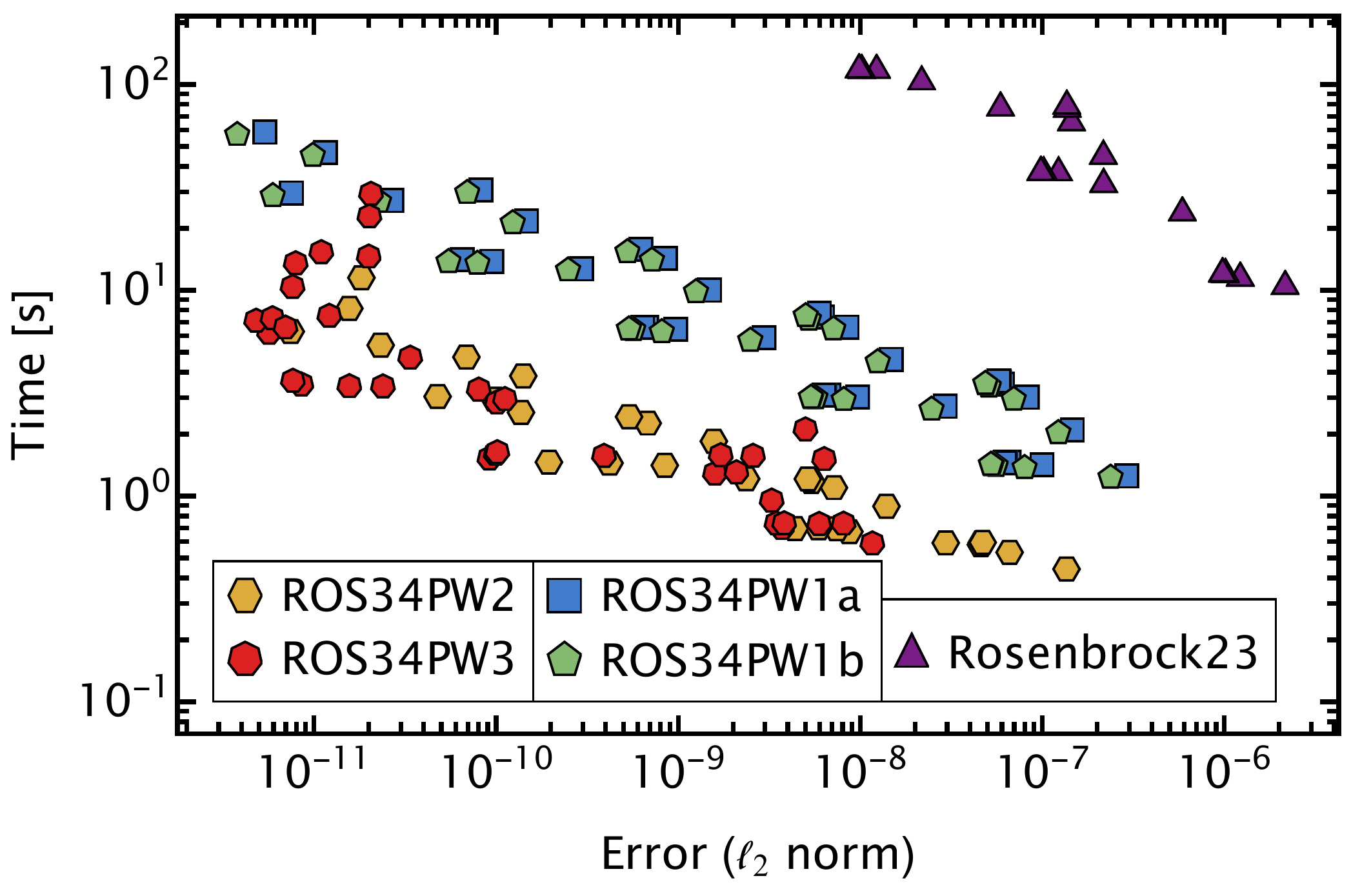}
		\caption{Rosenbrock-W}
		\label{fig:msq_WP_RBW}
	\end{subfigure}	~
	\begin{subfigure}[t]{0.45\textwidth}
		\centering
		\includegraphics[width=\linewidth]{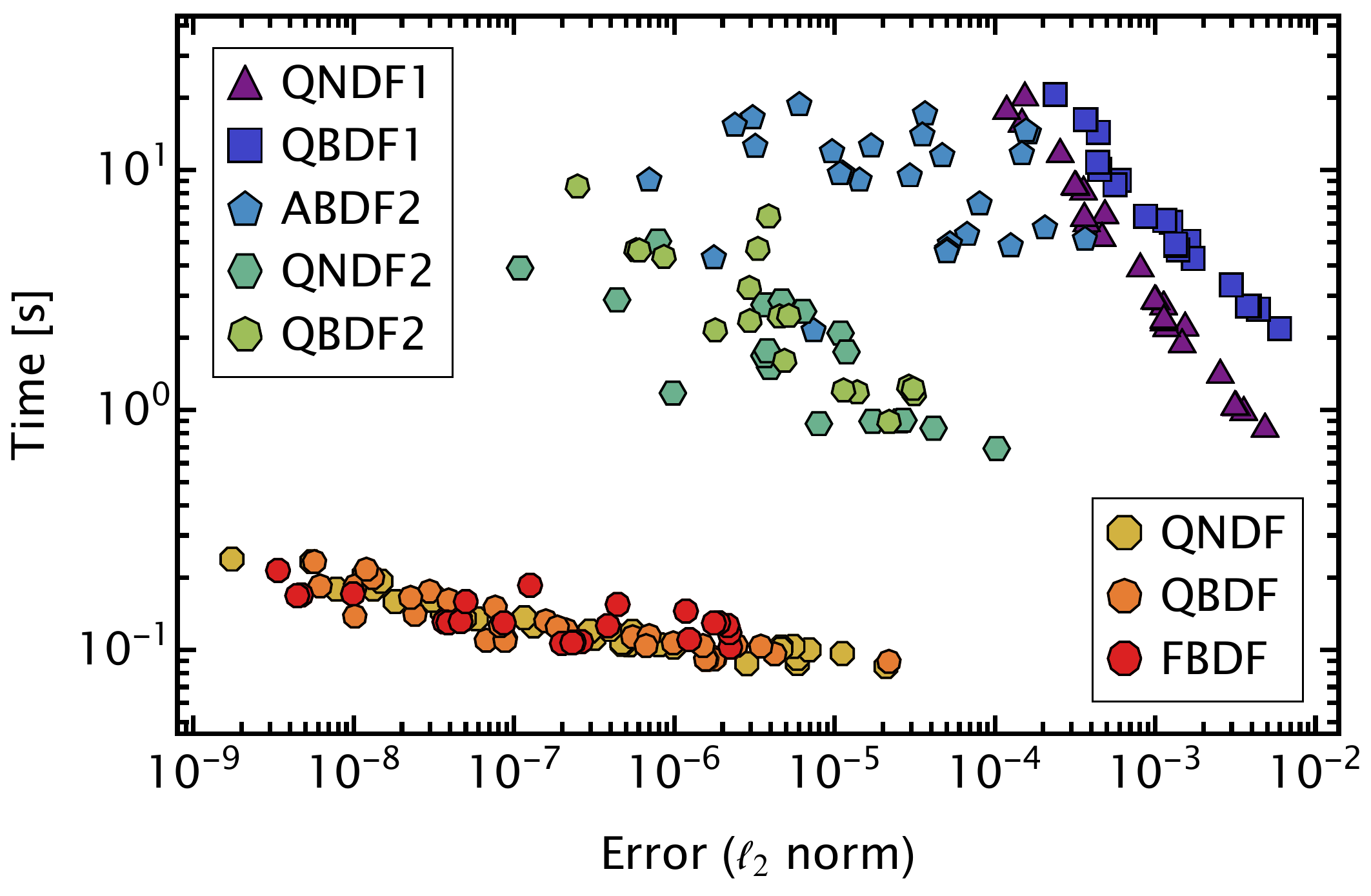}
		\caption{Implicit multistep}
		\label{fig:msq_WP_IM}
	\end{subfigure}
	\begin{subfigure}[t]{0.45\textwidth}
		\centering
		\includegraphics[width=\linewidth]{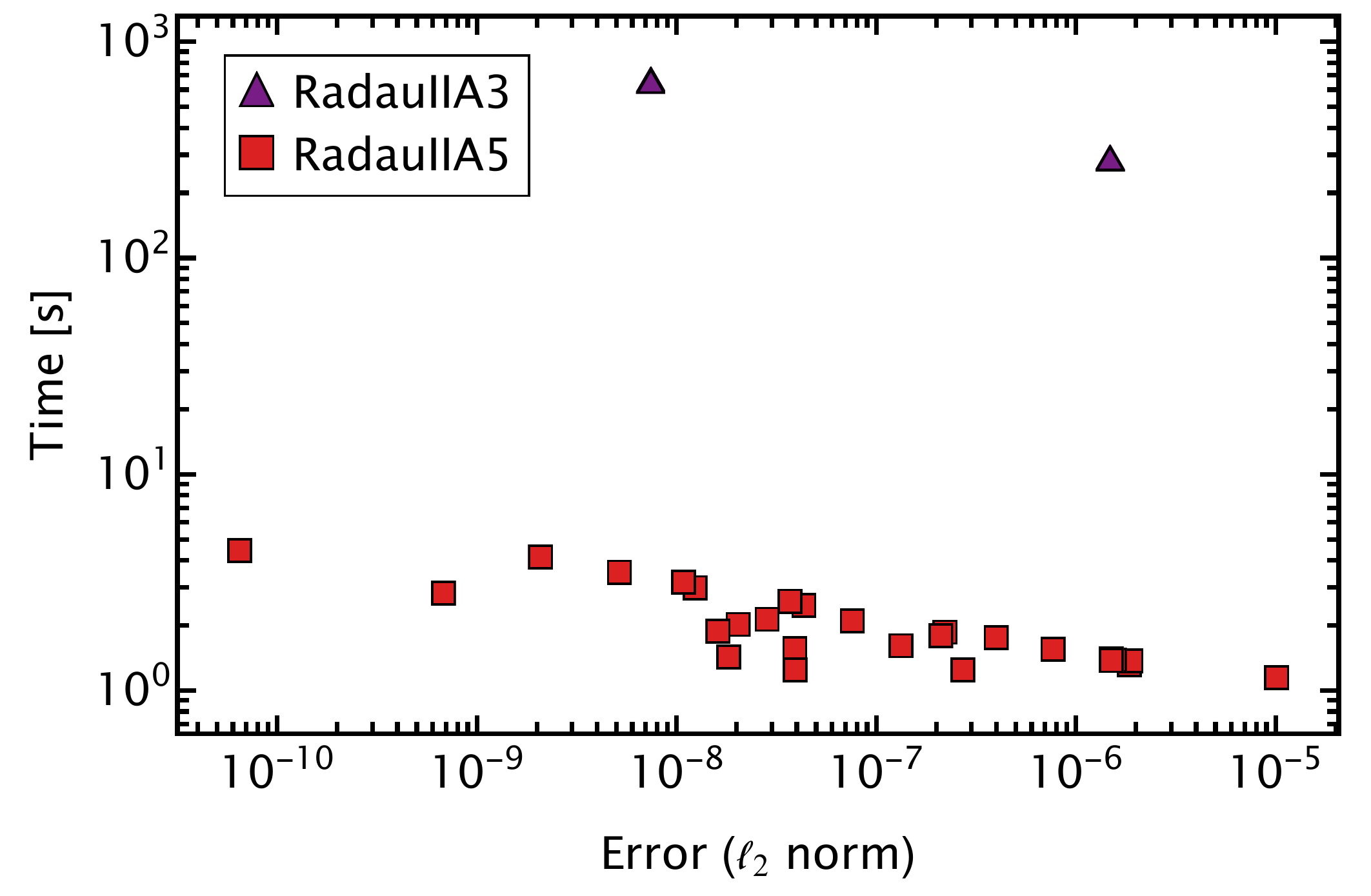}
		\caption{RadauII}
		\label{fig:msq_WP_RadauII}
	\end{subfigure}	~
	\begin{subfigure}[t]{0.45\textwidth}
		\centering
		\includegraphics[width=\linewidth]{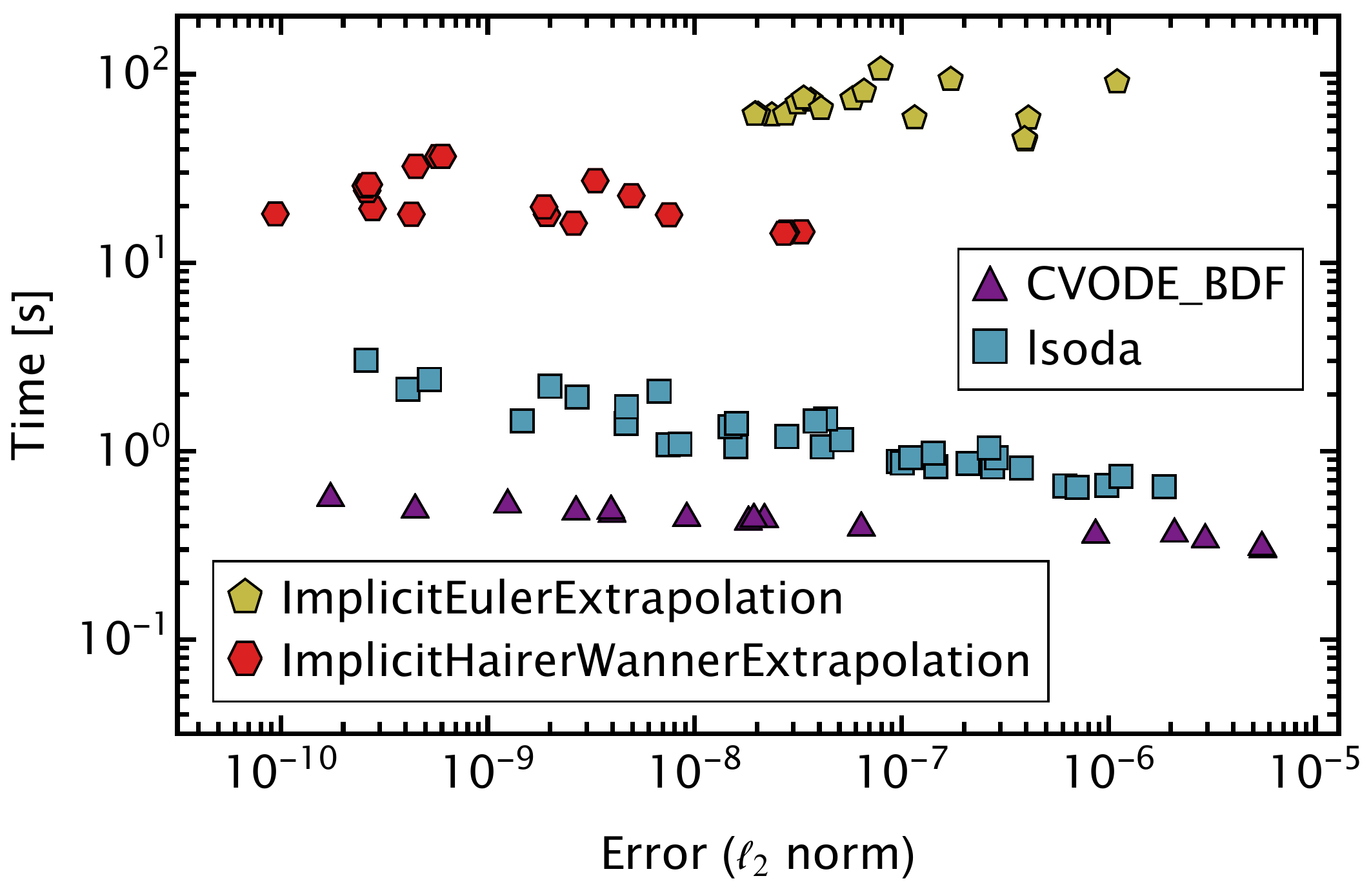}
		\caption{External libraries \& implicit extrapolation}
		\label{fig:msq_WP_ext}
	\end{subfigure}%
	\caption{Work-precision survey for the mass formulation \labelcref{eq:flow_msqi}.}
	\label{fig:msq_WP_all}
	\vspace{-30pt} 
\end{figure*}
%
\begin{figure*}[h!]
	\centering
	\begin{subfigure}[t]{0.45\textwidth}
		\centering
		\includegraphics[width=\linewidth]{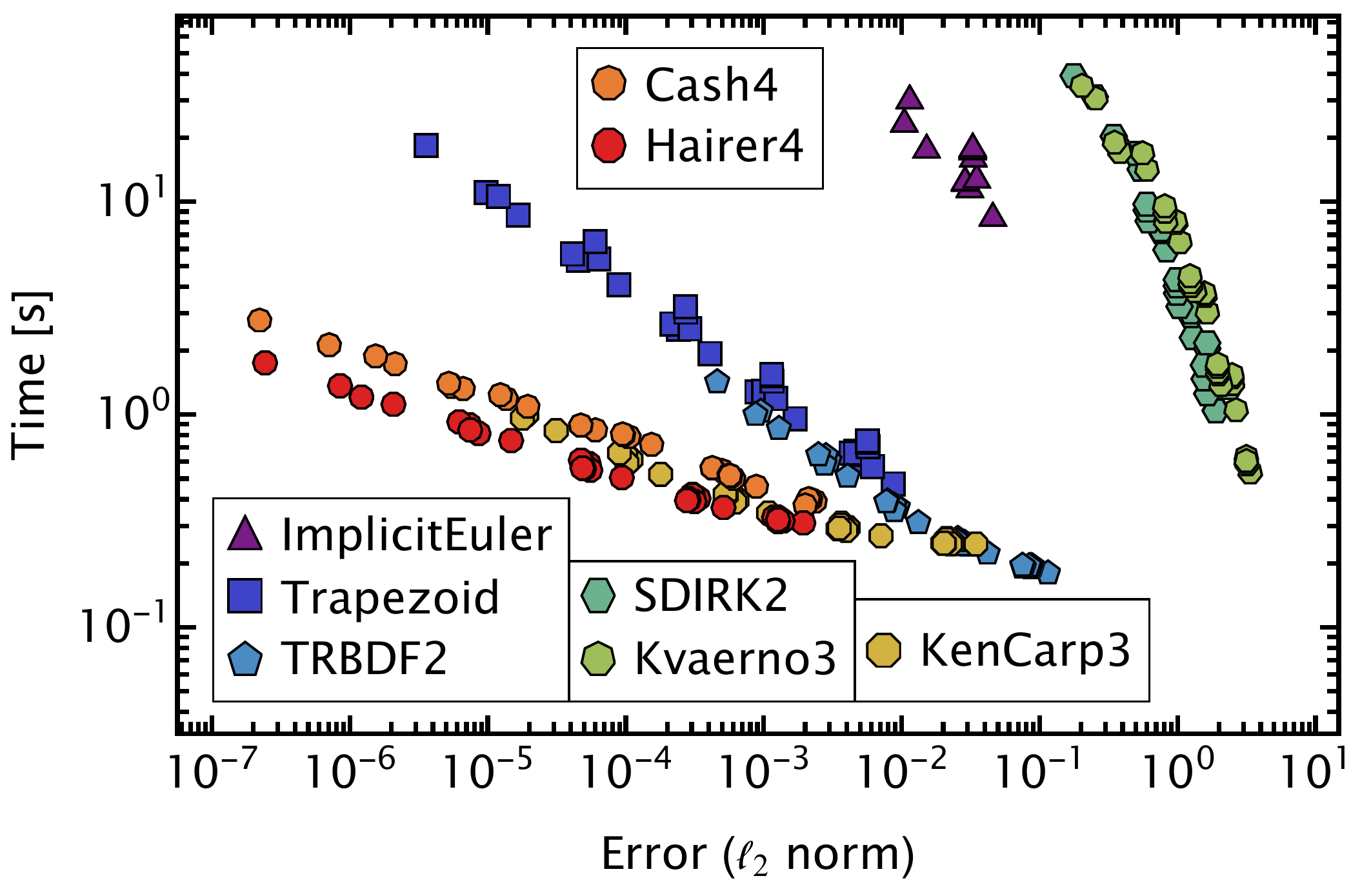}
		\caption{DIRK 1}
		\label{fig:log_WP_DIRK1}
	\end{subfigure}	~
	\begin{subfigure}[t]{0.45\textwidth}
		\centering
		\includegraphics[width=\linewidth]{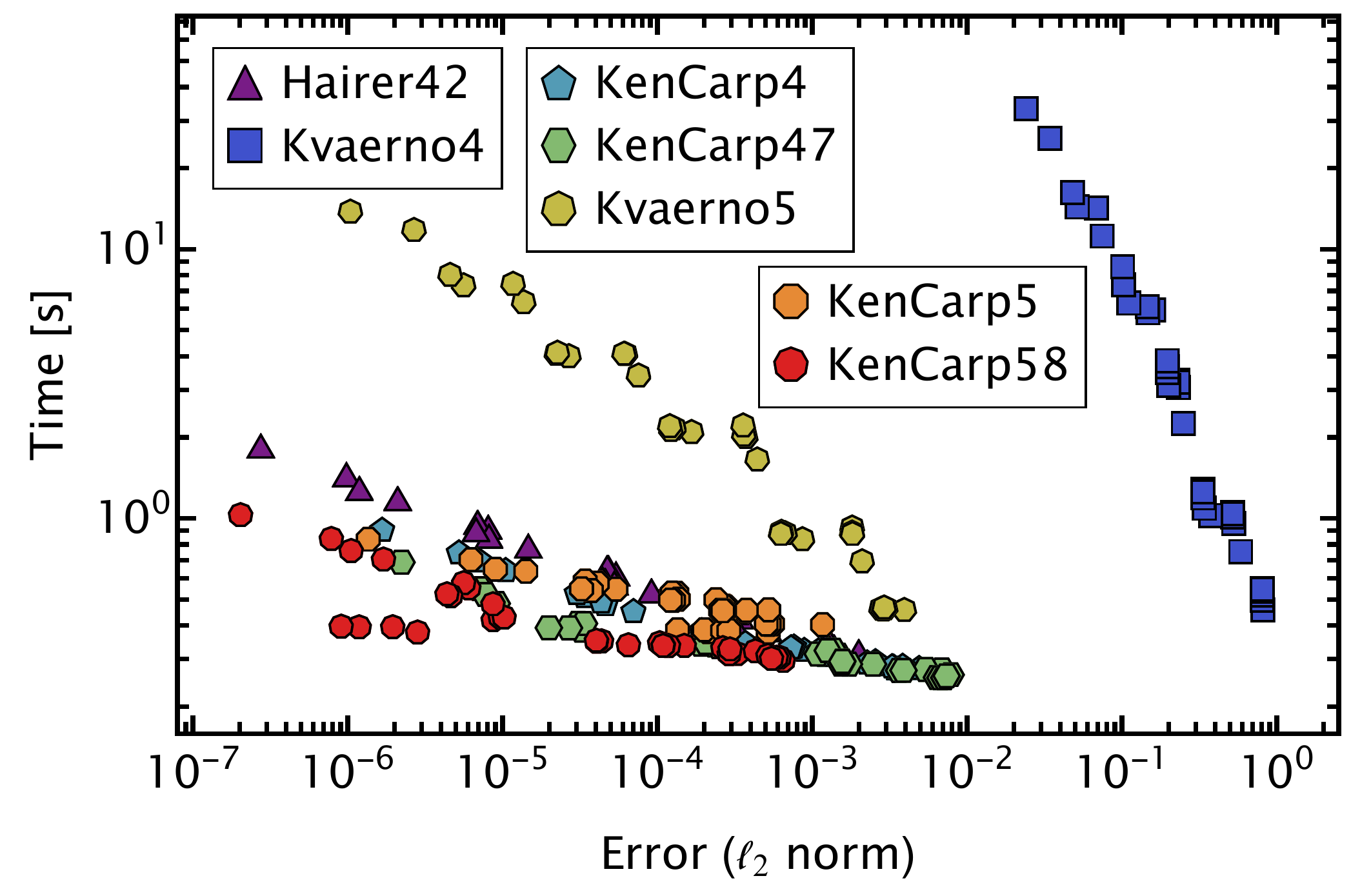}
		\caption{DIRK 2}
		\label{fig:log_WP_DIRK2}
	\end{subfigure}
	\begin{subfigure}[t]{0.45\textwidth}
		\centering
		\includegraphics[width=\linewidth]{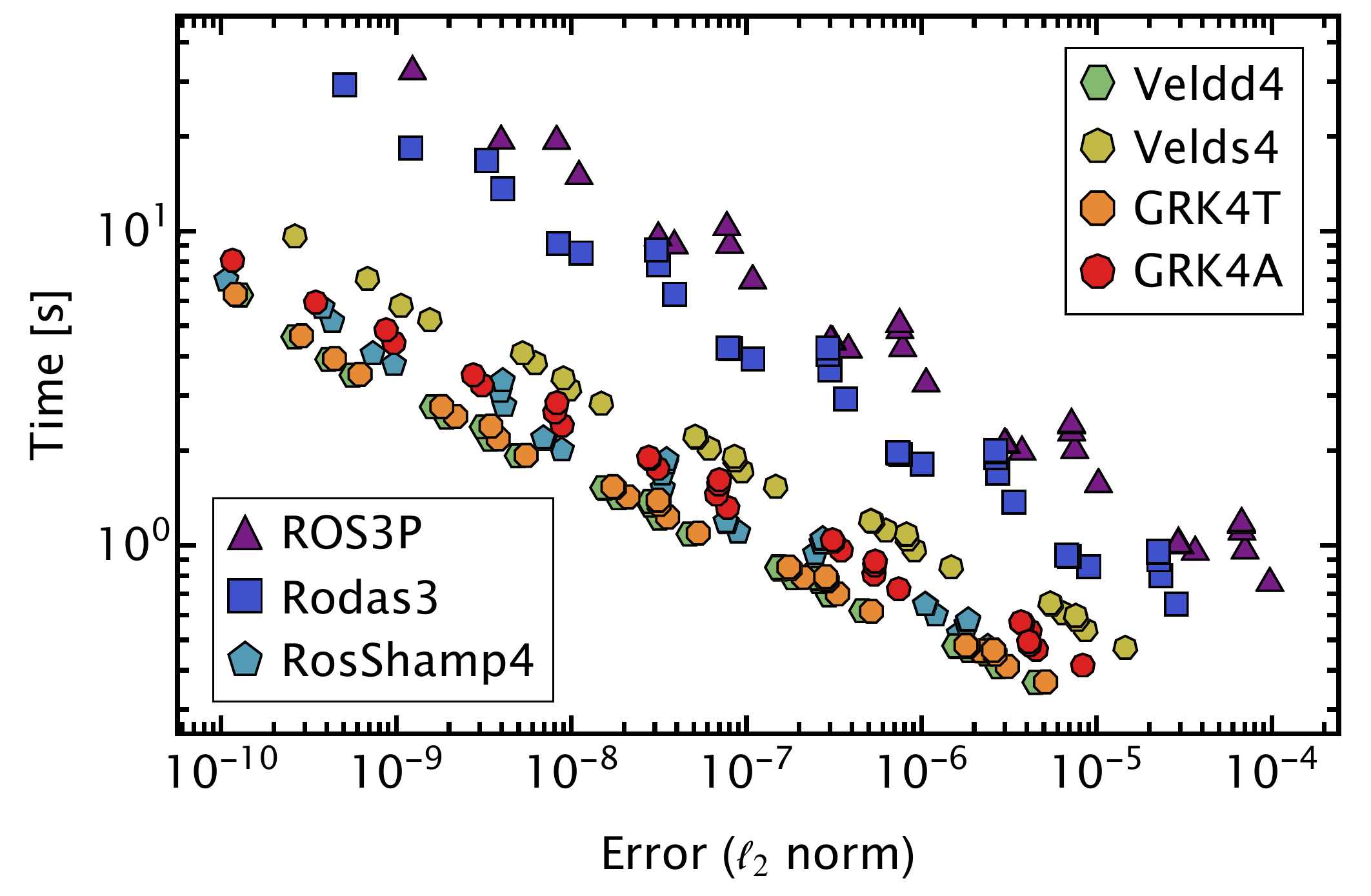}
		\caption{Rosenbrock 1}
		\label{fig:log_WP_RB1}
	\end{subfigure}	~
	\begin{subfigure}[t]{0.45\textwidth}
		\centering
		\includegraphics[width=\linewidth]{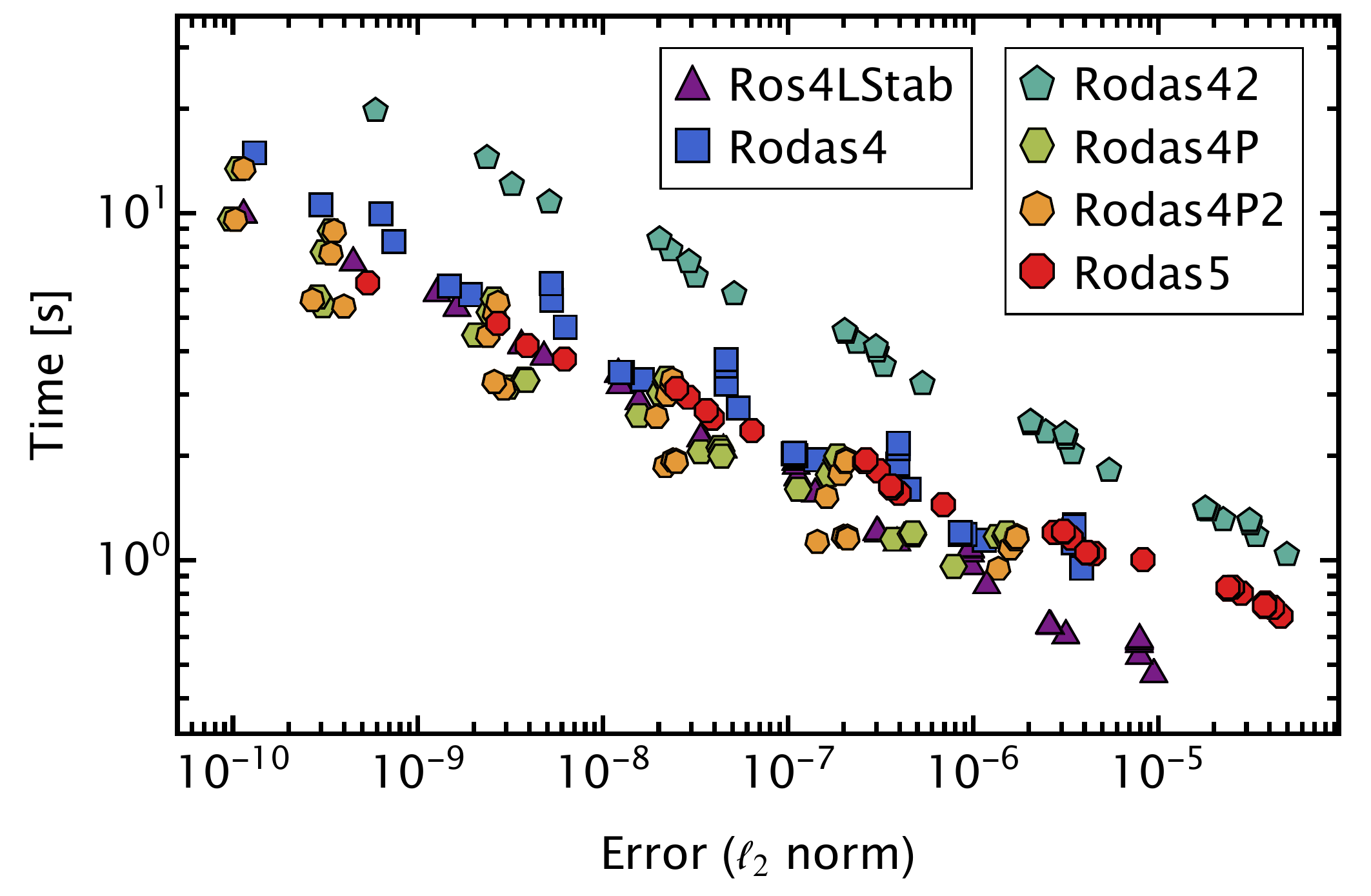}
		\caption{Rosenbrock 2}
		\label{fig:log_WP_RB2}
	\end{subfigure}
		\begin{subfigure}[t]{0.45\textwidth}
		\centering
		\includegraphics[width=\linewidth]{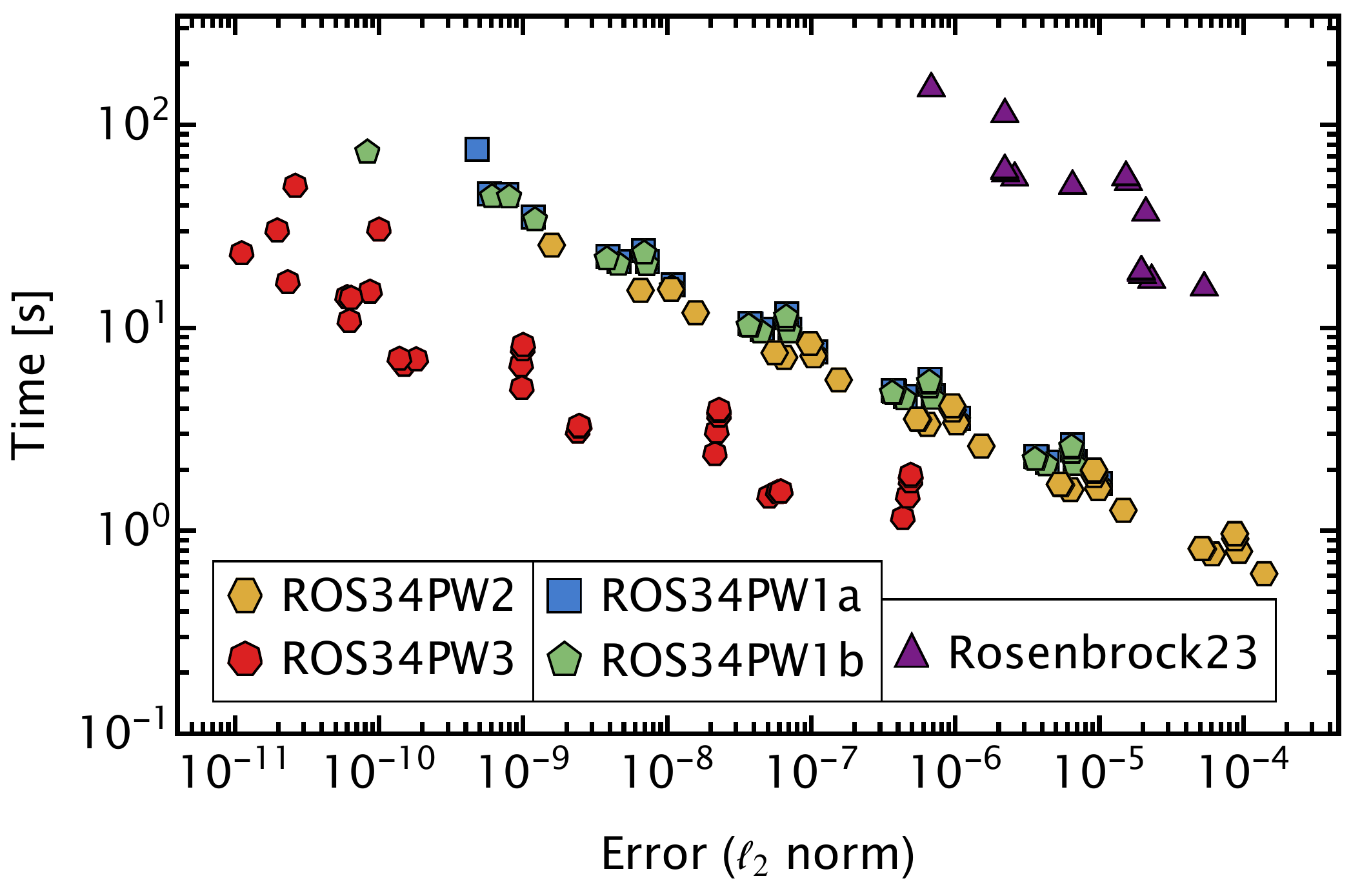}
		\caption{Rosenbrock-W}
		\label{fig:log_WP_RBW}
	\end{subfigure}	~
	\begin{subfigure}[t]{0.45\textwidth}
		\centering
		\includegraphics[width=\linewidth]{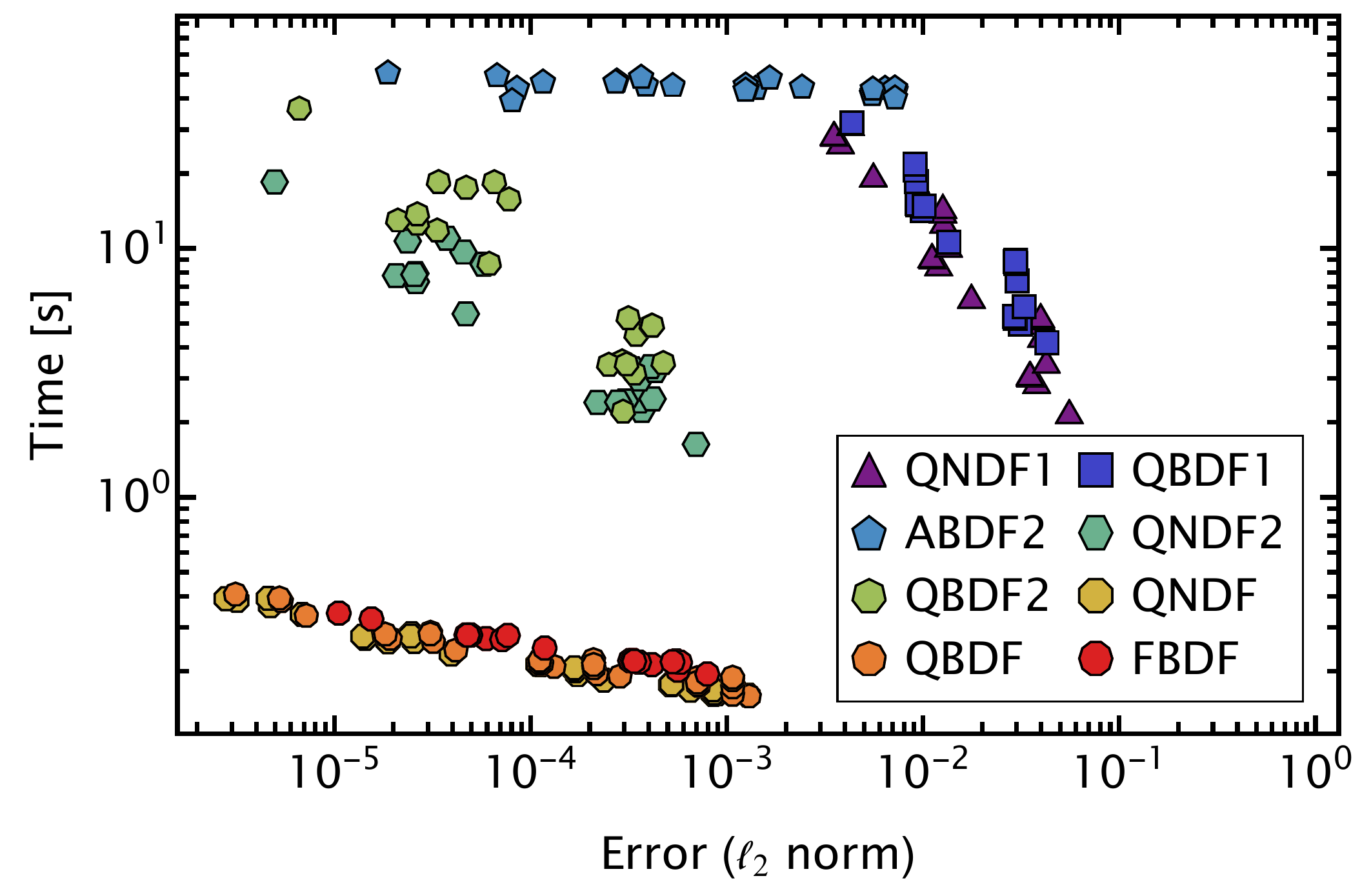}
		\caption{Implicit multistep}
		\label{fig:log_WP_IM}
	\end{subfigure}
	\begin{subfigure}[t]{0.45\textwidth}
		\centering
		\includegraphics[width=\linewidth]{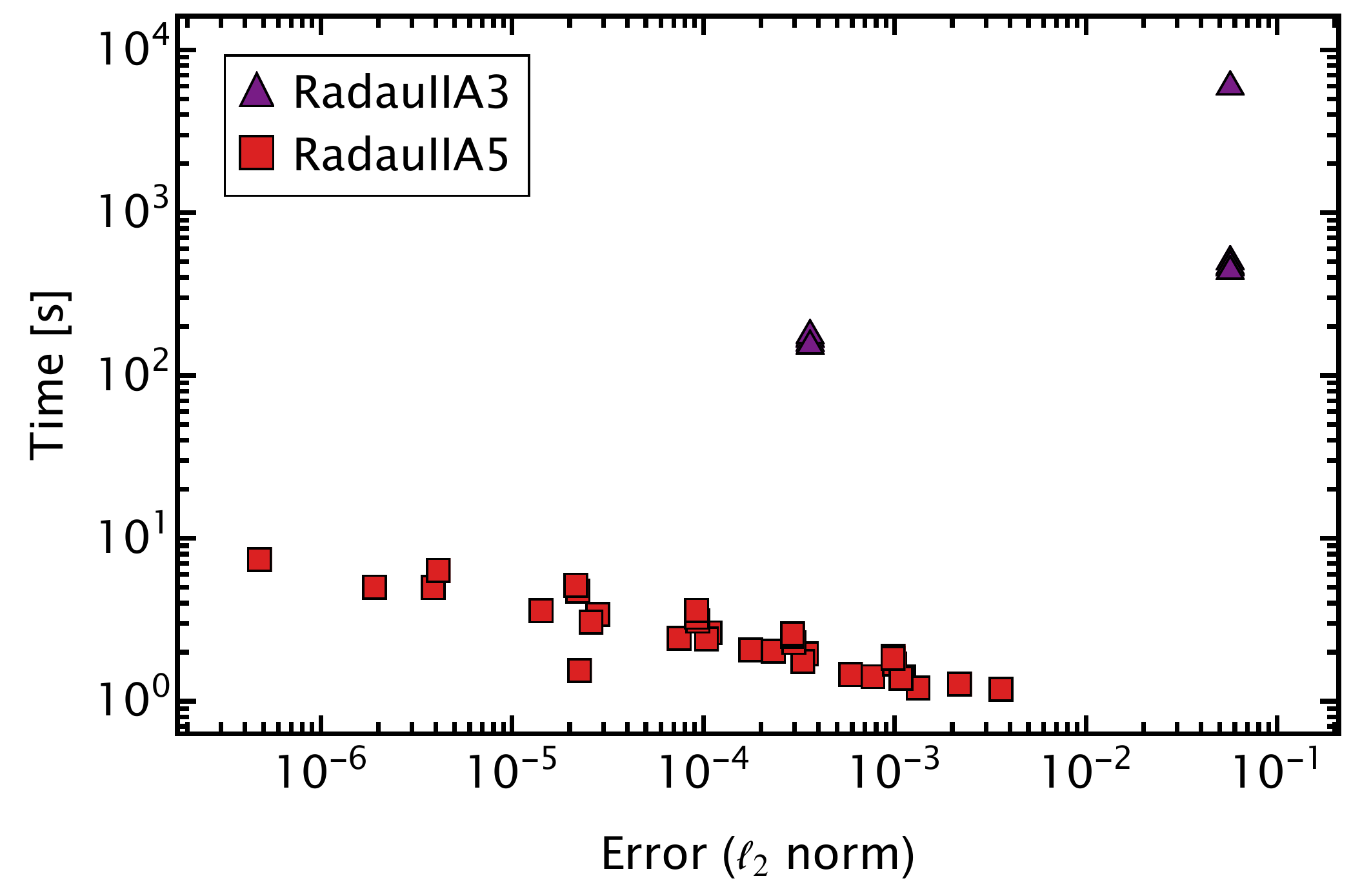}
		\caption{RadauII}
		\label{fig:log_WP_RadauII}
	\end{subfigure}	~
	\begin{subfigure}[t]{0.45\textwidth}
		\centering
		\includegraphics[width=\linewidth]{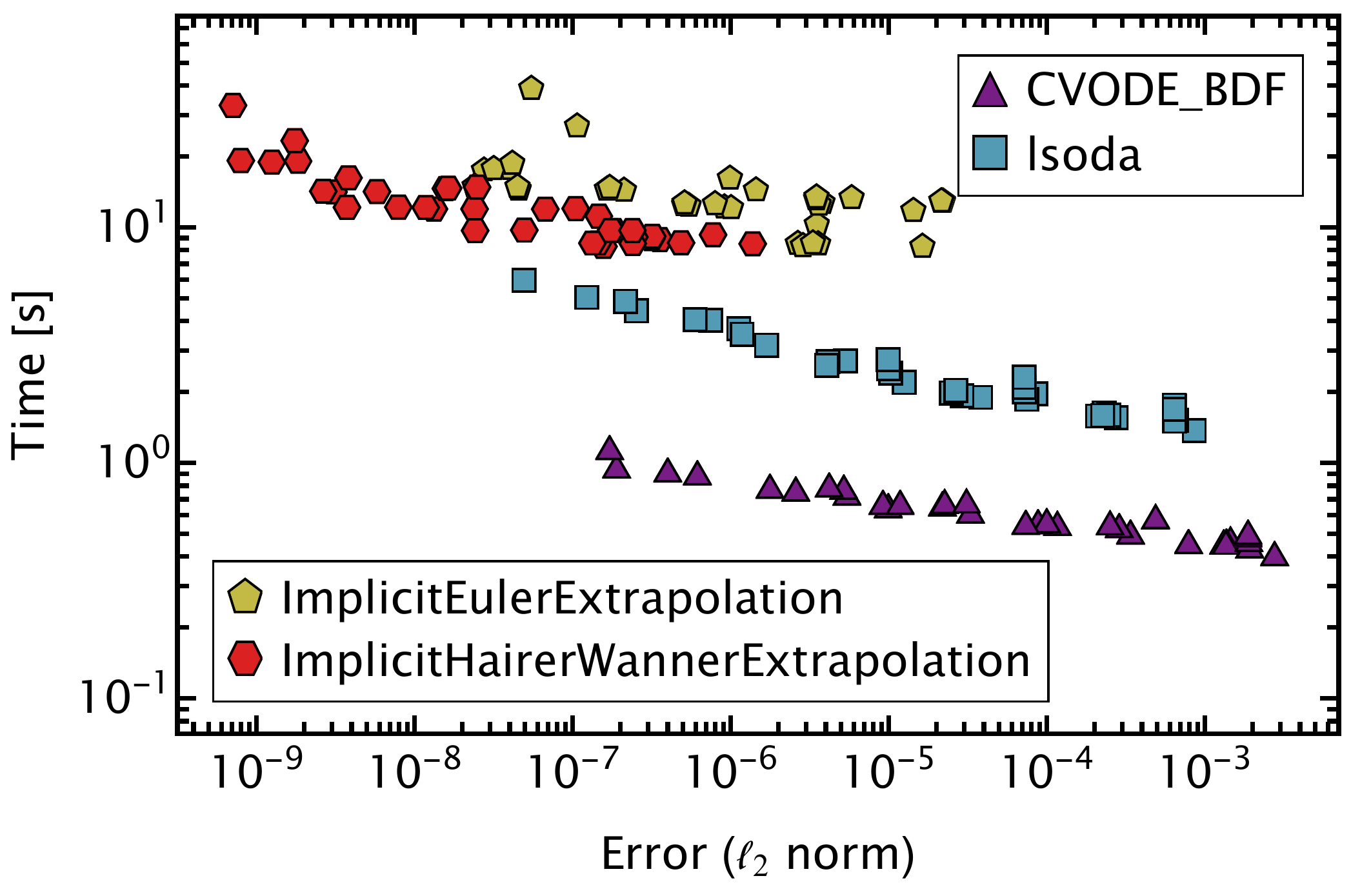}
		\caption{External libraries \& implicit extrapolation}
		\label{fig:log_WP_ext}
	\end{subfigure}%
	\caption{Work-precision survey for the logarithmic formulation \labelcref{eq:flow_logi}.}
	\label{fig:log_WP_all}
	\vspace{-30pt} 
\end{figure*}
%
\begin{figure*}[h!]
	\centering
	\begin{subfigure}[t]{0.45\textwidth}
		\centering
		\includegraphics[width=\linewidth]{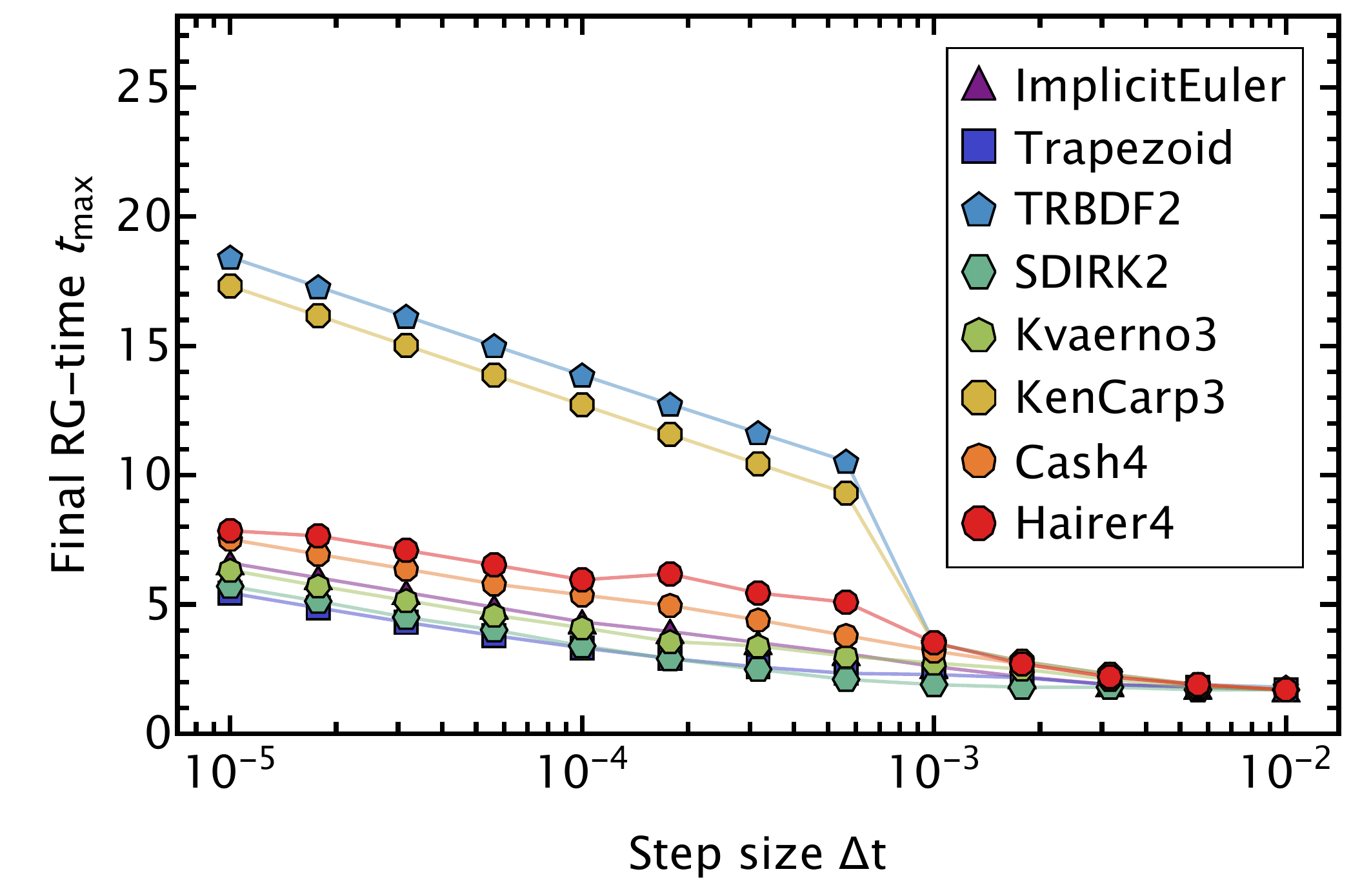}
		\caption{DIRK 1}
		\label{fig:std_tmax_DIRK1}
	\end{subfigure}	~
	\begin{subfigure}[t]{0.45\textwidth}
		\centering
		\includegraphics[width=\linewidth]{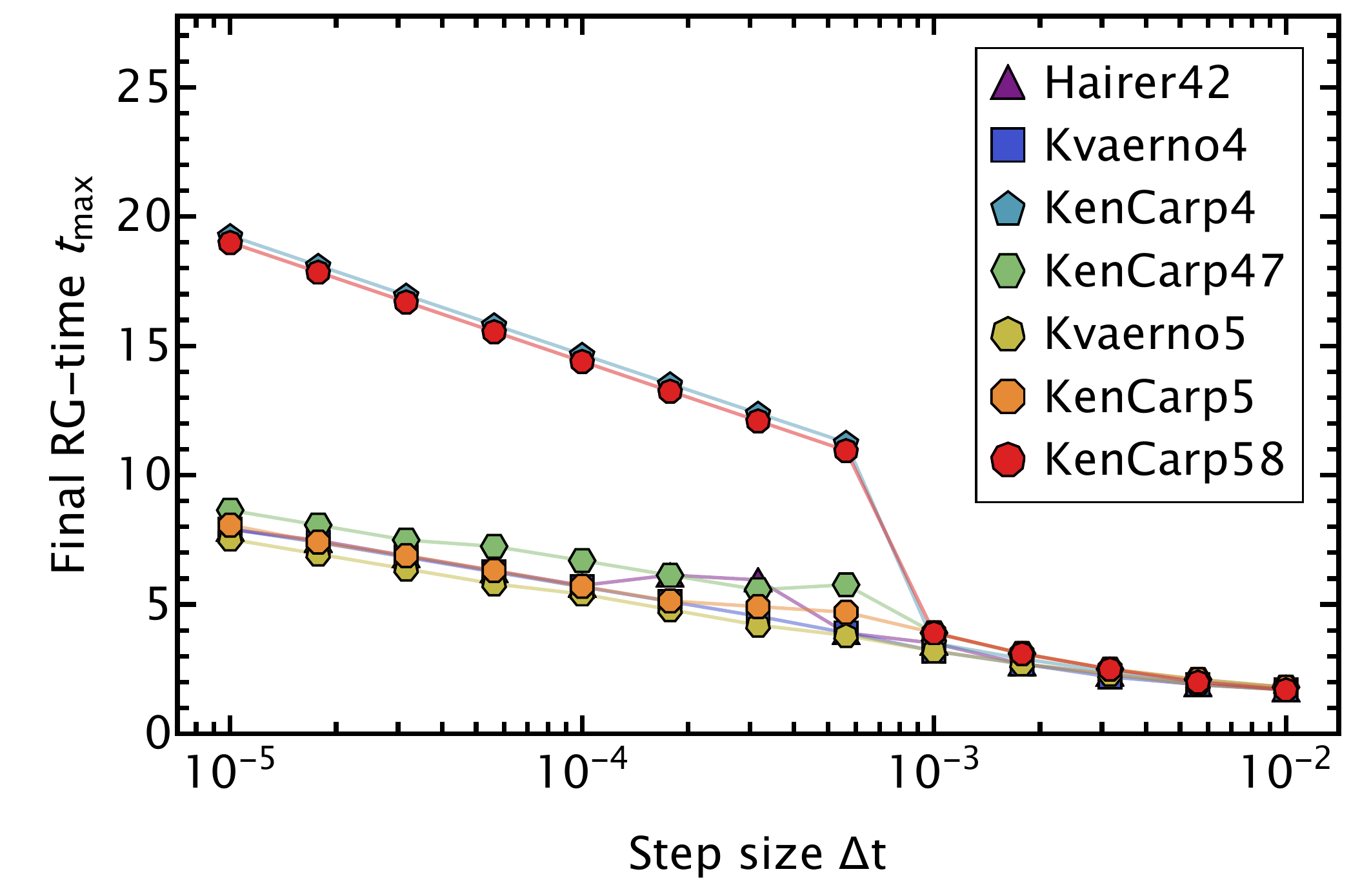}
		\caption{DIRK 2}
		\label{fig:std_tmax_DIRK2}
	\end{subfigure}
	\begin{subfigure}[t]{0.45\textwidth}
		\centering
		\includegraphics[width=\linewidth]{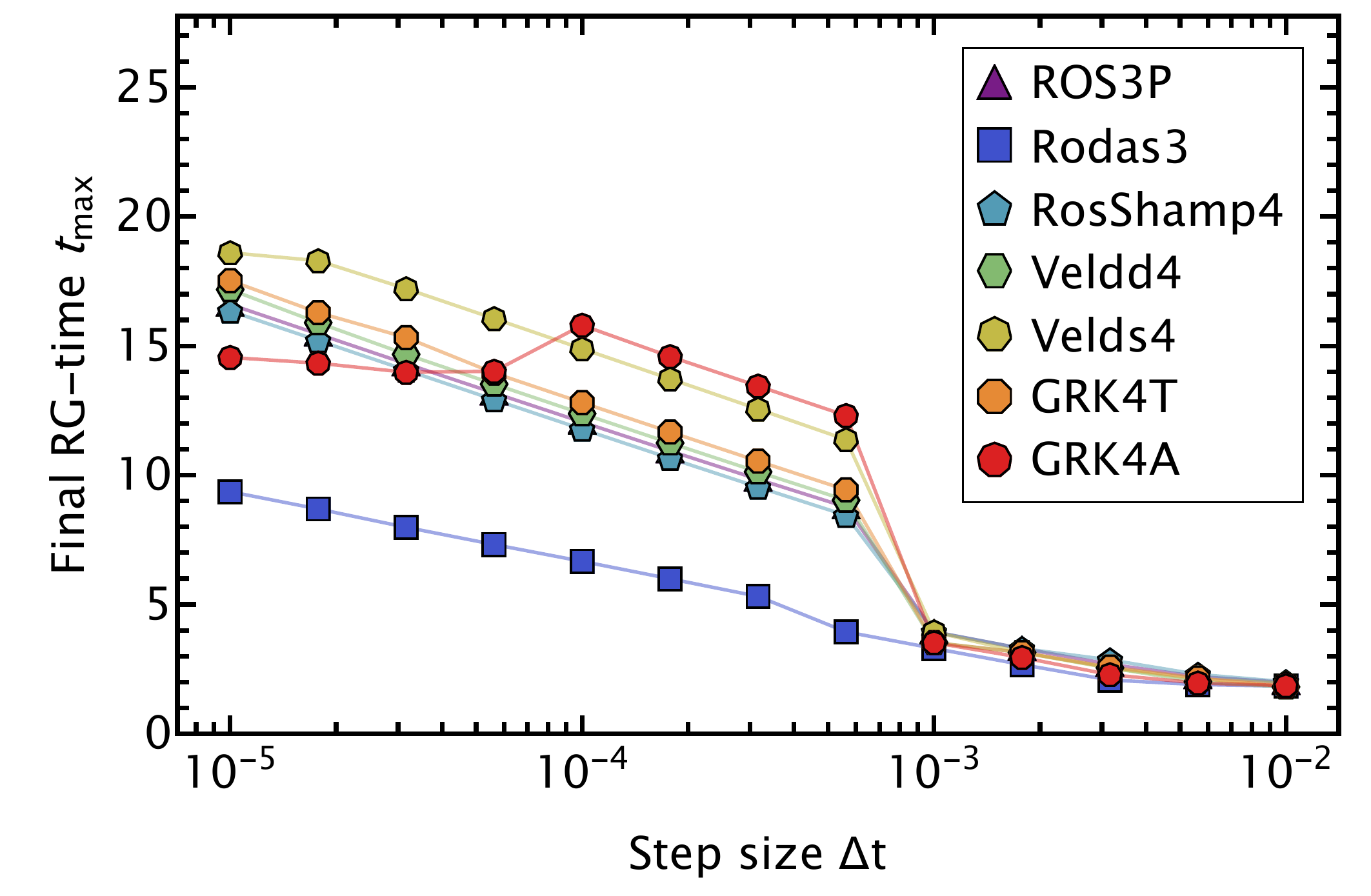}
		\caption{Rosenbrock 1}
		\label{fig:std_tmax_RB1}
	\end{subfigure}	~
	\begin{subfigure}[t]{0.45\textwidth}
		\centering
		\includegraphics[width=\linewidth]{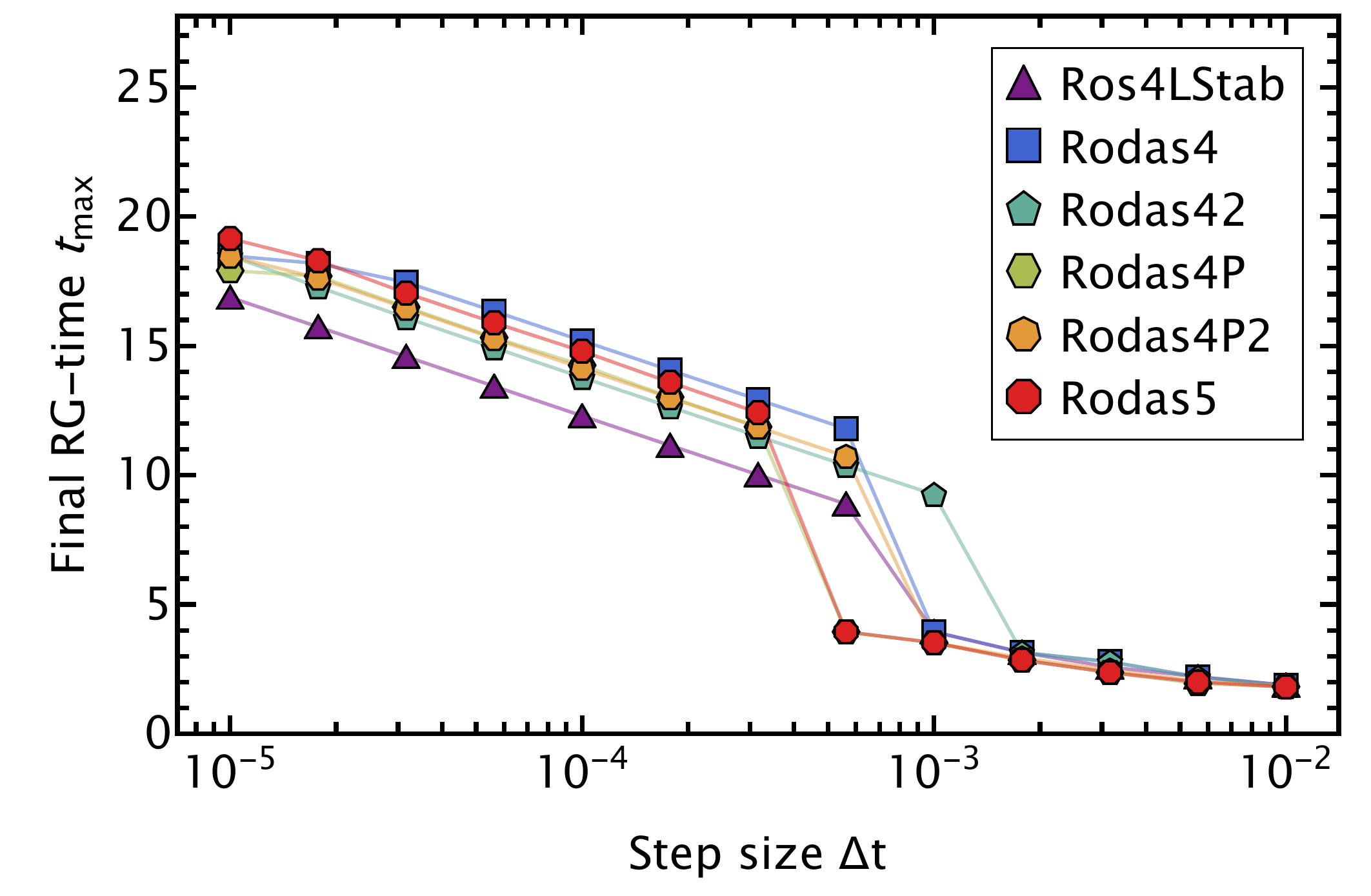}
		\caption{Rosenbrock 2}
		\label{fig:std_tmax_RB2}
	\end{subfigure}
		\begin{subfigure}[t]{0.45\textwidth}
		\centering
		\includegraphics[width=\linewidth]{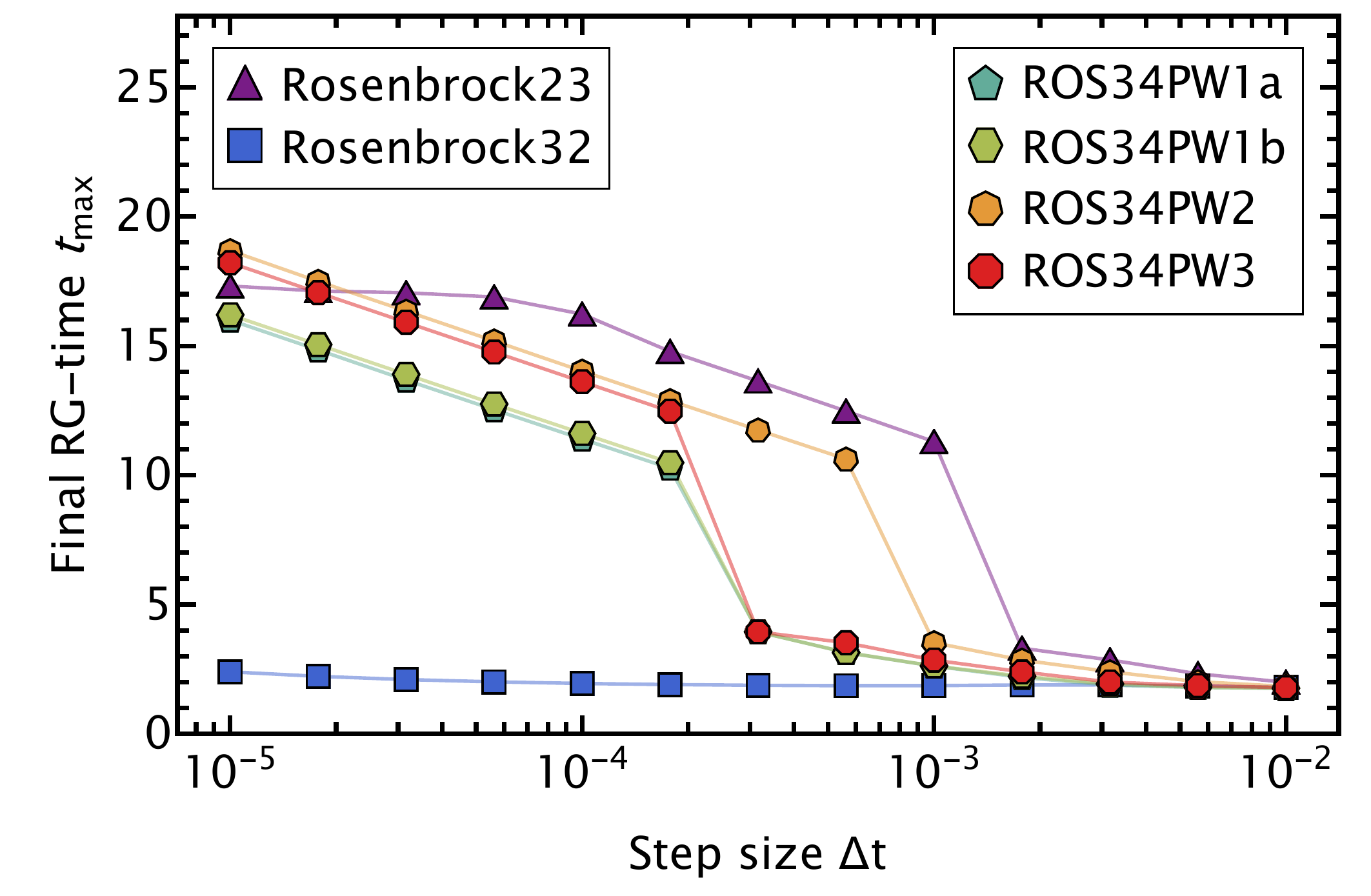}
		\caption{Rosenbrock-W}
		\label{fig:std_tmax_RBW}
	\end{subfigure}	~
	\begin{subfigure}[t]{0.45\textwidth}
		\centering
		\includegraphics[width=\linewidth]{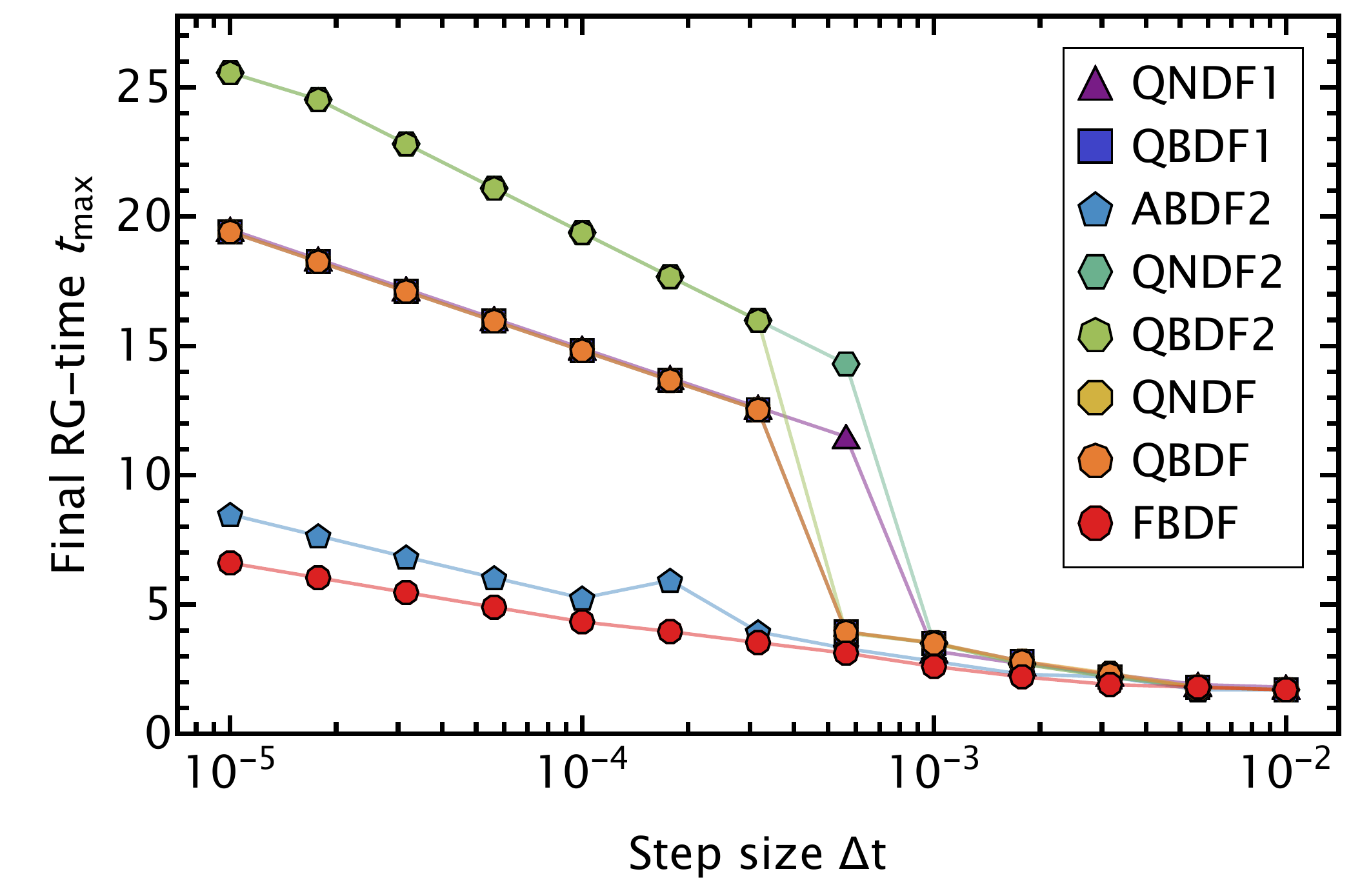}
		\caption{Implicit multistep}
		\label{fig:std_tmax_IM}
	\end{subfigure}
	\begin{subfigure}[t]{0.45\textwidth}
		\centering
		\includegraphics[width=\linewidth]{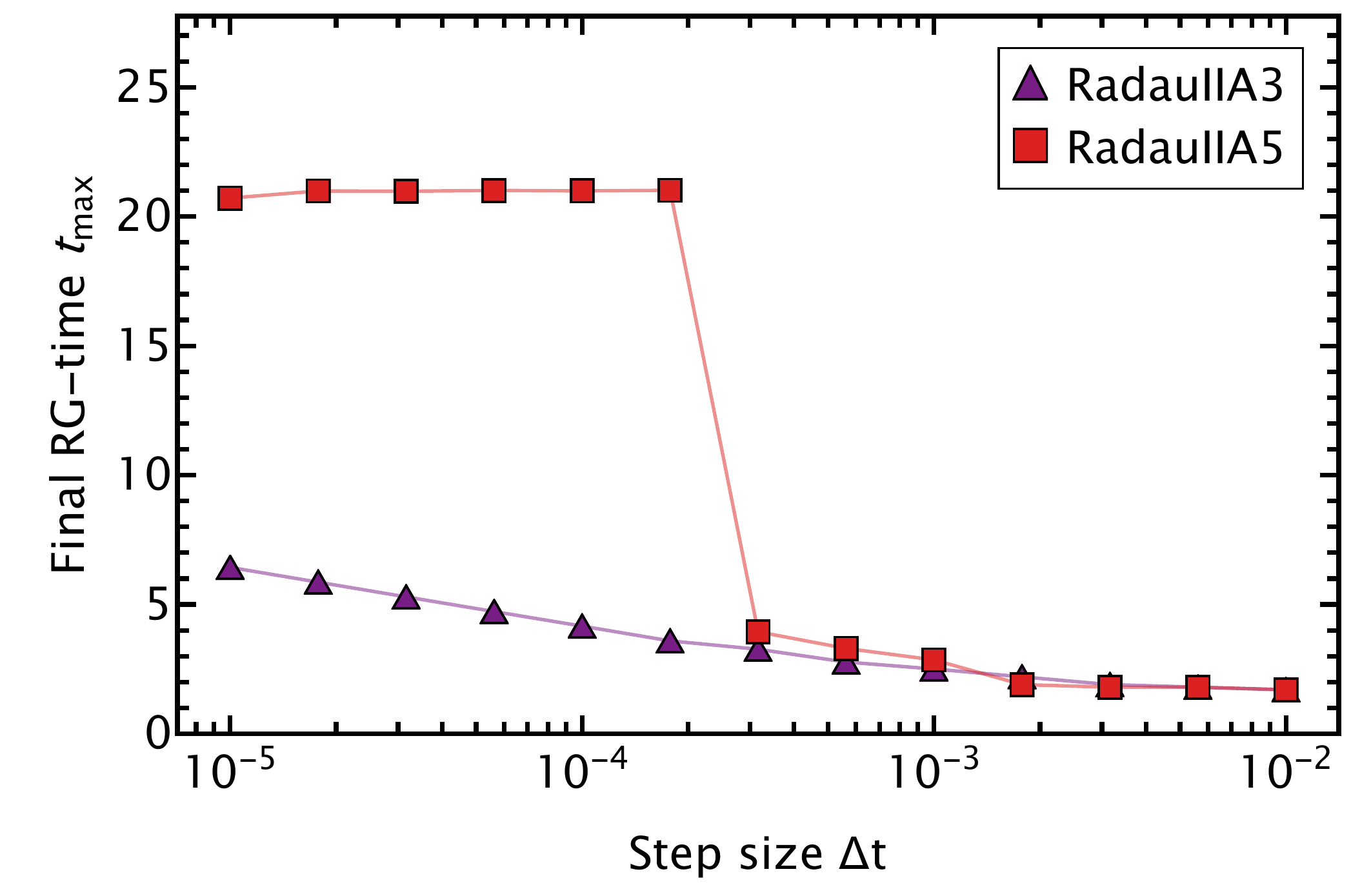}
		\caption{RadauII}
		\label{fig:std_tmax_RadauII}
	\end{subfigure}	~
	\begin{subfigure}[t]{0.45\textwidth}
		\centering
		\includegraphics[width=\linewidth]{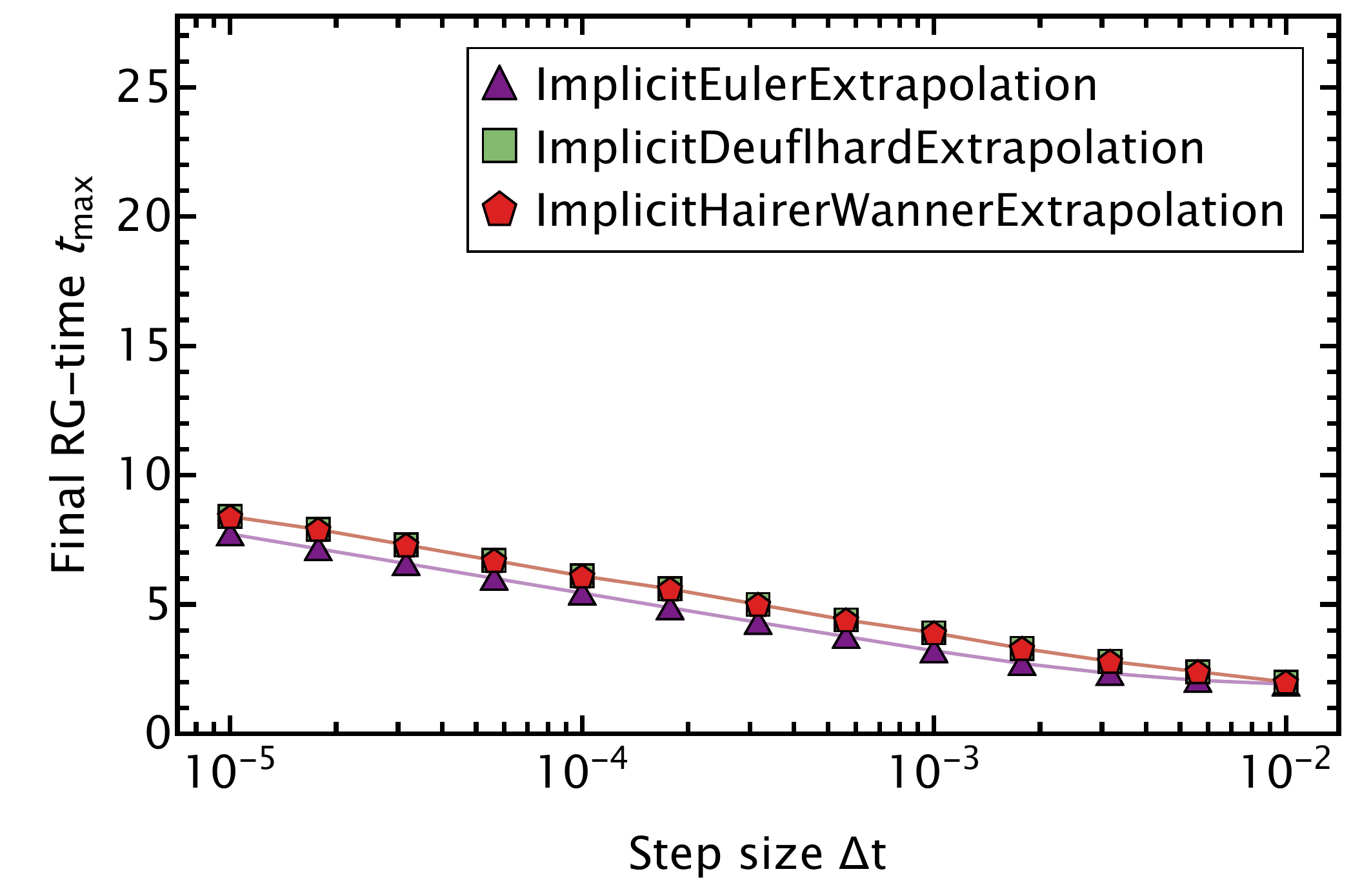}
		\caption{Implicit extrapolation}
		\label{fig:std_tmax_IE}
	\end{subfigure}%
	\caption{Fixed step size RG-time stability survey for the standard formulation \labelcref{eq:flow_ui}.}
	\label{fig:std_tmax_all}
	\vspace{-30pt}
\end{figure*}
%
\begin{figure*}[h!]
	\centering
	\begin{subfigure}[t]{0.45\textwidth}
		\centering
		\includegraphics[width=\linewidth]{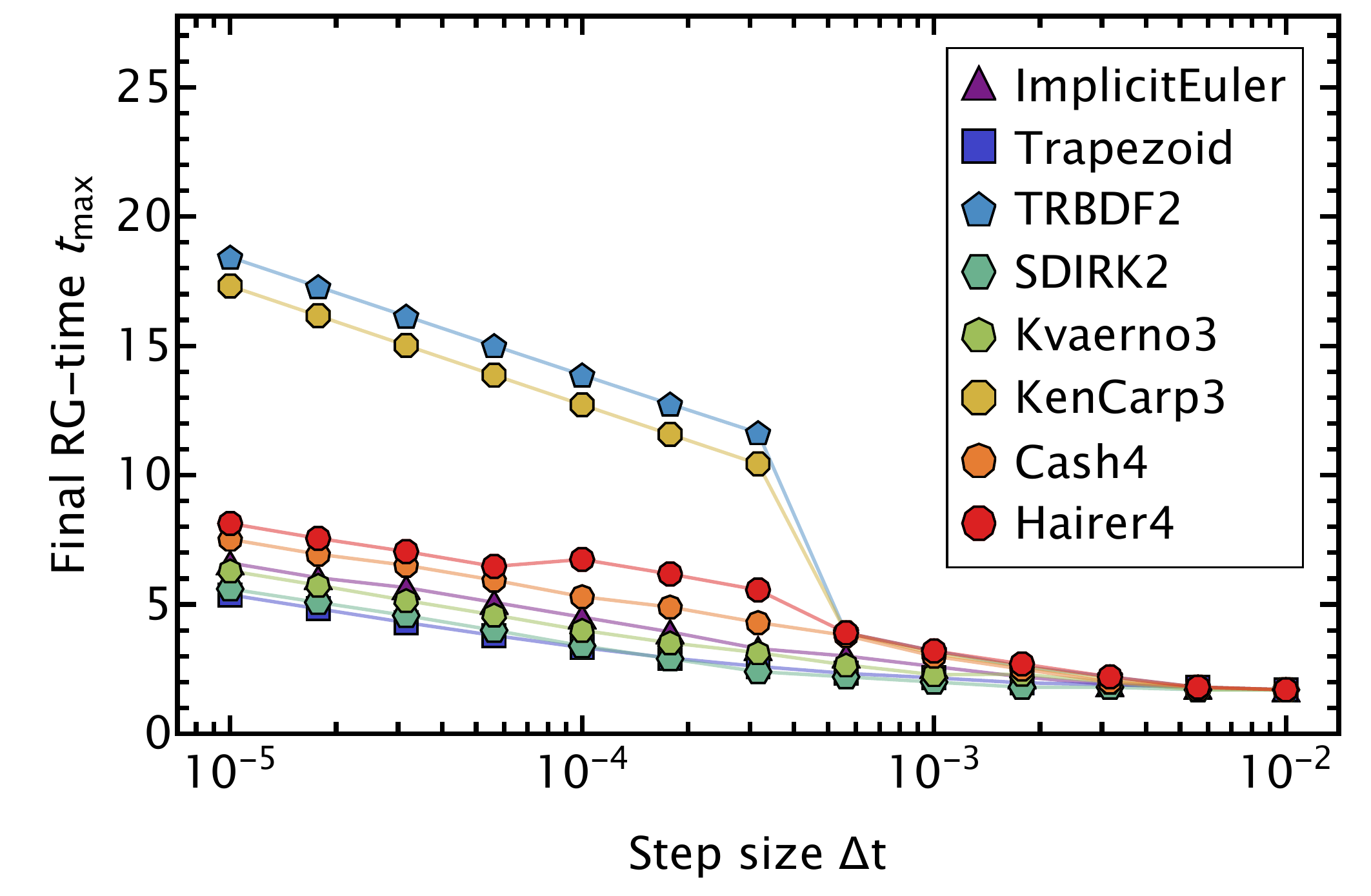}
		\caption{DIRK 1}
		\label{fig:msq_tmax_DIRK1}
	\end{subfigure}	~
	\begin{subfigure}[t]{0.45\textwidth}
		\centering
		\includegraphics[width=\linewidth]{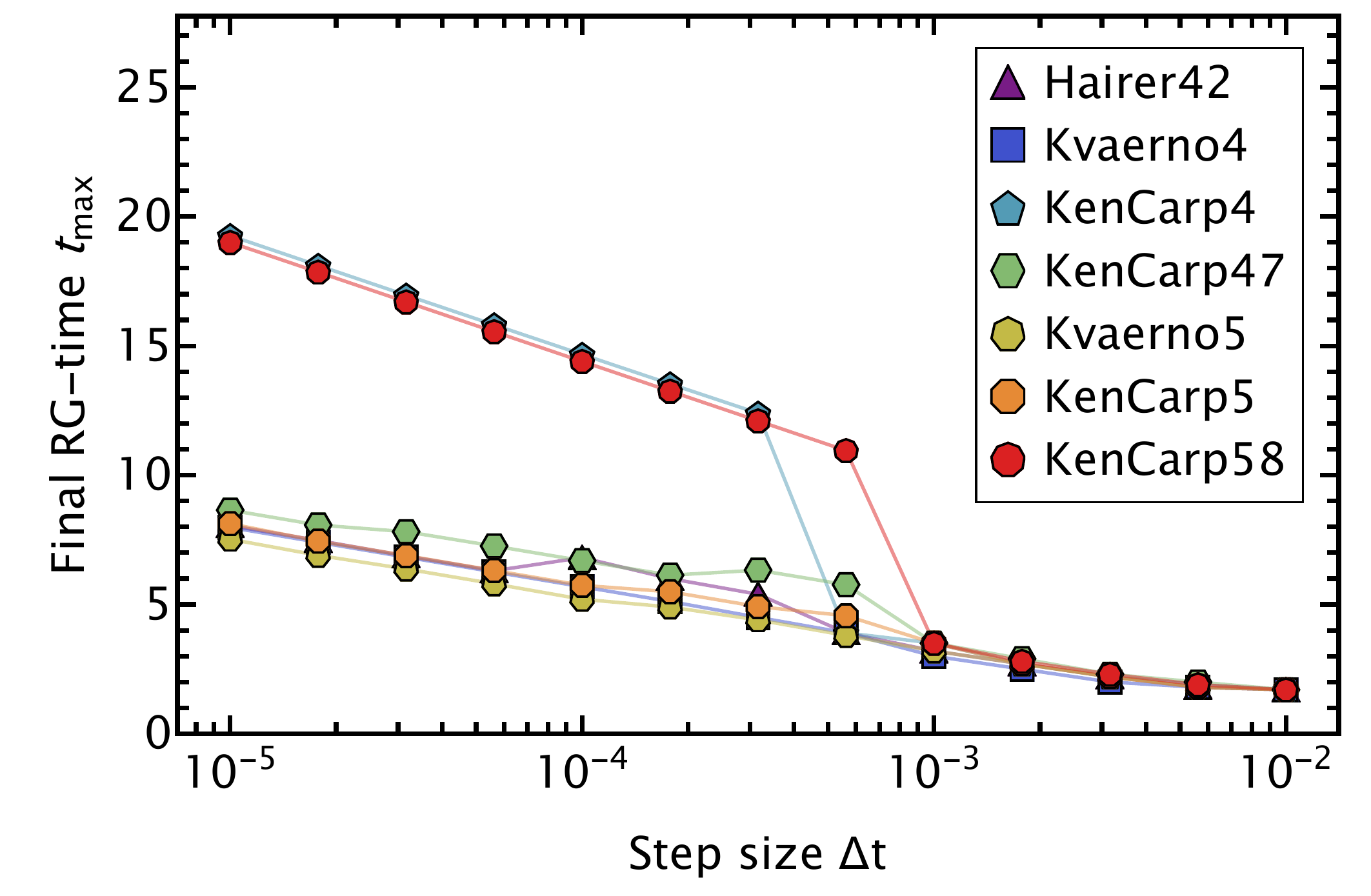}
		\caption{DIRK 2}
		\label{fig:msq_tmax_DIRK2}
	\end{subfigure}
	\begin{subfigure}[t]{0.45\textwidth}
		\centering
		\includegraphics[width=\linewidth]{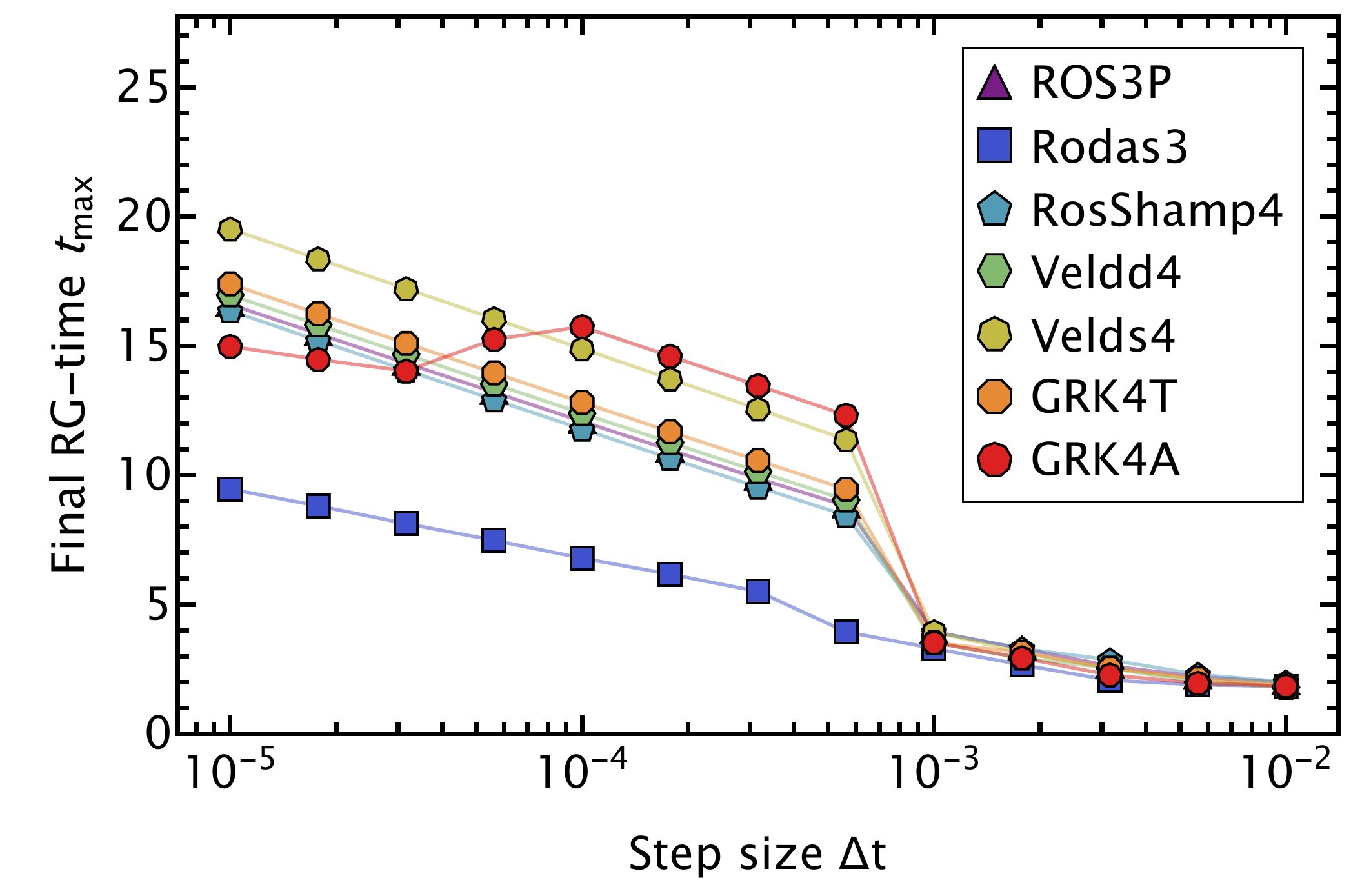}
		\caption{Rosenbrock 1}
		\label{fig:msq_tmax_RB1}
	\end{subfigure}	~
	\begin{subfigure}[t]{0.45\textwidth}
		\centering
		\includegraphics[width=\linewidth]{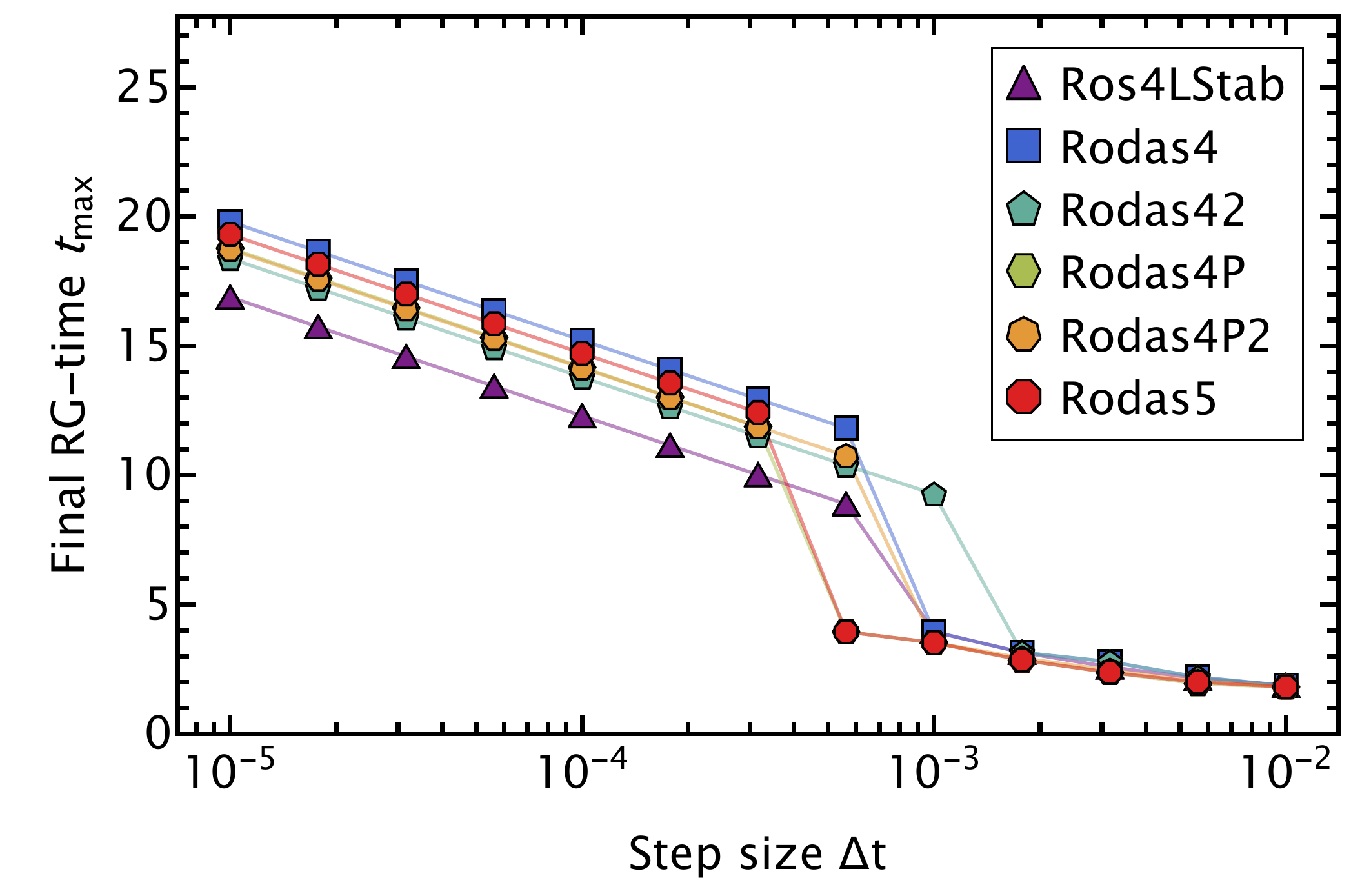}
		\caption{Rosenbrock 2}
		\label{fig:msq_tmax_RB2}
	\end{subfigure}
		\begin{subfigure}[t]{0.45\textwidth}
		\centering
		\includegraphics[width=\linewidth]{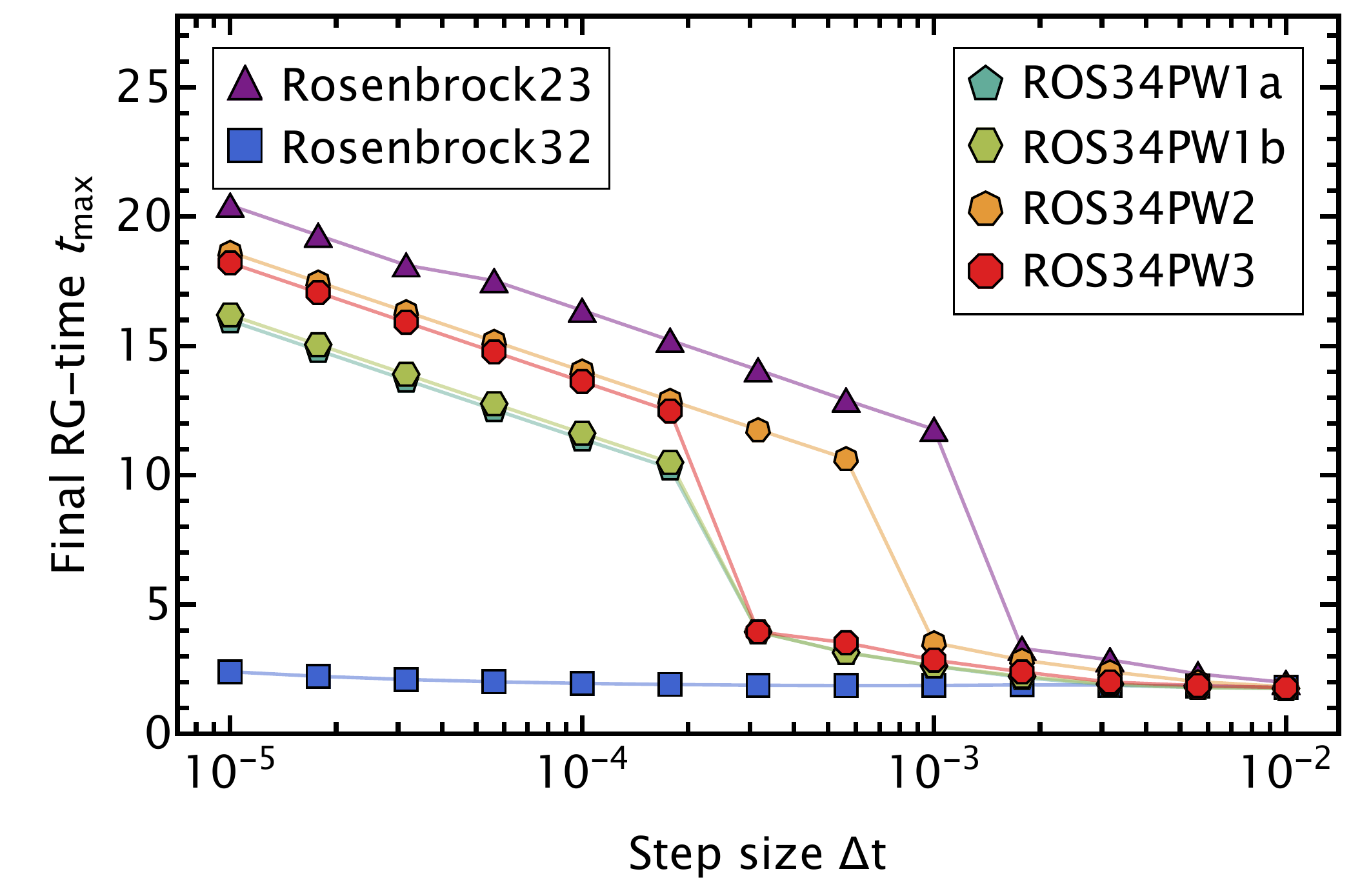}
		\caption{Rosenbrock-W}
		\label{fig:msq_tmax_RBW}
	\end{subfigure}	~
	\begin{subfigure}[t]{0.45\textwidth}
		\centering
		\includegraphics[width=\linewidth]{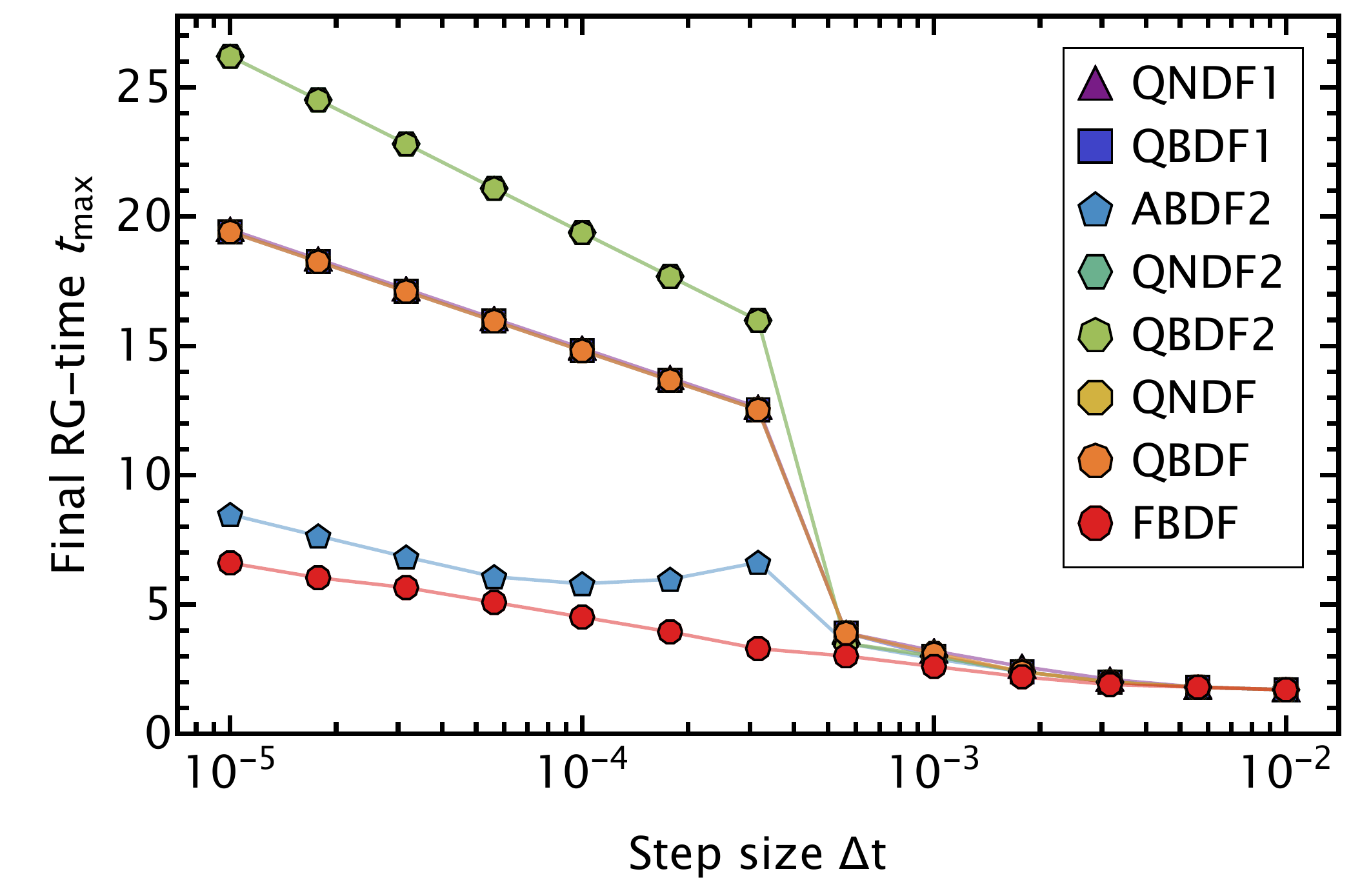}
		\caption{Implicit multistep}
		\label{fig:msq_tmax_IM}
	\end{subfigure}
	\begin{subfigure}[t]{0.45\textwidth}
		\centering
		\includegraphics[width=\linewidth]{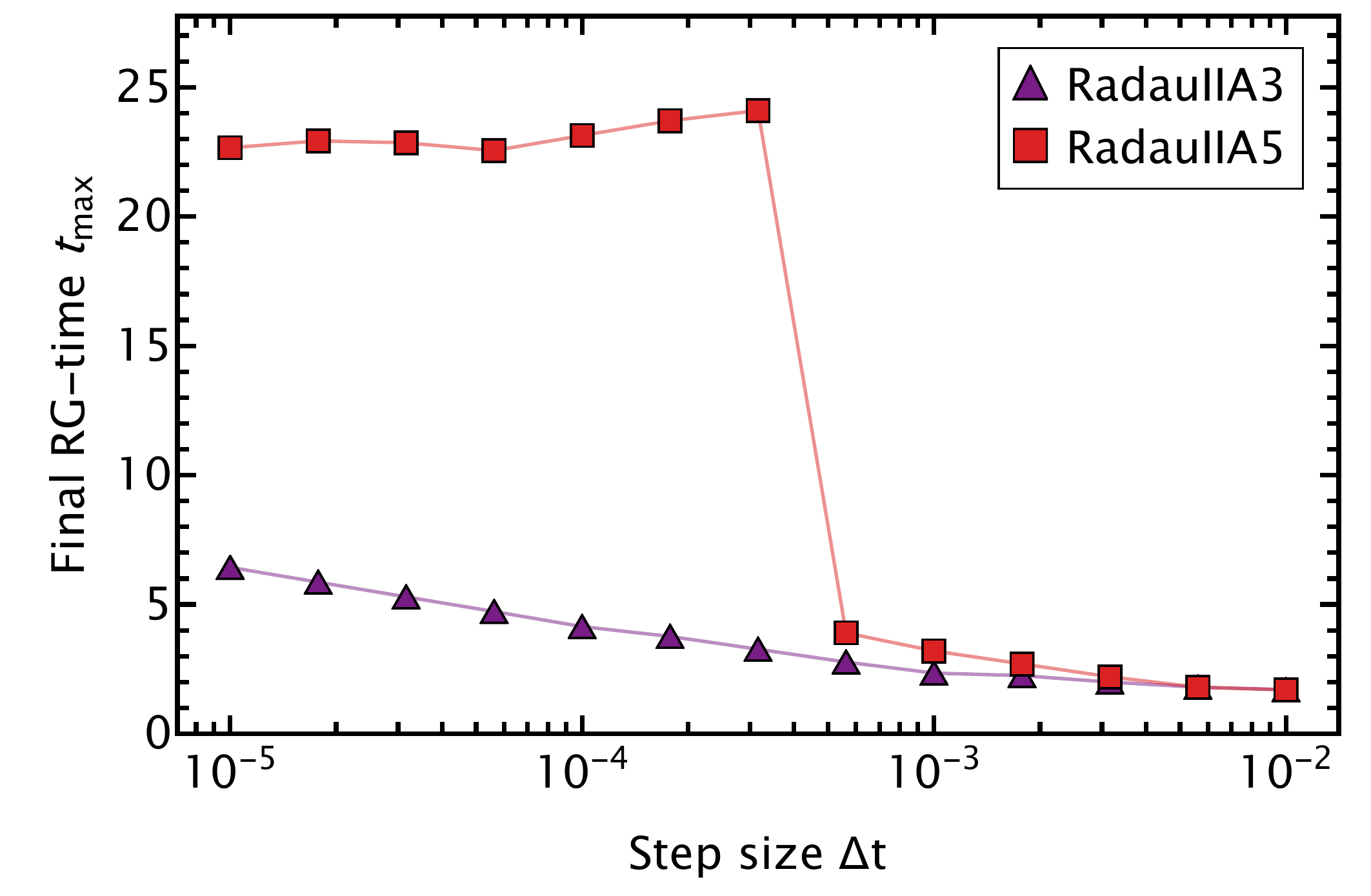}
		\caption{RadauII}
		\label{fig:msq_tmax_RadauII}
	\end{subfigure}	~
	\begin{subfigure}[t]{0.45\textwidth}
		\centering
		\includegraphics[width=\linewidth]{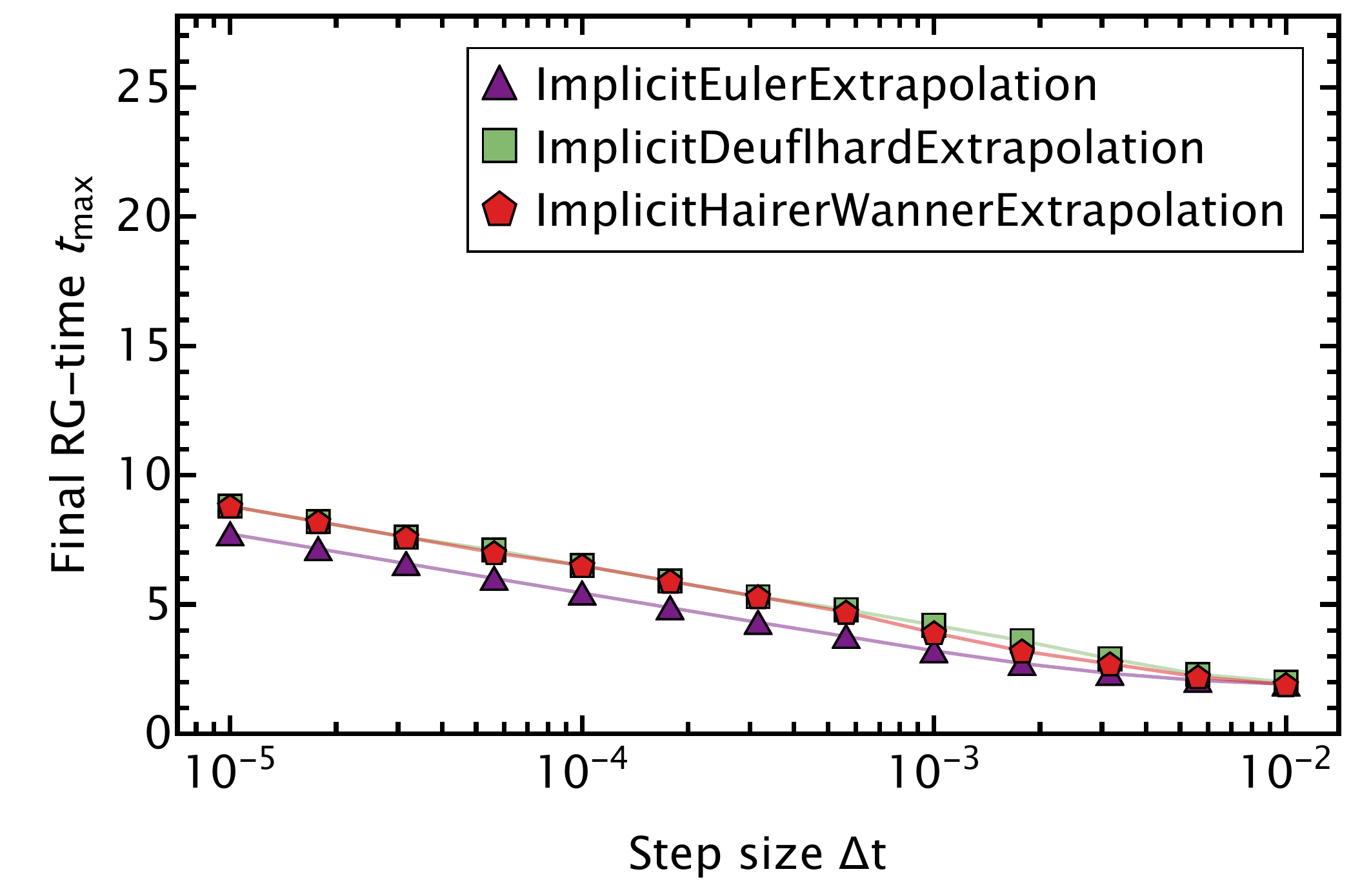}
		\caption{Implicit extrapolation}
		\label{fig:msq_tmax_IE}
	\end{subfigure}%
	\caption{Fixed step size RG-time stability survey for the mass formulation \labelcref{eq:flow_msqi}.}
	\label{fig:msq_tmax_all}
	\vspace{-30pt}
\end{figure*}
%
\begin{figure*}[h!]
	\centering
	\begin{subfigure}[t]{0.45\textwidth}
		\centering
		\includegraphics[width=\linewidth]{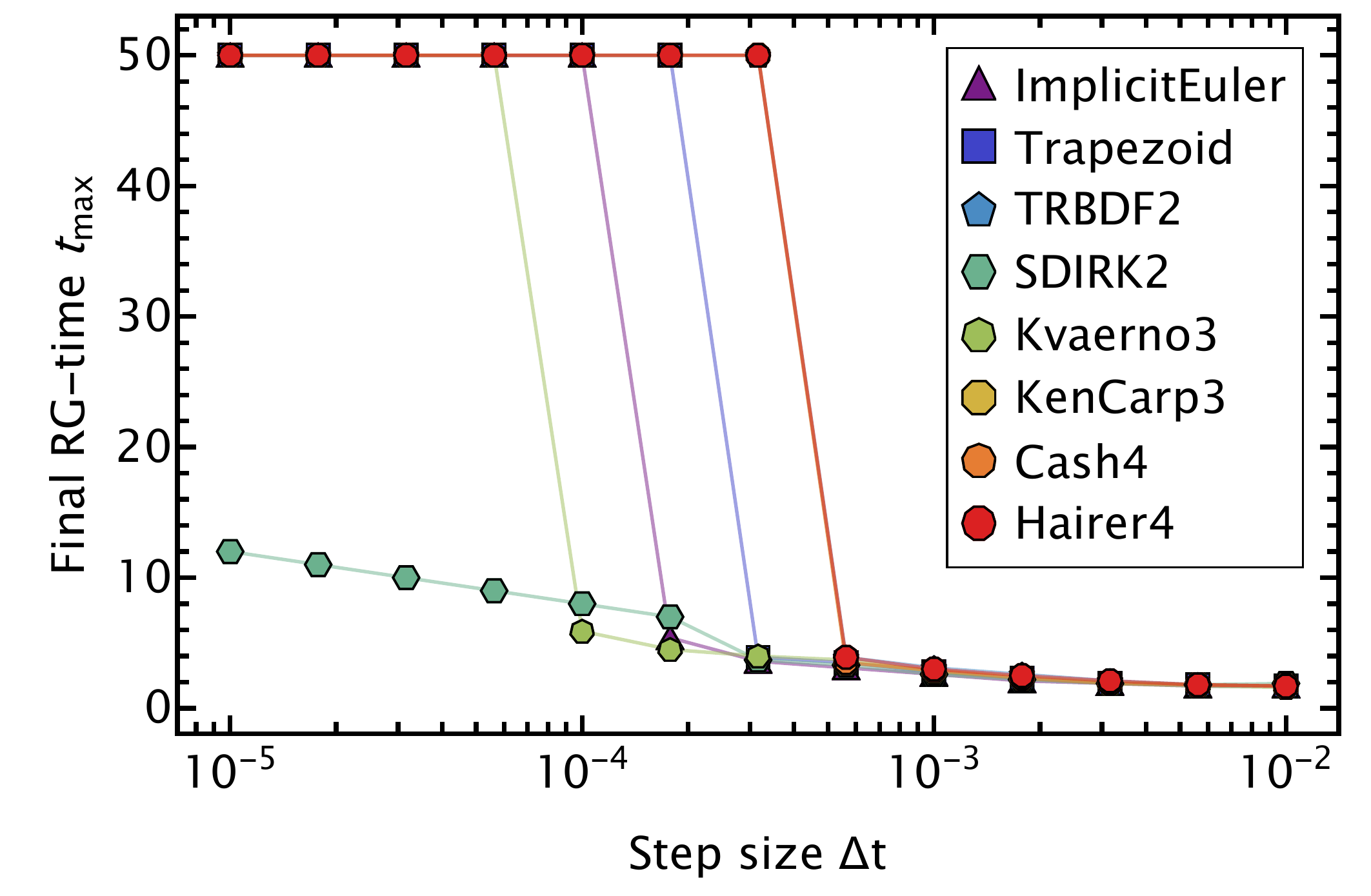}
		\caption{DIRK 1}
		\label{fig:log_tmax_DIRK1}
	\end{subfigure}	~
	\begin{subfigure}[t]{0.45\textwidth}
		\centering
		\includegraphics[width=\linewidth]{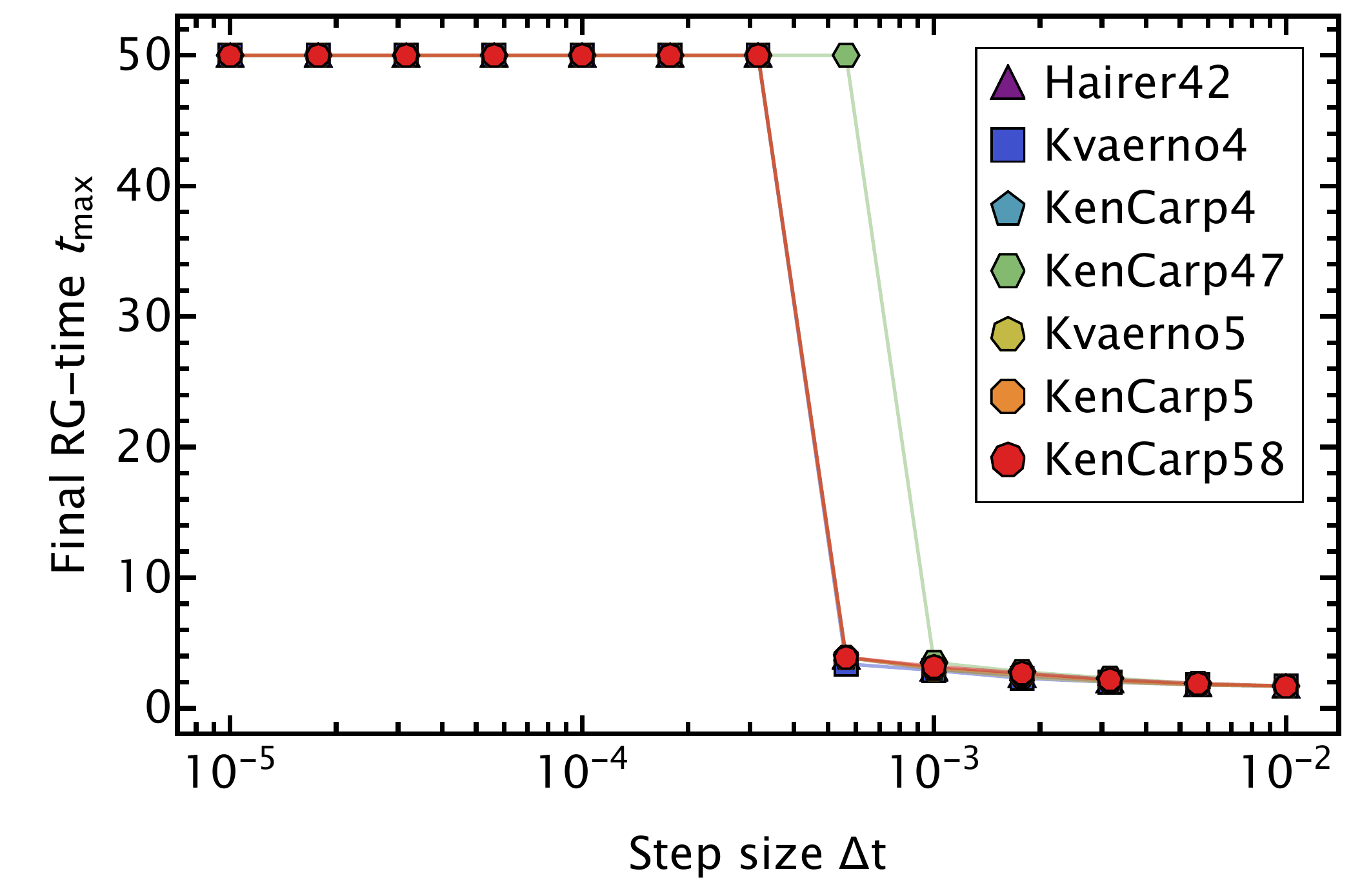}
		\caption{DIRK 2}
		\label{fig:log_tmax_DIRK2}
	\end{subfigure}
	\begin{subfigure}[t]{0.45\textwidth}
		\centering
		\includegraphics[width=\linewidth]{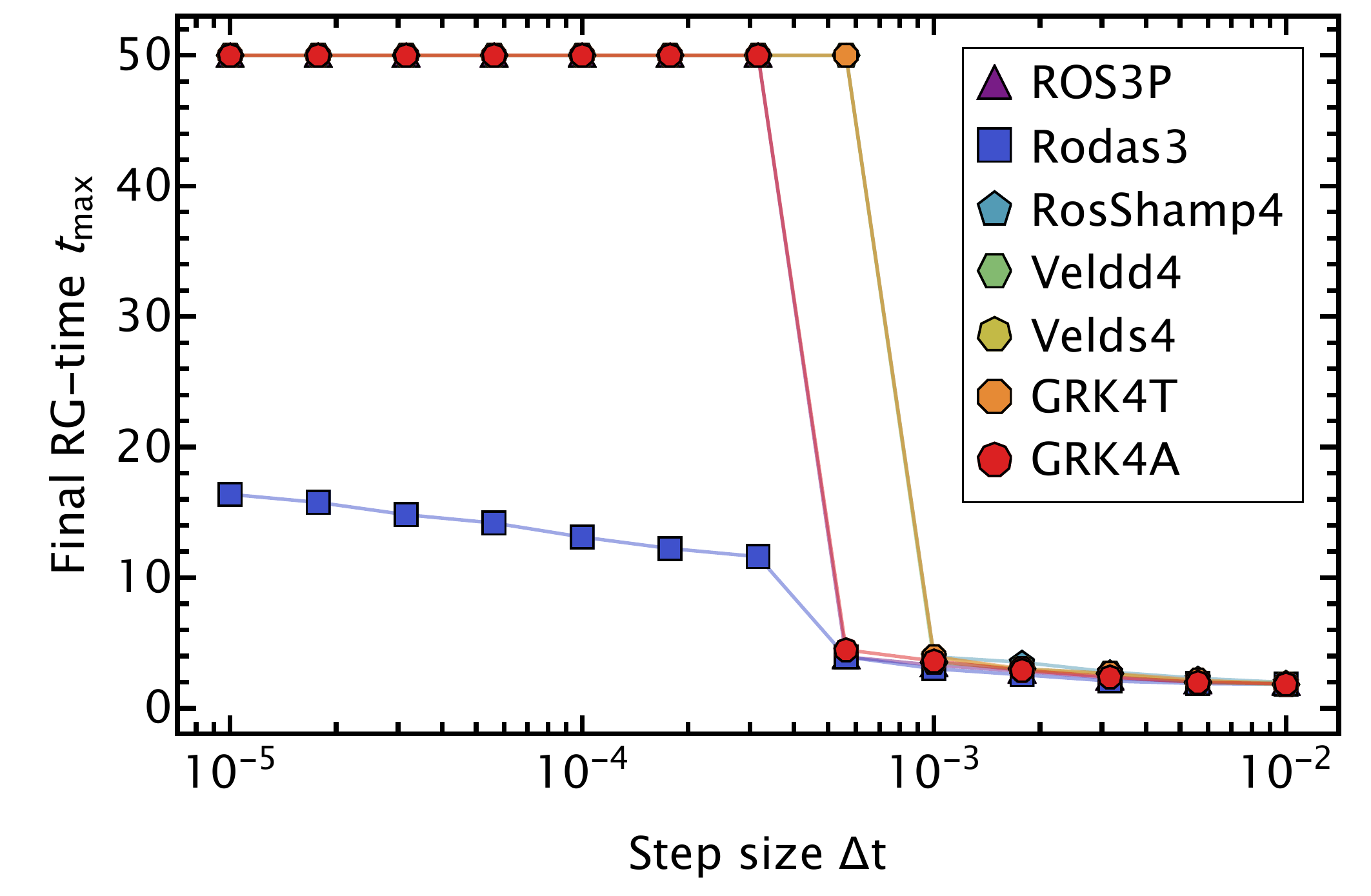}
		\caption{Rosenbrock 1}
		\label{fig:log_tmax_RB1}
	\end{subfigure}	~
	\begin{subfigure}[t]{0.45\textwidth}
		\centering
		\includegraphics[width=\linewidth]{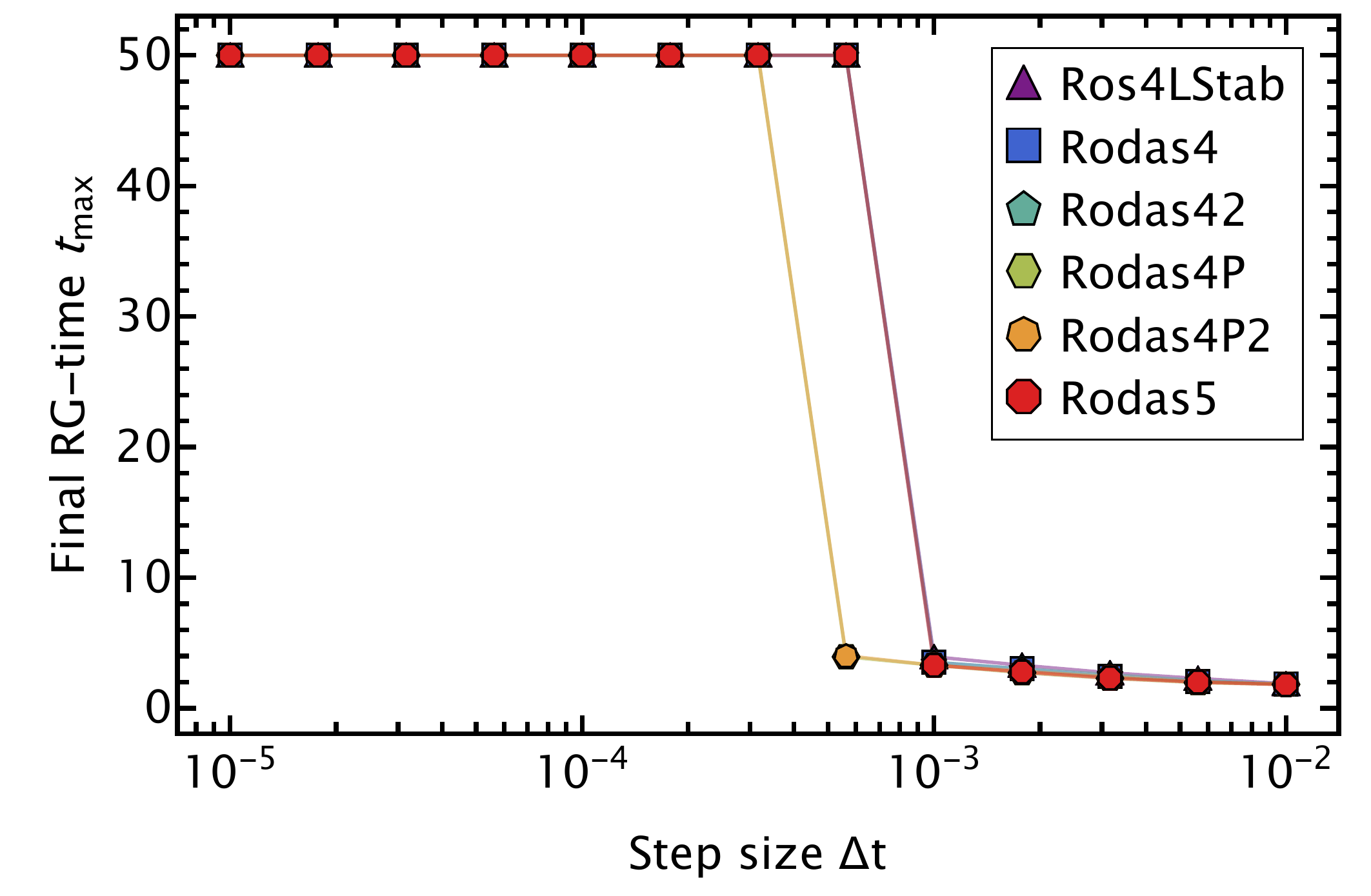}
		\caption{Rosenbrock 2}
		\label{fig:log_tmax_RB2}
	\end{subfigure}
		\begin{subfigure}[t]{0.45\textwidth}
		\centering
		\includegraphics[width=\linewidth]{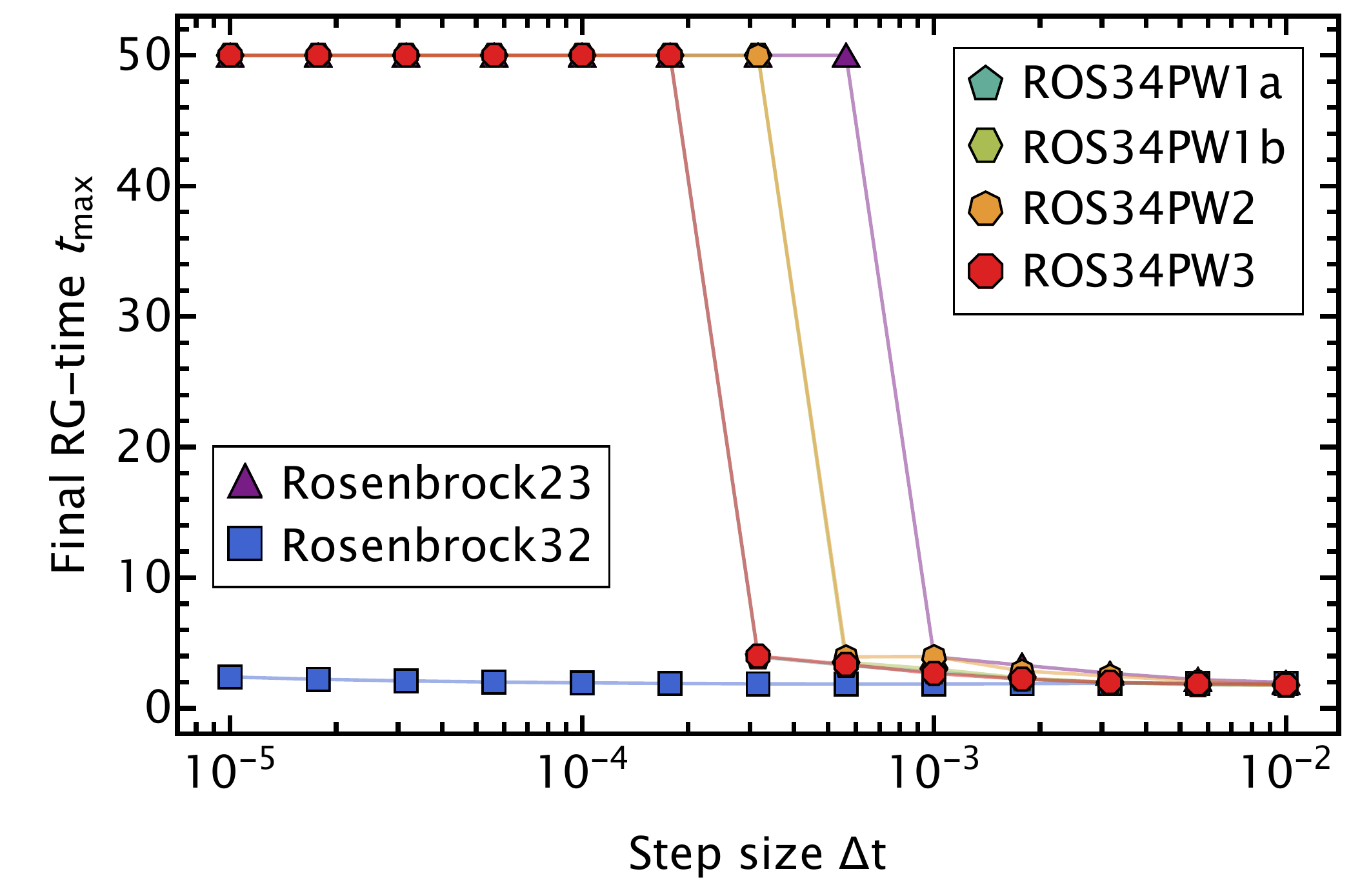}
		\caption{Rosenbrock-W}
		\label{fig:log_tmax_RBW}
	\end{subfigure}	~
	\begin{subfigure}[t]{0.45\textwidth}
		\centering
		\includegraphics[width=\linewidth]{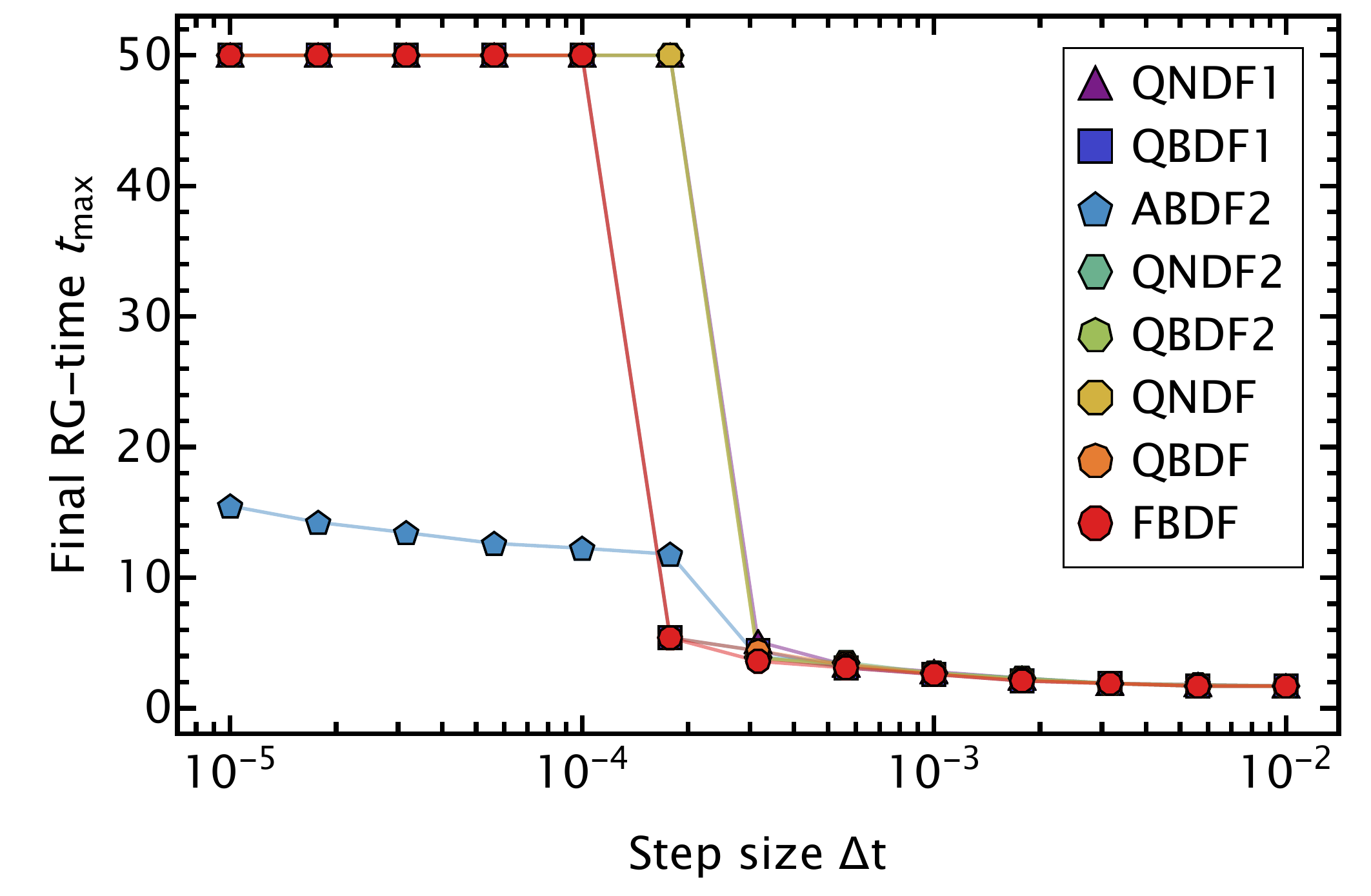}
		\caption{Implicit multistep}
		\label{fig:log_tmax_IM}
	\end{subfigure}
	\begin{subfigure}[t]{0.45\textwidth}
		\centering
		\includegraphics[width=\linewidth]{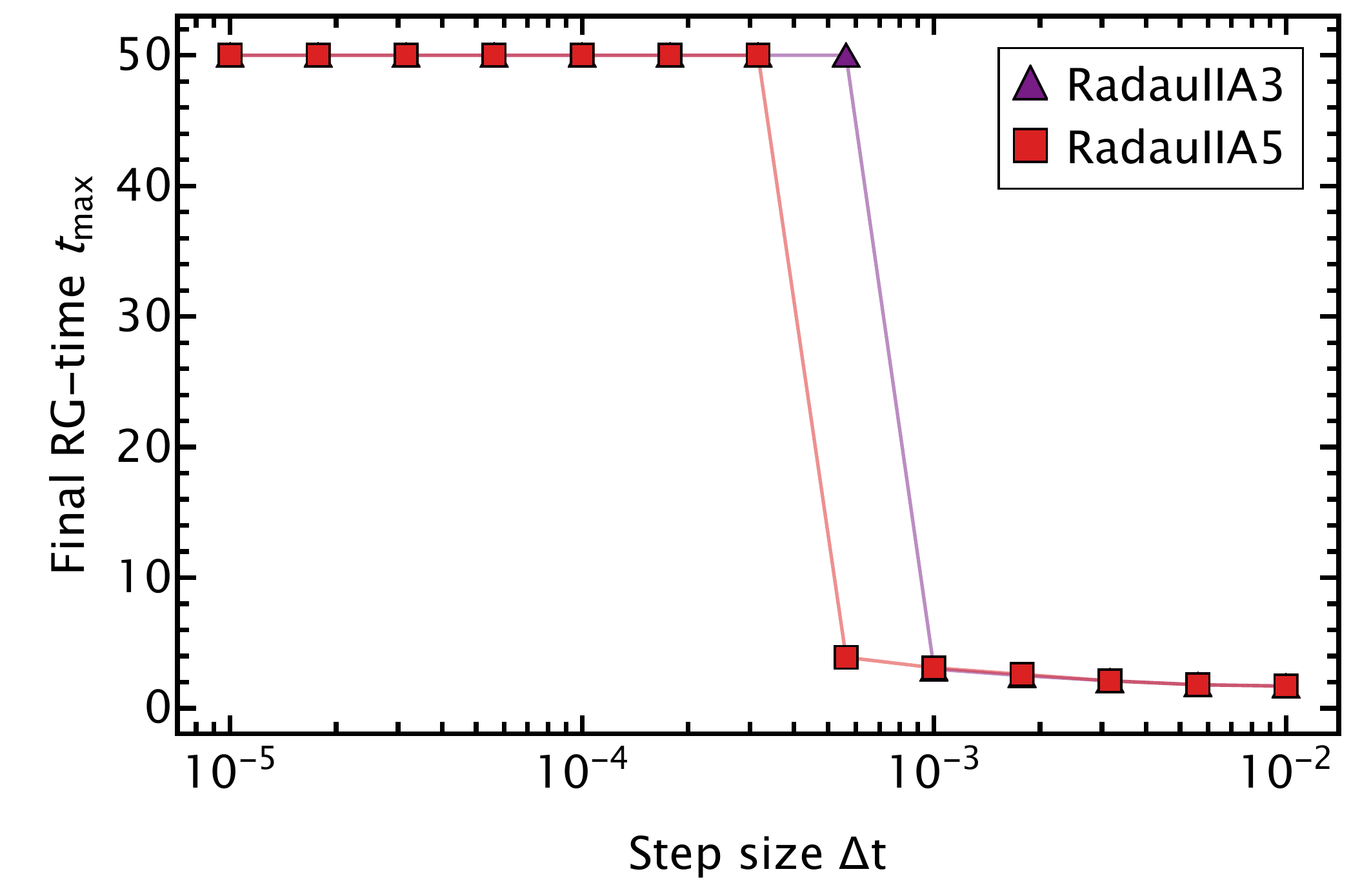}
		\caption{RadauII}
		\label{fig:log_tmax_RadauII}
	\end{subfigure}	~
	\begin{subfigure}[t]{0.45\textwidth}
		\centering
		\includegraphics[width=\linewidth]{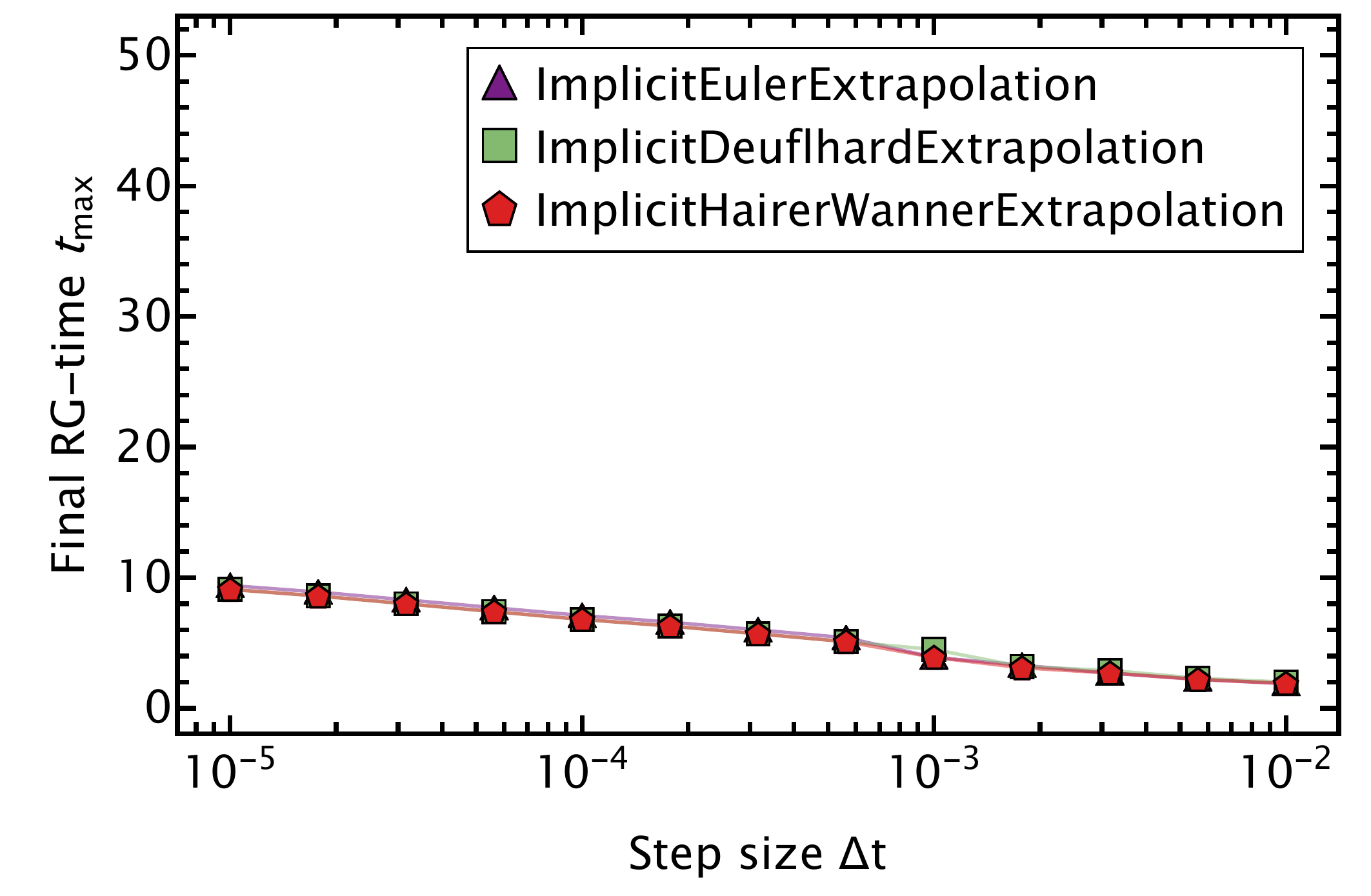}
		\caption{Implicit extrapolation}
		\label{fig:log_tmax_IE}
	\end{subfigure}%
	\caption{Fixed step size RG-time stability survey for the log formulation \labelcref{eq:flow_logi}. The maximum RG-time was set to $t=50$, which can, for all practical purposes, be identified with infinity.}
	\label{fig:log_tmax_all}
	\vspace{-30pt}
\end{figure*}

\clearpage
\bibliography{bib_master}

\end{document}